%% file: main.tex
\documentclass[11pt,dvipsnames,svgnames,x11names]{article}

\usepackage{pdfpages}
\usepackage{fullpage}
\usepackage{textcomp}
\usepackage{amsmath}
\usepackage{amsfonts}
\usepackage{graphicx}
\usepackage{amssymb}
\usepackage{multirow}
\usepackage{geometry}
\usepackage{soul}
\usepackage{colortbl}
\usepackage{wrapfig}
\usepackage{algorithm}
\usepackage{algorithmicx}
\usepackage{algpseudocode}
\usepackage{todonotes}
\usepackage{amsthm}
\usepackage{thmtools} 
\usepackage{thm-restate}

\usepackage{xcolor}
\usepackage{soul}

\algtext*{EndWhile}
\algtext*{EndIf}
\algtext*{EndFor}

\newtheorem{lemma}{Lemma}
\numberwithin{lemma}{section}
\newtheorem{theorem}[lemma]{Theorem}
\newtheorem{corollary}[lemma]{Corollary}
\newtheorem{definition}[lemma]{Definition}
\newtheorem{observation}[lemma]{Observation}
\newtheorem{fact}[lemma]{Fact}
\newtheorem{proposition}[lemma]{Proposition}
\newtheorem{claim}[lemma]{Claim}

\usepackage{wrapfig}
\usepackage{bm}

\usepackage{booktabs}
\usepackage[pdftex, plainpages = false, pdfpagelabels, 
                 bookmarks=false,
                 bookmarksopen = true,
                 bookmarksnumbered = true,
                 breaklinks = true,
                 linktocpage,
                 pagebackref,
                 colorlinks = true,  
                 linkcolor = blue,
                 urlcolor  = blue,
                 citecolor = red,
                 anchorcolor = green,
                 hyperindex = true,
                 hyperfigures
                 ]{hyperref}

\input{saketPreamble}

\newcommand{\dist}{\ensuremath{\mathsf{dist}}}
\newcommand{\tw}{\ensuremath{\mathsf{tw}}}

\title{A Framework for Approximation Schemes on Disk Graphs}

\author{
Daniel Lokshtanov\footnote{University of California, Santa Barbara, USA. Email: \texttt{daniello@ucsb.edu}} \and Fahad Panolan\footnote{IIT Hyderabad, India. Email: \texttt{fahad@cse.iith.ac.in}} \and Saket Saurabh\footnote{Institute of Mathematical Sciences, Chennai, India. Email: \texttt{saket@imsc.res.in}} \and Jie Xue\footnote{New York University Shanghai, China. Email: \texttt{jiexue@nyu.edu}} \and Meirav Zehavi\footnote{Ben-Gurion University, Israel. Email: \texttt{meiravze@bgu.ac.il}}}

\date{}

\begin{document}


\maketitle

\begin{abstract}

We initiate a systematic study of approximation schemes for fundamental optimization problems on \textit{disk graphs}, a common generalization of both planar graphs and unit-disk graphs.
Our main contribution is a general framework for designing efficient polynomial-time approximation schemes (EPTASes) for vertex-deletion problems on disk graphs, which results in EPTASes for many fundamental problems including \textsc{Vertex Cover}, \textsc{Feedback Vertex Set}, \textsc{Small Cycle Hitting} (in particular, \textsc{Triangle Hitting}), \textsc{$P_k$-Vertex Deletion} for $k \in \{3,4,5\}$, \textsc{Path Deletion}, \textsc{Pathwidth $1$-Deletion}, \textsc{Component Order Connectivity}, \textsc{Bounded Degree Deletion}, \textsc{Pseudoforest Deletion}, \textsc{Finite-Type Component Deletion}, etc.
All EPTASes obtained using our framework are \textit{robust} in the sense that they do not require a realization of the input disk graph (in fact, we allow the input to be any graph, and our algorithms either output a correct approximation solution for the problem or conclude that the input graph is not a disk graph).
To the best of our knowledge, prior to this work, the only problems known to admit PTASes or EPTASes on disk graphs are \textsc{Maximum Clique}, \textsc{Independent Set}, \textsc{Dominating set}, and \textsc{Vertex Cover}, among which the existing PTAS [Erlebach et al., SICOMP'05] and EPTAS [Leeuwen, SWAT'06] for \textsc{Vertex Cover} require a realization of the input disk graph (while ours does not).

The core of our framework is a reduction for a broad class of (approximation) vertex-deletion problems from (general) disk graphs to disk graphs of bounded \textit{local radius}, which is a new invariant of disk graphs introduced in this work.
Disk graphs of bounded local radius can be viewed as a ``mild'' generalization of planar graphs, which preserves certain nice properties of planar graphs.
Specifically, we prove that disk graphs of bounded local radius admit the Excluded Grid Minor property and have locally bounded treewidth. This allows existing techniques for designing approximation schemes on planar graphs (e.g., bidimensionality and Baker's technique) to be directly applied to disk graphs of bounded local radius.
\end{abstract}



\section{Introduction}
Designing efficient approximation algorithms for NP-hard graph optimization problems is a central topic in algorithmic graph theory. The ``best'' type of approximation algorithm one can hope for is a polynomial time approximation scheme (PTAS)~\cite{hochbaum1997approximation,vazirani2001approximation,williamson2011design}. A PTAS is an algorithm which takes an instance $I$ of an optimization problem and a parameter $\varepsilon >0$, runs in time $n^{O(f(1/\varepsilon))}$, and produces a solution that is within a factor of $(1+\varepsilon)$ of being optimal. A PTAS with running time 
$f(1/\varepsilon) \cdot n^{O(1)}$ is called an efficient PTAS (EPTAS). On general graphs, most NP-hard optimization problems are known to be APX-hard; that is, they do not admit (E)PTASes unless P=NP. 

Thus, the study of approximation schemes for graph problems is largely focused on restricted graph classes whose structural properties can be exploited for efficient approximation. A particularly fruitful research direction has been the study of approximation schemes for problems on planar graphs, and, more generally, on geometric intersection graphs. 
Geometric intersection graphs refer to the graphs whose vertices correspond to a set of geometric objects such that two vertices are connected by an edge if and only if the two corresponding geometric objects intersect. Due to their beautiful structural properties and numerous real-world applications, geometric intersection graphs with various types of geometric objects have been extensively studied in graph theory, computational geometry, and other related areas~\cite{AgarwalKS98,KuhnWZ08,XuB06}. 

Perhaps the two most well-studied classes of geometric intersection graphs are {\em planar graphs} and {\em unit disk graphs}. 
For both of these graph classes a broad array of algorithm design techniques have been developed~\cite{Baker94,ChanH12,DawarGKS06,DemaineH05,EisenstatKM12,Eppstein00,DBLP:journals/jacm/FominLS18,Grohe03,HarPeledQ17,HochbaumM85,Klein08,MustafaR10}, simultaneously giving EPTASes for broad swaths of graph problems.
Algorithm design techniques that work on planar graphs can often be used to obtain approximation schemes on more general {\em sparse} graph classes, such as bounded genus graphs or $H$-minor free graphs~\cite{DemaineH05,DBLP:journals/jacm/FominLS18,Grohe03,HarPeledQ17}.
Similarly, algorithm design techniques that work on unit disk graphs often generalize to intersection graphs of ``similarly sized fat objects" (such as unit square graphs)~\cite{ChanH12,HarPeledQ17,MustafaR10}.

The techniques for designing approximation algorithms on planar graphs and unit disk graphs appear to fall short on intersection graphs of objects whose size can vary arbitrarily. Arguably, disk graphs, the intersection graphs of arbitrarily sized closed disks in the plane, is the simplest and most natural such class.
Trivially, every unit disk graph is a disk graph. Less trivially, every planar graph is a disk graph as well~\cite{thurston1979geometry}. Thus, disk graphs are a common generalization of planar graphs and unit disk graphs.
Being a natural graph class that simultaneously generalizes planar graphs and unit disk graphs, one would expect a wealth of approximation algorithms and approximation schemes to be known for problems on disk graphs. Surprisingly, this is not the case. 
To the best of our knowledge, the only problems for which approximation schemes are known on disk graphs are {\sc Vertex Cover}~\cite{ErlebachJS05,Leeuwen06}, {\sc Dominating Set}~\cite{GibsonP10}, {\sc Independent Set}~\cite{ErlebachJS05} and {\sc Maximum Clique}~\cite{bonamy2021eptas}. In fact, it remains a tantalizing open problem whether {\sc Maximum Clique} can be solved in polynomial time on disk graphs~\cite{bonamy2021eptas}. 

One might wonder whether the lack of approximation schemes for problems on disk graphs is rooted at hardness results that refute their existence.
However, it appears that even less is known about APX-hardness on disk graphs than what is known about approximation schemes. It is perfectly consistent with current knowledge that there exist general techniques that yield (E)PTASes on disk graphs for all or most of the problems that admit (E)PTASes on planar graphs and on unit disk graphs. Such techniques would have to simultaneously handle the ``denseness'' of unit disk graphs and the ``non-locality'' of planar graphs. Motivated by this, we initiate a systematic study of approximation schemes for optimization problems on disk graphs.

\subsection{Our results}
We establish a general framework for designing EPTASes for a broad class of \textit{vertex-deletion} problems on disk graphs, and use it to obtain EPTASes for a number of well-studied problems on disk graphs. Each vertex-deletion problem is defined by a \textit{target graph class} $\mathcal{G}$: the input is a graph $G$ and the goal is to delete a smallest subset $S$ of vertices from $G$ such that the resulting graph $G-S$ is in the class $\mathcal{G}$. Our main result is the following theorem.


\begin{restatable}{theorem}{combined}\label{thm:combinedThm}
Each of the following problems admits a robust EPTAS on disk graphs: 

\begin{itemize}\setlength\itemsep{-.9pt}
\item {\sc Feedback Vertex Set}. (Here, $\mathcal{G}$ is the set of acyclic graphs.)
\item {\sc Pseudoforest Deletion}. (Each component of every $G\in\mathcal{G}$ contains at most one cycle.)
\item {\sc Finite-Type Component Deletion}. (Here,  $\mathcal{G}$ is specified by a finite list ${\cal L}$ of graphs and a graph $G$ belongs to $\mathcal{G}$ if every component of $G$ is in ${\cal L}$.) This generalizes well-known problems such as  {\sc Vertex Cover} and {\sc $\ell$-Component Order Connectivity} for every fixed $\ell\in\mathbb{N}_0$.
\item {\sc $P_t$-Vertex Deletion} for every $2\leq t\leq 5$. (Here, $\mathcal{G}$ is the set of graphs excluding a path on $t$ vertices as a subgraph.)
\item {\sc Path Deletion} (Each component of every $G \in \mathcal{G}$ is a path.)
\item {\sc Pathwidth $1$-Deletion}, which is also known as {\sc Caterpillar Deletion}. (Every $G \in \mathcal{G}$ has pathwidth at most $1$.)
\item {\sc $\ell$-Small Cycle Deletion} for every fixed $\ell\in\mathbb{N}$, $\ell\geq 3$. (Here, $\mathcal{G}$ is the set of graphs with no cycle of length at most $\ell$.)
\item {\sc $d$-Bounded Degree Vertex Deletion} for every fixed $d\in\mathbb{N}_0$. (Here, $\mathcal{G}$ is the set of graphs of maximum degree at most $d$.)
\end{itemize}
\end{restatable}

The exact {\em class} of vertex deletion problems to which our methods apply is not easy to state succinctly, and is discussed in more details in Sections~\ref{sec:prelims} and \ref{sec:application}. We note that all of the EPTASes of Theorem~\ref{thm:combinedThm} are {\em robust}, in the sense that they do not require a disk realization of the input graph $G$ to be given.
In fact, we allow the input to be any graph $G$, and our algorithms either output a correct approximation solution for the problem or conclude that $G$ is not a disk graph.
Prior to our work, the only problem mentioned in Theorem~\ref{thm:combinedThm} for which an EPTAS (or PTAS) was already known on disk graphs was {\sc Vertex Cover}~\cite{ErlebachJS05,Leeuwen06}.
As opposed to the EPTAS from Theorem~\ref{thm:combinedThm}, the previously known approximation schemes for {\sc Vertex Cover} on disk graphs given by Erlebach et al. \cite{ErlebachJS05} and Leeuwen \cite{Leeuwen06} are not robust.
We remark the computation of a realization of a given disk graph is at least as hard as the decision of whether a given graph is a disk graph, which is an NP-hard problem~\cite{BREU19983}.


\subsection{Framework for EPTASes on disk graphs} \label{sec-contribution} 
\begin{figure}
    \centering
    \includegraphics[height=4.8cm]{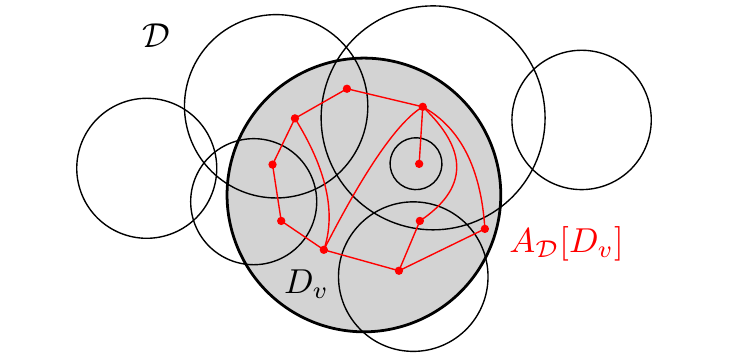}
    \caption{Illustration of the graph $A_\mathcal{D}[D_v]$ (the red graph) where $D_v$ is the grey disk.}
    \label{fig-localradius}
\end{figure}

The main conceptual contribution in our framework is a new invariant of disk graphs, called \textit{local radius}, which is defined (roughly) as follows.
Consider a disk graph $G$ with a realization $\mathcal{D}$ (i.e., a set of disks in the plane whose intersection graph is $G$), and let $D_v \in \mathcal{D}$ be the disk representing the vertex $v \in V(G)$.
The boundaries of the disks in $\mathcal{D}$ subdivide the plane into regions, called \textit{faces}.
Two faces are \textit{adjacent} if their boundaries share an arc.
The \textit{arrangement graph} $A_\mathcal{D}$ of $\mathcal{D}$ is the planar graph whose vertices are the faces contained in at least one disk in $\mathcal{D}$ with two faces connected by an edge if they are adjacent.
For each $v \in V(G)$, the faces inside $D_v$ induce a subgraph of $A_\mathcal{D}$, which we denote by $A_\mathcal{D}[D_v]$.
See Figure~\ref{fig-localradius} for an illustration.
The \textit{local radius} of the disk set $\mathcal{D}$ is the \textit{maximum} radius\footnote{Recall that the \textit{radius} of a graph is the smallest integer $r \geq 0$ such that there exists a vertex $v$ satisfying that every vertex in the graph is within distance $r$ from $v$.} of $A_\mathcal{D}[D_v]$ among all $v \in V(G)$.
We define the \textit{local radius} of $G$ as the \textit{minimum} local radius of a disk set that realizes $G$.
It can be shown that planar graphs are exactly disk graphs of local radius at most $1$ (Fact~\ref{fact-planarlocalradius}).
Thus the local radius of a disk graph is a parameter that smoothly interpolates between planar graphs (local radius $1$) and disk graphs in general (local radius $O(n)$). 

Our framework consists of two parts. The first part, which is our main technical contribution (and is discussed in more detail in Sections~\ref{overview:secBLR} and \ref{sec:redBLR}) is a $(1+\varepsilon)$-approximation preserving reduction of sufficiently well-behaved vertex deletion problems on disk graphs to the same problem on disk graphs of local radius at most $(\frac{1}{\varepsilon})^{O(1)}$. 
We now state this reduction as a theorem, using $\mathbf{P}_\mathcal{G}$ to denote the vertex-deletion problem with target graph class $\mathcal{G}$. We note that all the problems in Theorem~\ref{thm:combinedThm} are ``well behaved'' in the sense of Theorem~\ref{thm-reduction-informal}.
\begin{theorem}[Informal version of Theorem~\ref{thm-reduction}]\label{thm-reduction-informal}
Let $\mathcal{G}$ be a ``well behaved'' graph class.
Given an approximation factor $\varepsilon > 0$ and a graph $G$ of $n$ vertices, one can compute in $n^{O(1)}$ time an induced subgraph $G'$ of $G$ such that the following conditions hold.
\begin{itemize}
    \item If $G$ is a disk graph, then $G'$ is a disk graph of local radius $(\frac{1}{\varepsilon})^{O(1)}$.
    \item Given a $(1+\frac{\varepsilon}{4})$-approximate solution for $\mathbf{P}_\mathcal{G}$ on $G'$, one can compute in $n^{O(1)}$ time a $(1+\varepsilon)$-approximate solution for $\mathbf{P}_\mathcal{G}$ on $G$.
\end{itemize}
\end{theorem}

The second part of the framework is the observation that disk graphs with {\em constant} local radius retain most of the structural properties of planar graphs which are useful for the design of approximation schemes. Most prominently, just as planar graphs~\cite{gu2012improved}, disk graphs with constant local radius have grid minors of size linear in their treewidth (see Section~\ref{sec:prelims} for definitions of treewidth and grid minors).



\begin{lemma}[Informal version of Lemma~\ref{lem:diskSQGM}]
Every disk graph $G$ of local radius $r$ excluding the $t \times t$ grid as a minor has treewidth at most $f(r) \cdot t$ for some fixed function $f$.
\end{lemma}
\noindent
Using this property, we extend the bidimensionality approach~\cite{
demaine2005subexponential,DemaineH05,DBLP:journals/jacm/FominLS18} to disk graphs of bounded local radius, resulting in a meta-theorem (Theorem~\ref{cor:meta1}) that gives EPTASes for a broad class of problems to which the bidimensionality technique applies.
The second property is the ``locally'' bounded treewidth, again a well-known property of planar graphs.
\begin{lemma}[Informal version of Lemma~\ref{lem:twPiece}]
Let $G$ be a disk graph of local radius $r$, and consider a BFS procedure in $G$ from a vertex $v \in V(G)$.
Then the induced subgraph of $G$ consisting of any $k$ consecutive BFS-layers has treewidth at most $f(r) \cdot k$ for some fixed function $f$.
\end{lemma}
\noindent
This property enables us to extend another powerful tool, Baker's layering technique~\cite{Baker94}, to disk graphs of bounded local radius, resulting in a meta-theorem (Theorem~\ref{thm:meta2}) that gives EPTASes for a broad class of problems to which Baker's technique applies.

Theorem~\ref{thm:combinedThm} follows by combining the reduction from disk graphs to disk graphs of bounded local radius with the EPTASes on disk graphs of bounded local radius given by bidimensionality (Theorem~\ref{cor:meta1}) and Baker's layering (Theorem~\ref{thm:meta2}).

\paragraph{Connection to the existing EPTASes on unit-disk graphs.}
Our framework has an interesting connection to the existing EPTASes on unit-disk graphs given by Fomin et al. \cite{DBLP:journals/jacm/FominLS18}.
The EPTASes of \cite{DBLP:journals/jacm/FominLS18} first reduce the \textit{ply} of the input unit-disk graph (that is, the maximum number of unit disks whose mutual intersection is nonempty) to $O_\varepsilon(1)$ and then apply bidimensionality on the bounded-ply unit-disk graph.
It was shown in \cite{DBLP:journals/jacm/FominLS18} that for unit-disk graphs, bounded ply implies bounded degree.
Bounded local-radius is a condition stronger than bounded ply (see Observation~\ref{obs:boundedPly}) and weaker than bounded degree (which is obvious).
Therefore, on unit-disk graphs, the notions of bounded ply, bounded local-radius, and bounded degree are all equivalent.
In this sense, our framework can be viewed a generalization of the approach of \cite{DBLP:journals/jacm/FominLS18} to disk graphs (actually, such a generalization was thought to be unlikely in \cite{DBLP:journals/jacm/FominLS18}).
One can easily see that on disk graphs, bounded ply is \textit{strictly} weaker than bounded local-radius, which is in turn \textit{strictly} weaker than bounded degree.
Interestingly, the ply-reduction step in \cite{DBLP:journals/jacm/FominLS18} still applies to reduce (general) disk graphs to \textit{bounded-ply} disk graphs, and on the other hand, bidimensionality can be applied to design EPTASes on \textit{bounded-degree} disk graphs, as shown in \cite{DBLP:journals/jacm/FominLS18}.
Unfortunately, due to the big gap between bounded ply and bounded degree for disk graphs, this does not give us EPTASes on disk graphs.
Now the notion of bounded local-radius results in a graph class in-between, which is more general than bounded-degree disk graphs so that we can reduce general disk graphs to, and is more restricted than bounded-ply disk graphs so that bidimensionality (and Baker) applies to.


\subsection{Other related work on disk graphs}
Apart from the above mentioned (E)PTASes  for {\sc Vertex Cover}~\cite{ErlebachJS05,Leeuwen06}, {\sc Dominating Set}~\cite{GibsonP10}, {\sc Independent Set}~\cite{ErlebachJS05} and {\sc Maximum Clique}~\cite{bonamy2021eptas})  on disk graphs, there has also been some study for problems on disk graphs in the realm of moderately exponential time algorithms~\cite{FominK10} and parameterized complexity~\cite{CyganFKLMPPS15}. de Berg et al.~\cite{DBLP:journals/siamcomp/BergBKMZ20} developed a framework to design optimal (assuming the ETH) subexponential time algorithms for intersection graphs of similarly-sized fat objects. Among their results, they obtained $2^{O(\sqrt{n})}$ time algorithms for {\sc Vertex Cover} and {\sc Independent Set} on $n$-vertex disk graphs, and showed that, assuming the ETH, these algorithms are optimal up to constant factors in the exponent. In a recent article, Lokshtanov et al.~\cite{lokshtanov2022subexponential} gave methods to design subexponential time parameterized algorithms for problems on disk graphs. Using this they designed the first $2^{o(k)}n^{O(1)}$ time algorithm for several problems, including {\sc Feedback Vertex Set} and {\sc Odd Cycle Transversal}. 

\subsection{Roadmap}

In Section \ref{sec:overview}, we present an overview of our proof ideas. Afterwards, in Section~\ref{sec:prelims}, we present the notation and formal definitions of concepts used in this paper. Our main technical section is Section~\ref{sec:redBLR}, where we present our procedure to reduce the local radius of a given disk graph. Next, in Section~\ref{sec:boundedLocalRadius}, we show how to design EPTASes for problems on bounded local radius disk graphs. Section~\ref{sec:application} then combines the results in the two aforementioned sections to attain EPTASes for problems on (general) disk graphs. Lastly, in Section \ref{sec:conclusion}, we further discuss our results as well as present open questions for future research.

\section{Overview of our techniques}\label{sec:overview}
In this section, we give an informal overview of the techniques used for obtaining our results.
As mentioned in Section~\ref{sec-contribution}, our framework consists two main parts: \textbf{(i)} reducing from general disk graphs to disk graphs of bounded local radius and \textbf{(ii)} designing approximation schemes on disk graphs of bounded local radius.

\subsection{Reduction to bounded local radius}\label{overview:secBLR}

We now give a sketch of the proof of Theorem~\ref{thm-reduction-informal}.
Our reduction in Theorem~\ref{thm-reduction-informal} applies to a broad class of vertex-deletion problems.
However, in order to give a more intuitive exposition, in this overview, we fix a typical problem, \textsc{Feedback Vertex Set} (FVS), and describe our basic ideas in the context of FVS.
Recall that in FVS, the goal is to delete a smallest set $S$ of vertices from a graph $G$ such that $G-S$ is acyclic.
Also, for simplicity, we only consider the case where $G$ is a disk graph and assume a realization $\mathcal{D}$ of $G$ is also given.
For each vertex $v \in V(G)$, let $D_v \in \mathcal{D}$ denote the disk representing $v$.
While our actual reduction is robust, the presentation can be cleaner and more intuitive based on the realization $\mathcal{D}$.

Our reduction begins with $G$ and consists of three steps.
Each step removes some vertices from the graph obtained in the previous step, and hence results in an induced subgraph of the previous graph.
Through the three steps, the graph becomes more and more structured, and finally of bounded local radius after the last step.
Let $\varepsilon>0$ be the approximation factor.
In this overview, we do not care about the $(\frac{1}{\varepsilon})^{O(1)}$ bound in Theorem~\ref{thm-reduction-informal} and only aim to reduce to a disk graph of local radius $O_\varepsilon(1)$, some constant depending on $\varepsilon$.

\vspace{-0.3cm}

\paragraph{Step 1.}
To illustrate the first step of our reduction, we first recall a basic notion called \textit{ply}.
The ply of a disk set is the maximum number of disks in the set that can be stabbed by a point in the plane (or equivalently, the maximum number of disks that have a common intersection).
The ply of a disk graph is simply defined as the minimum ply of a disk set that realizes it.
In the first step, we reduce the input disk graph $G$ to an induced subgraph $G_1$ of bounded \textit{ply}.
This reduction is standard, and was also used in many other algorithms on (unit-)disk graphs as a preprocessing step.
The observation is the following.
If the ply of $G$ is large, say larger than $\frac{1}{\varepsilon}$, then we can find a point $p \in \mathbb{R}^2$ that stabs (at least) $\frac{1}{\varepsilon}$ disks in $\mathcal{D}$.
The disks stabbed by $p$ pairwise intersect, so they correspond to a clique $K$ of size at least $\frac{1}{\varepsilon}$ in $G$.
But for FVS, a solution must hit all cycles in $G$ and thus must contain all but at most 2 vertices in $K$ (for otherwise there are 3 vertices survive which form a cycle).
As $|K| \geq \frac{1}{\varepsilon}$, for the purpose of approximation, we can simply include all vertices of $K$ in our solution and remove $K$ from $G$ (and the corresponding disks from $\mathcal{D}$).
By doing this we only create a multiplicative error of at most $\frac{|K|}{|K|-2} = 1+O(\varepsilon)$, since the optimal solution must include at least $|K|-2$ vertices in $K$.
We can keep doing this until the ply of $G$ is at most $\frac{1}{\varepsilon}$.
Let $G_1$ be the resulting graph.
By the above argument, one can see that if $S_1$ is a $(1+\varepsilon)$-approximation solution for FVS on $G_1$, then $(V(G) \backslash V(G_1)) \cup S_1$ is a $(1+O(\varepsilon))$-approximation solution for FVS on $G$.
So in what follows, it suffices to solve FVS on $G_1$.

\vspace{-0.3cm}

\begin{figure}
    \centering
    \includegraphics[height=5.7cm]{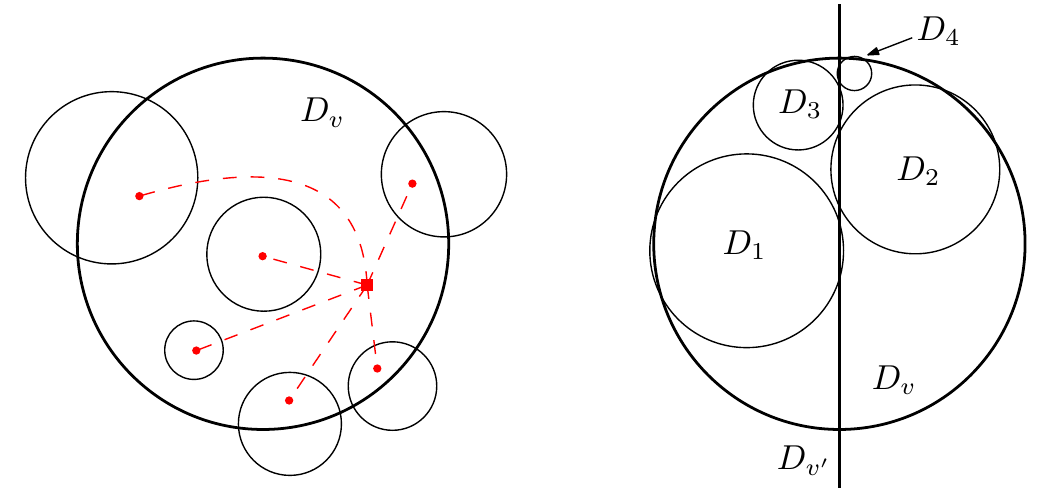}
    \caption{Illustration of a disk with (almost) independent neighbors. Left part is the case where all disks intersecting $D_v$ are disjoint, in which the faces inside $D_v$ form a star (the red part). Right part is an example where all but one (the ``infinitely'' large disk $D_{v'}$) disks intersecting $D_v$ are disjoint, in which the local radius inside $D_v$ can be made unbounded.}
    \label{fig-independent}
\end{figure}

\paragraph{Step 2.}
Consider the graph $G_1$ obtained in Step~1, and let $\mathcal{D}_1 \subseteq \mathcal{D}$ be the subset consisting of disks representing the vertices of $G_1$.
In this step, we shall reduce from $G_1$ to an induced subgraph $G_2$.
The desired property of $G_2$ is the following: the neighbors of every vertex of $G_2$ are ``almost'' an independent set in $G_2$.
Formally, let $N_{G_2}(v)$ denote the set of neighbors of a vertex $v \in V(G_2)$ in $G_2$ (excluding $v$ itself).
What we want is that for every $v \in V(G_2)$, $N_{G_2}(v)$ can be partitioned into two subsets $S(v)$ and $I(v)$ such that $S(v)$ is small (specifically, of size $O_\varepsilon(1)$) and $I(v)$ is an independent set in $G_2$.
Before discussing the reduction step, let us first briefly explain why achieving such a graph helps.
Our final goal is to achieve bounded local radius.
Suppose now we were in an ideal situation where the neighbors of every vertex of $G_2$ are \textit{exactly} an independent set, i.e., $S(v) = \emptyset$ for all $v \in V(G_2)$.
Then the disks corresponding to the neighbors of a vertex $v \in V(G_2)$ are disjoint, and it turns out that the part of the arrangement graph contained in $D_v$ forms a star whose center is the face $D_v \backslash (\bigcup_{u \in N_{G_2}(v)} D_u)$.
See the left part of Figure~\ref{fig-independent}.
Thus, the local radius of $G_2$ is at most 1 in this case.
Unfortunately, this is not the case when $S(v) \neq \emptyset$, even if $S(v)$ is small; in fact, by the construction in the right part of Figure~\ref{fig-independent} (which will be discussed in Step~3), even if $|S(v)| = 1$, the part of the arrangement graph contained in $D_v$ can have unbounded radius.
However, making the neighbors of of every vertex ``almost'' independent still helps a lot.
Based on this, with some additional work in Step 3, we can finally achieve bounded local radius.

Next, we discuss this reduction step.
Like Step~1, here we shall keep deleting vertices from $G_1$ and including the deleted vertices in our solution.
The resulting graph will be our $G_2$.
For correctness, we need to have that the deleted vertices (i.e., $V(G_1) \backslash V(G_2)$) together with a $(1+\varepsilon)$-approximation FVS solution of $G_2$ form a $(1+O(\varepsilon))$-approximation FVS solution of $G_1$.
We guarantee this by requiring that $V(G_1) \backslash V(G_2)$ contains at most $\varepsilon \cdot \mathsf{opt}(G_1)$ ``wrong'' vertices, where $\mathsf{opt}(G_1)$ denote the size of an optimal FVS solution of $G_1$.
More precisely, for some optimal FVS solution $S_\mathsf{opt} \subseteq V(G_1)$ of $G_1$, $V(G_1) \backslash V(G_2)$ contains at most $\varepsilon \cdot \mathsf{opt}(G_1)$ vertices that are not in $S_\mathsf{opt}$.
If this is the case, one can easily see that $(V(G_1) \backslash V(G_2)) \cup S_2$ is a $(1+2\varepsilon)$-approximation FVS solution of $G_1$, for any $(1+\varepsilon)$-approximation FVS solution $S_2 \subseteq V(G_2)$ of $G_2$.

We begin with a simple but useful observation, which was also used in the parameterized FVS algorithm by Lokshtanov et al. \cite{lokshtanov2022subexponential} to guide a branching procedure.
The observation is the following: for a vertex $v \in V(G_1)$, if $M$ is a matching in the induced subgraph $G_1[N_{G_1}(v)]$, then any FVS solution on $G_1$ either contains $v$ or contains at least one endpoint of each edge in $M$.
The reason is simply that if a solution $S$ contains neither $v$ nor the two endpoints of an edge in $M$, then the three vertices form a cycle in $G_1 - S$, which is impossible.
To see why this observation helps, consider a vertex $v \in V(G_1)$ and a maximal matching $M$ in $G_1[N_{G_1}(v)]$.
If $|M| < \frac{1}{\varepsilon}$, then the neighbors of $v$ are almost an independent set, because we can set $S(v)$ to be the endpoints of the edges in $M$ (which is of size at most $\frac{2}{\varepsilon}$) and $N_{G_1}(v) \backslash S(v)$ is an independent set.
If $|M| \geq \frac{1}{\varepsilon}$, then an optimal solution $S_\mathsf{opt}$ must contain either $v$ or at least $|M|$ endpoints of the edges in $M$.
It turns out that in this case, we can safely delete $v$ from $G_1$.
Indeed, if $S_\mathsf{opt}$ contains $v$, we are happy, because we are deleting is a ``right'' vertex.
If $S_\mathsf{opt}$ does not contain $v$, we are deleting a ``wrong'' vertex.
But we can charge the cost of doing this to the vertices in $S_\mathsf{opt} \cap M$\footnote{With a bit abuse of notation, here we denote by $S_\mathsf{opt} \cap M$ the set of endpoints of edges in $M$ contained in $S_\mathsf{opt}$}.
As argued before, when $v \notin S_\mathsf{opt}$, we have $|S_\mathsf{opt} \cap M| \geq |M| \geq \frac{1}{\varepsilon}$.
So one wrong vertex deleted is charged to $\frac{1}{\varepsilon}$ vertices in $S_\mathsf{opt}$, which will be used to ensure that at most $\varepsilon \cdot \mathsf{opt}(G_1) = \varepsilon \cdot |S_\mathsf{opt}|$ wrong vertices are deleted.
In this way, we can get rid of any ``bad'' vertex (i.e., a vertex whose neighbors are \textit{not} almost an independent set) in $G_1$.
However, one thing we need to be careful about here is that we cannot simply repeat this to get rid of \textit{all} ``bad'' vertices in $G_1$.
The reason is simple: when deleting a wrong vertex, we need to charge the cost to $\frac{1}{\varepsilon}$ vertices in $S_\mathsf{opt}$, but these vertices may get charged again in the future when deleting other wrong vertices, which makes our argument for bounding the number of wrong vertices deleted fail.
A natural idea to avoid repeated charging is to \textit{mark} the vertices that have already been charged, and ignore the marked vertices in the future rounds.
Specifically, we maintain a set $U \subseteq V(G_1)$ of unmarked vertices, which is initially $V(G_1)$.
When considering a vertex $v \in V(G_1)$, we construct a maximal matching $M$ in $G_1[N_{G_1}(v) \cap U]$, i.e., only among the unmarked neighbors of $v$.
If $|M| \geq \frac{1}{\varepsilon}$, we delete $v$ from $G_1$ and mark the endpoints of $\frac{1}{\varepsilon}$ edges in $M$ (i.e., remove those endpoints from $U$); if $v \notin S_\mathsf{opt}$, then at least $\frac{1}{\varepsilon}$ vertices in $S_\mathsf{opt}$ get marked and we can charge the cost of deleting $v$ to those vertices.
Since the vertices in $S_\mathsf{opt}$ getting charged will be marked and ignored in the future, we can safely repeat this procedure and the number of wrong vertices deleted will be bounded by $\varepsilon \cdot |S_\mathsf{opt}|$.
However, now the issue is that we may \textit{not} end up with a situation where all ``bad'' vertices are deleted from $G_1$: in the resulting $G_1$, there can exist vertices $v$ where $G_1[N_{G_1}(v)]$ has a large matching (so the neighbors of $v$ are not almost an independent set) but $G_1[N_{G_1}(v) \cap U]$ does not (so we cannot further delete $v$ from $G_1$).

We resolve this issue by exploiting geometric properties of disks.
In fact, what we are going to do is fairly simple.
Recall that $D_v \in \mathcal{D}_1$ is the disk corresponding to the vertex $v \in V(G_1)$.
We order the vertices in $G_1$ as $v_1,\dots,v_n$ such that $D_{v_1} \geq \cdots \geq D_{v_n}$ (here the disks are compared by their sizes), and then apply the above delete-and-mark procedure along this order.
Interestingly, as we will see shortly, such a simple twist makes the resulting $G_1$ satisfy the desired property, i.e., the neighbors of each vertex are almost an independent set.
For convenience, let $G_2$ denote the resulting graph.
Consider a vertex $v_i \in V(G_2)$, and the round we consider $v_i$ in the procedure.
We have the set $U$ of unmarked vertices and a maximal matching $M$ of $G_1[N_{G_1}(v_i) \cap U]$.
Let $R$ be the set of marked vertices at this point.
Note that $|M| < \frac{1}{\varepsilon}$ as $v_i$ is not deleted from $G_1$.
We simply let $S(v_i)$ consist of all endpoints of edges in $M$ (that survive in $G_2$) and all vertices in $N_{G_2}(v_i) \cap R$ (i.e., neighbors of $v_i$ in $G_2$ that are marked at this point).
Clearly, $I(v_i) = N_{G_2}(v_i) \backslash S(v_i)$ is an independent set, as $M$ is a maximal matching of the unmarked neighbors of $v_i$ (in $G_1$) and the marked neighbors are either in $S(v_i)$ or not in $V(G_2)$.
So it suffices to show $|S(v_i)| = O_\varepsilon(1)$.

\begin{figure}
    \centering
    \includegraphics[height=5.2cm]{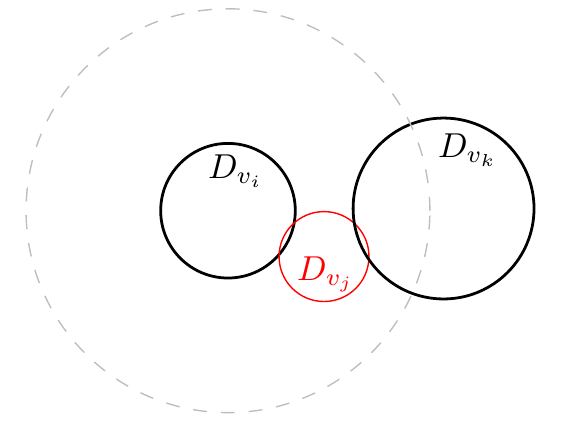}
    \caption{Illustration of the disks $D_{v_i}, D_{v_j}, D_{v_k}$. Here $D_{v_j}$ is the smallest one, which connects $D_{v_k}$ and $D_{v_i}$. So $D_{v_k}$ must intersect a 3-expansion of $D_{v_i}$ (i.e., disk bounded by the grey dashed circle).}
    \label{fig-3disks}
\end{figure}

We can ignore the endpoints of $M$ in $S(v_i)$ as the number of them is at most $\frac{1}{\varepsilon}$.
The remaining vertices in $S(v_i)$ are those in $N_{G_2}(v_i) \cap R$, i.e., the marked neighbors of $v_i$.
Consider a vertex $v_j \in N_{G_2}(v_i) \cap R$.
Assume $v_j$ got marked when we consider some vertex $v_k$.
We must have $k<i$ as we proceed along the order $v_1,\dots,v_n$.
If $j < i$, we charge $v_j$ to itself, otherwise we charge $v_j$ to $v_k$.
We do this charging for every vertex in $N_{G_2}(v_i) \cap R$.
The following two facts are clear: \textbf{(i)} only the vertices $v_1,\dots,v_{i-1}$ can get charged and \textbf{(ii)} each vertex gets charged at most $1+\frac{1}{\varepsilon}$ times (because it can only get charged from itself and vertices marked when considering it, where the number of the latter is at most $\frac{1}{\varepsilon}$).
We then claim that only $O_\varepsilon(1)$ vertices among $v_1,\dots,v_{i-1}$ get charged, which implies $|S(v_i)| = O_\varepsilon(1)$ by property \textbf{(ii)} above.
The key observation is that if a vertex $v_k \in \{v_1,\dots,v_{i-1}\}$ gets charged, the disk $D_{v_k}$ must be locationally ``close'' to $D_{v_i}$.
If $v_k$ gets charged from itself, then $v_k \in N_{G_2}(v_i) \cap R$ and thus $D_{v_k}$ intersects $D_{v_i}$.
If $v_k$ gets charged from another vertex $v_j \in N_{G_2}(v_i) \cap R$, by our charging rule we must have $j > i$ and $v_j$ got marked when considering $v_k$.
In this case, $v_j$ is neighboring to both $v_i$ and $v_k$, so $D_{v_j}$ intersects both $D_{v_i}$ and $D_{v_k}$.
Note that $k<i<j$ and $D_{v_k} \geq D_{v_i} \geq D_{v_j}$.
So the two larger disks $D_{v_k}$ and $D_{v_i}$ are ``connected'' by the smallest disk $D_{v_j}$, as shown in Figure~\ref{fig-3disks}.
As $D_{v_j} \leq D_{v_i}$, $D_{v_k}$ must be close to $D_{v_i}$ in the sense that a constant-factor expansion (more precisely, $3$-expansion) of $D_{v_i}$ intersects $D_{v_k}$.
To summarize, for every $v_k \in \{v_1,\dots,v_{i-1}\}$ getting charged, $D_{v_k}$ intersects the $3$-expansion of $D_{v_i}$.
Now recall that $\mathcal{D}_1$ is of $O_\varepsilon(1)$ ply.
Bounded-ply disk sets admits a well-known sparseness property (which was repeatedly used in the literature): any disk of radius $r$ in the plane can only intersect $O(p)$ disks of radius $\Omega(r)$ in a disk set of ply $p$.
Thus, the $3$-expansion of $D_{v_i}$ only intersects $O_\varepsilon(1)$ disks in $\{D_{v_1},\dots,D_{v_{i-1}}\}$, as $D_{v_k} \geq D_{v_i}$ for all $v_k \in \{v_1,\dots,v_{i-1}\}$.
Therefore, only $O_\varepsilon(1)$ vertices among $v_1,\dots,v_{i-1}$ get charged and $|S(v_i)| = O_\varepsilon(1)$.
So we finally reduce to an induced subgraph $G_2$ in which the neighbors of every vertex are almost an independent set.

\vspace{-0.3cm}

\paragraph{Step 3.}
In the last step, we have achieved a graph $G_2$ in which $N_{G_2}(v) = S(v) \cup I(v)$ for every vertex $v \in V(G_2)$ such that $|S(v)| = O_\varepsilon(1)$ and $I(v)$ is an independent set.
Let $\mathcal{D}_2 \subseteq \mathcal{D}$ be the subset of disks representing the vertices of $G_2$.
We have seen in the left part of Figure~\ref{fig-independent} that if $S(v) = \emptyset$ for all $v \in V(G_2)$, then $G_2$ is already of bounded local radius.
However, this no longer holds even when $|S(v)| = 1$.
We give such an example in the right part of Figure~\ref{fig-independent}.
Here $S(v)$ only contains one vertex $v'$, whose disk $D_{v'}$ is ``infinitely'' large (and hence becomes a halfplane).
The boundary of $D_{v'}$ cuts $D_v$ into two halves.
The vertices $I(v)$ correspond to the sequence $D_1,D_2,D_3,\dots$ of disjoint disks in the figure, which appear alternately in the left/right half of $D_v$ and intersect the boundaries of both $D_v$ and $D_{v'}$.
As shown in the figure, in the sequence $D_1,D_2,D_3,\dots$, the disks become smaller and smaller, and get closer and closer to the top intersection point of the boundaries of $D_v$ and $D_{v'}$.
Although we only show the first four disks $D_1,D_2,D_3,D_4$ here, this sequence of disks can be made arbitrarily long, in which case the local radius inside $D_v$ is unbounded (one can easily see that going from a face in $D_1$ to a face in $D_k$ takes $\Omega(k)$ steps).

The goal of Step~3 is further reducing to an induced subgraph $G_3$ of $G_2$ which has bounded local radius (and hence the situation above cannot happen).
Again, the reduction itself is simple, while the analysis is nontrivial.
Recall that two vertices $v,v' \in V(G_2)$ are \textit{false twins} if they have the same neighbors, i.e., $N_{G_2}(v) = N_{G_2}(v')$.
This notion induces an equivalent relation $\sim$ on $V(G_2)$ where $v \sim v'$ iff $v$ and $v'$ are false twins.
Let $\mathcal{X}$ be the set of equivalence classes in $V(G_2)$ for $\sim$, which is a partition of $V(G_2)$.
Each class $X \in \mathcal{X}$ is a maximal set of false twins in $V(G_2)$.
Denote by $N_{G_2}(X)$ the set of neighbors of $X$ in $G_2$ (excluding the vertices in $X$) and $d_X = |N_{G_2}(X)|$.
We have $N_{G_2}(X) = N_{G_2}(v)$ for every $v \in X$ since $X$ consists of false twins.
The construction of $G_3$ is simply the following.
In every class $X \in \mathcal{X}$, we (arbitrarily) pick $(1+\frac{1}{\varepsilon}) \cdot d_X$ vertices.
Then let $G_3$ be the subgraph of $G_2$ induced by the picked vertices.
Now we need to answer two questions: \textbf{(i)} why we can reduce from $G_2$ to $G_3$ and \textbf{(ii)} why $G_3$ is of bounded local radius.

To answer question \textbf{(i)}, we need to show that given a $(1+\varepsilon)$-approximation FVS solution $S \subseteq V(G_3)$ on $G_3$, we can compute a $(1+O(\varepsilon))$-approximation FVS solution on $G_2$ (in polynomial time).
We shall begin with the graph $G_2 - S$ and additionally delete a few vertices from $G_2 - S$ to make the resulting graph acyclic.
Since $G_3 - S$ is acyclic, every cycle in $G_2 - S$ must contain some vertices in $V(G_2) \backslash V(G_3)$.
So it suffices to guarantee that every vertex in $V(G_2) \backslash V(G_3)$ is not involved in any cycle in the resulting graph.
For convenience, let $\bar{S} = V(G_3) \backslash S$.
Then $G_3[\bar{S}]$ is acyclic.
We have the following two simple observations.
\begin{itemize}
    \item For each $X \in \mathcal{X}$, either $X \subseteq V(G_3)$ or $N_{G_2}(X) \subseteq V(G_3)$.
    To see this, suppose $X \nsubseteq V(G_3)$, then $|X| > (1+\frac{1}{\varepsilon}) \cdot d_X > d_X$.
    Consider $v \in N_{G_2}(X)$ and let $X' \in \mathcal{X}$ such that $v \in X'$.
    We must have $X' \subseteq N_{G_2}(X)$ and $X \subseteq N_{G_2}(X')$ as both $X$ and $X'$ are false-twin classes.
    Thus, $d_X' = |N_{G_2}(X')| \geq |X| > d_X = |N_{G_2}(X)| \geq |X'|$, which implies $v \in X' \subseteq V(G_3)$.
    \item For each $X \in \mathcal{X}$, either $|X \cap \bar{S}| \leq 1$ or $|N_{G_2}(X) \cap \bar{S}| \leq 1$.
    Indeed, if $|X \cap \bar{S}| \geq 2$ and $|N_{G_2}(X) \cap \bar{S}| \geq 2$, two vertices in $X \cap \bar{S}$ and two vertices in $N_{G_2}(X) \cap \bar{S}$ form a 4-cycle (as all vertices in $X$ are neighboring to all vertices in $N_{G_2}(X)$), but $G_3[\bar{S}]$ must be acyclic.
\end{itemize}
Consider a class $X \in \mathcal{X}$.
We want to make the vertices in $(V(G_2) \backslash V(G_3)) \cap X$ be not involved in any cycle.
If $X \subseteq V(G_3)$, then $(V(G_2) \backslash V(G_3)) \cap X = \emptyset$.
Assume $X \nsubseteq V(G_3)$.
Then $|X \cap V(G_3)| = (1+\frac{1}{\varepsilon}) \cdot d_X$.
By the first observation above, we have $N_{G_2}(X) \subseteq V(G_3)$ and thus $N_{G_2}(X) \backslash S = N_{G_2}(X) \cap \bar{S}$.
If $|N_{G_2}(X) \backslash S| \leq 1$, we are done because all vertices in $X$ are of degree 0 or 1 in $G_2 - S$ and thus not involved in any cycle.
Otherwise, $|N_{G_2}(X) \cap \bar{S}| = |N_{G_2}(X) \backslash S| > 1$, which implies $|X \cap \bar{S}| \leq 1$ by the second observation above.
Then $|X \cap S| = |X \cap V(G_3)| - |X \cap \bar{S}| \geq (1+\frac{1}{\varepsilon}) \cdot d_X - 1 \geq \frac{1}{\varepsilon} \cdot d_X$.
In this case, we simply delete all vertices in $N_{G_2}(X)$ so that all vertices in $X$ become isolated.
We can charge the cost of doing this to the vertices in $X \cap S$.
As $|X \cap S| \geq \frac{1}{\varepsilon} \cdot d_X$, we delete at most $d_X = |N_{G_2}(X)|$ vertices and charge at least $\frac{1}{\varepsilon} \cdot d_X$ vertices in $S$.
After doing this for every $X \in \mathcal{X}$, no vertices in $V(G_2) \backslash V(G_3)$ are involved in a cycle and hence the resulting graph is acyclic.
The number of additionally deleted vertices is at most $\varepsilon \cdot |S|$ by our charging argument.
Therefore, we obtained a $(1+O(\varepsilon))$-approximation FVS solution on $G_2$.

To answer question \textbf{(ii)} is more difficult.
The first key observation is that each false-twin class in $G_3$ is of size $O_\varepsilon(1)$.
To see this, first note that any two vertices $v,v' \in V(G_3)$ are false twins iff they are false twins in $G_2$, i.e., $v,v' \in X$ for some $X \in \mathcal{X}$ (the ``if'' part is clear and the ``only if'' part holds mainly because $G_3$ includes at least one vertex in every $X \in \mathcal{X}$).
Next we observe that for each $X \in \mathcal{X}$, either $|X| = O_\varepsilon(1)$ or $d_X = O_\varepsilon(1)$.
This follows from the bounded ply of $G_2$.
Every vertex in $X$ is neighboring to every vertex in $N_{G_2}(X)$, so the induced subgraph $G_2[X \cup N_{G_2}(X)]$ contains at least $|X| \cdot d_X$ edges.
But it can be shown that a disk graph $H$ of ply $p$ has at most $O(p \cdot |V(H)|)$ edges.
Thus, $G_2[X \cup N_{G_2}(X)]$ has $O_\varepsilon(|X|+d_X)$ edges, implying $|X| \cdot d_X = O_\varepsilon(|X|+d_X)$ and thus either $|X| = O_\varepsilon(1)$ or $d_X = O_\varepsilon(1)$.
If $|X| = O_\varepsilon(1)$, then $|X \cap V(G_3)| = O_\varepsilon(1)$.
If $d_X = O_\varepsilon(1)$, we also have $|X \cap V(G_3)| = O_\varepsilon(1)$ as $G_3$ contains $(1+\frac{1}{\varepsilon}) \cdot d_X$ vertices in $X$.
As argued above, the set of false-twin classes in $G_3$ is $\{X \cap V(G_3): X \in \mathcal{X}\}$, so each class is of size $O_\varepsilon(1)$.

Now let us look at again the example in the right part of Figure~\ref{fig-independent}.
Ideally, if these disks are the only vertices in $G_2$, then we are happy because the sequence of disks $D_1,D_2,\dots$ correspond to vertices that are false twins in $G_2$ and hence only $O_\varepsilon(1)$ of them will remain in $G_3$, making $G_3$ have $O_\varepsilon(1)$ local radius.
But the actual situation is more complicated: the disks $D_1,D_2,\dots$ may also intersect disks other than $D_v,D_{v'}$ and need not to be false twins.
So we need more efforts to bound the local radius of $G_3$.
Recall that for each $v \in V(G_2)$, we have a ``small-independent'' partition $N_{G_2}(v) = S(v) \cup I(v)$ where $S(v)$ is small and $I(v)$ is an independent set.
We remove from $S(v)$ and $I(v)$ the vertices that are not in $G_3$.
Now for each $v \in V(G_3)$, the sets $S(v)$ and $I(v)$ form a small-independent partition of $N_{G_3}(v)$; for convenience, we call the vertices in $S(v)$ \textit{singular neighbors} of $v$ and those in $I(v)$ \textit{independent neighbors} of $v$.
In what follows, let us fix a vertex $v \in V(G_3)$ and investigate the radius of the arrangement graph of $\mathcal{D}_3$ inside $D_v$ (where $\mathcal{D}_3 \subseteq \mathcal{D}$ consists of the disks corresponding to the vertices in $G_3$).

We have the small-independent partition $N_{G_3}(v) = S(v) \cup I(v)$ of $N_{G_3}(v)$.
However, this partition seems not that tractable when bounding the local radius inside $D_v$.
So our first step is to derive another small-independent partition $N_{G_3}(v) = S^* \cup I^*$ which satisfies an \textit{additional} property: any disk $D_u$ for $u \in I^*$ is not contained in $U = \bigcup_{w \in \{v\} \cup S^*} D_w$, the union of all disks corresponding to vertices in $\{v\} \cup S^*$.
This additional property will later allow us to use geometric arguments to bound the local radius inside $D_v$.
We construct $S^*$ and $I^*$ as follows.
Let $S^2(v) = \bigcup_{w \in S(v)} S(w)$ consist of the \textit{singular neighbors of singular neighbors} of $v$.
A vertex $u \in N_{G_3}(v)$ is included in $S^*$ if $u \in S(v) \cup S^2(v)$ or $N_{G_3}(u) \subseteq \{v\} \cup S(v) \cup S^2(v)$.
Then simply let $I^* = N_{G_3}(v) \backslash S^*$.
Since $S^* \supseteq S(v)$ and $I^* \subseteq I(v)$, clearly $I^*$ is an independent set.
To see $|S^*| = O_\varepsilon(1)$, note that $|S(v) \cup S^2(v)| = O_\varepsilon(1)$ as every vertex in $G_3$ has $O_\varepsilon(1)$ singular neighbors.
It follows that the number of vertices $u$ with $N_{G_3}(u) \subseteq \{v\} \cup S(v) \cup S^2(v)$ is also bounded by $O_\varepsilon(1)$, because each false-twin class in $G_3$ is of size $O_\varepsilon(1)$.
To see the additional property, let us consider a vertex $u \in N_{G_3}(v)$ with $D_u \subseteq U$.
We must show $u \in S^*$.
If $u \in S(v) \cup S^2(v)$, this is the case.
Assume $u \notin S(v) \cup S^2(v)$, and we claim $N_{G_3}(u) \subseteq \{v\} \cup S(v) \cup S^2(v)$.
Let $z \in N_{G_3}(u)$ such that $z \neq v$.
One can show $z \in S(v) \cup S^2(v)$ by the following three facts.
\begin{itemize}
    \item $z \notin I(v)$.
    The reason for this is $u \notin S(v) \cup S^2(v)$.
    So $u \in I(v)$.
    But $I(v)$ is an independent set.
    Thus, $z$ and $u$ cannot be both in $I(v)$ because they are neighbors.
    \item $z \notin I(w)$ for any $w \in S(v)$ such that $z,u \in N_{G_3}(w)$.
    The reason for this is $u \notin S^2(v)$.
    So $u \notin S(w)$ and $u \in I(w)$.
    Again, $z$ and $u$ cannot be both in $I(w)$.
    \item $z,u \in N_{G_3}(w)$ for some $w \in \{v\} \cup S(v)$.
    The reason for this is $D_u \subseteq U$.
    Since $D_z$ intersects $D_u$, we can pick a point $p \in D_z \cap D_u$.
    The $p \in U$ and thus $p \in D_w$ for some $w \in \{v\} \cup S(v)$.
    Now the disks $D_z,D_u,D_w$ intersect at the point $p$, which implies $z,u \in N_{G_3}(w)$.
\end{itemize}
By the last fact above, we can find $w \in \{v\} \cup S(v)$ such that $z,u \in N_{G_3}(w)$.
If $w = v$, then $z \in S(v)$ by the first fact.
Otherwise, $z \in S(w) \subseteq S^2(v)$ by the second fact.

\begin{figure}
    \centering
    \includegraphics[height=5cm]{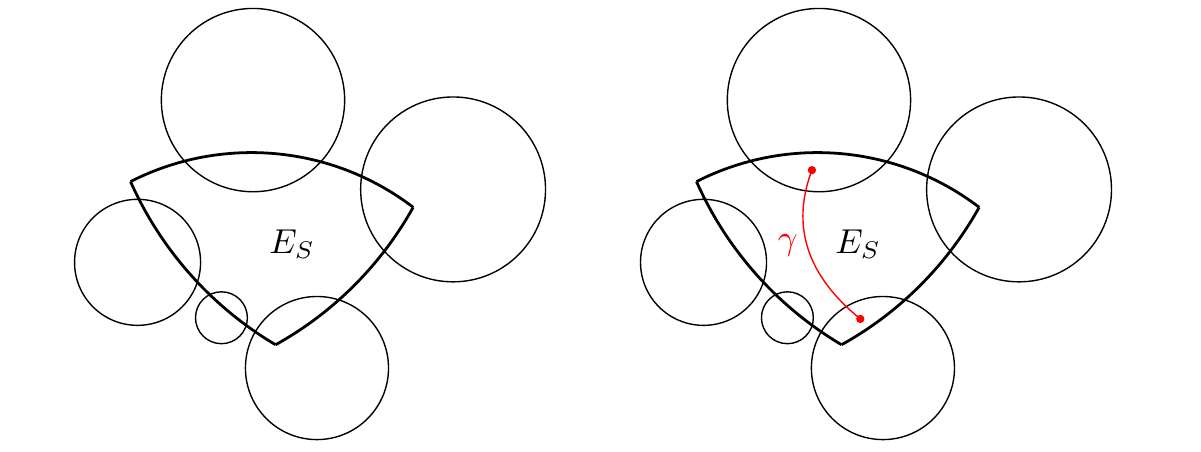}
    \caption{Illustration of the intersection pattern of $E_S$ and the disks $D_u$ for $u \in I^*$.}
    \label{fig-ESintersection}
\end{figure}

With the new small-independent partition $N_{G_3}(v) = S^* \cup I^*$ in hand, we finally bound the local radius inside $D_v$ via geometric arguments.
This part is loosely inspired by a proof in \cite{lokshtanov2022subexponential} for bounding the treewidth of a certain disk graph.
We first briefly explain why the additional property $D_u \nsubseteq U$ for all $u \in I^*$ helps.
For each $S \subseteq S^*$, denote by $E_S = \bigcap_{w \in \{v\} \cup S} D_w$.
Clearly, $E_S$ is a convex region in $D_v$ consisting of some faces of the arrangement of $\mathcal{D}_3$.
A geometric observation shown in \cite{lokshtanov2022subexponential} is that if a disk $D$ is not contained in the union of a set of disks, then the boundary of $D$ crosses the boundary of the intersection of the disks in the set at most twice.
Based on this, we know that the boundary of a disk $D_u$ for $u \in I^*$ crosses the boundary of $E_S$ at most twice for any $S \subseteq S^*$, because $D_u \nsubseteq U$ and thus $D_u \nsubseteq \bigcup_{w \in \{v\} \cup S} D_w$.
Also, $D_u$ cannot be contained in $E_S$ because $D_u \nsubseteq U$.
Since the disks $D_u$ for $u \in I^*$ are disjoint, the intersection pattern of $E_S$ and the disks $D_u$ for $u \in I^*$ looks like the left part of Figure~\ref{fig-ESintersection}: each $D_u$ contains a connected portion of the boundary of $E_S$ (if it intersects $E_S$) and these portions are disjoint.
Therefore, if we only look at the arrangement of the disks corresponding to the vertices $\{v\} \cup S \cup I^*$, then in the part contained in $E_S$, we have a face $E_S \backslash (\bigcup_{u \in I^*} D_u)$ adjacent to the other faces $E_S \cap D_u$ for $u \in I^*$.
This is a nice structure, which guarantees the following property: for any two points in $E_S$, there exists a curve $\gamma$ connecting them which crosses the boundaries of the disks $D_u$ for $u \in I^*$ at most twice (see the right part of Figure~\ref{fig-ESintersection}).
Using this, we sketch a short proof for the local radius inside $D_v$.

Let $\mathcal{S} = \{S \subseteq S^*: E_S \neq \emptyset\}$.
For convenience, for each $S \in \mathcal{S}$, we denote by $A[E_S]$ the induced subgraph of the arrangement graph of $\mathcal{D}_3$ consisting of the faces contained in $E_S$.
We shall prove the following statement for any $d$: if the radius of $A[E_S]$ is at most $r$ for any $S \in \mathcal{S}$ with $|S^*|-|S| = d$, then the radius of $A[E_S]$ is at most $f(\varepsilon,r)$ for any $S \in \mathcal{S}$ with $|S^*|-|S| = d+1$ (here $f$ is some fixed function).
If the statement holds, we can apply induction from the base case $d = -1$ where the radius of $A[E_S]$ is at most $r_{-1} = 0$ for any $S \in \mathcal{S}$ with $|S^*|-|S| = -1$ simply because no such $S \in \mathcal{S}$ exists.
The induction gives us a sequence of number $r_{-1},r_0,r_1,\dots,r_{|S^*|}$ satisfying $r_i = f(\varepsilon,r_{i-1})$ such that the radius of $A[E_S]$ is at most $r_d$ for any $S \in \mathcal{S}$ with $|S^*|-|S| = d$.
Since $|S^*| = O_\varepsilon(1)$, we have $r_{|S^*|} = O_\varepsilon(1)$ and hence the radius of $A[E_\emptyset]$ is at most $r_{|S^*|} = O_\varepsilon(1)$.
Note that $E_\emptyset = D_v$, so it directly follows the local radius of the arrangement graph inside $D_v$ is $O_\varepsilon(1)$.

To see the statement, assume the radius of $A[E_S]$ is at most $r$ for any $S \in \mathcal{S}$ with $|S^*|-|S| = d$.
Consider a set $S \in \mathcal{S}$ with $|S^*|-|S| = d+1$.
If $E_S$ does not intersect $D_w$ for any $w \in S^* \backslash S$, then it only intersects the disks $D_u$ for $u \in I^*$.
In this case, we have the situation in Figure~\ref{fig-ESintersection} and the radius of $A[E_S]$ is 1.
If $E_S$ intersects $D_w$ for some $w \in S^* \backslash S$, then $S \cup \{w\} \in \mathcal{S}$ and $|S^*|-|S \cup \{w\}| = d$.
So the radius of $A[E_{S \cup \{w\}}]$ is at most $r$.
We can have at most $|S^* \backslash S| = O_\varepsilon(1)$ vertices $w \in S^* \backslash S$ with $E_S \cap D_w \neq \emptyset$, and each $A[E_{S \cup \{w\}}]$ is an induced subgraph of $A[E_S]$ or radius $r$.
We further observe that each vertex in $A[E_S]$ is within distance 3 from $A[E_{S \cup \{w\}}]$ for some $w \in S^* \backslash S$ with $E_S \cap D_w \neq \emptyset$.
To see this, consider a face $F$ in $E_S$, which is a vertex in $A[E_S]$.
Pick an arbitrary face $F_0$ in $E_{S \cup \{w_0\}}$ for some $w_0 \in S^* \backslash S$ with $E_S \cap D_{w_0} \neq \emptyset$, which is a vertex in $E_{S \cup \{w_0\}}$.
As argued before and shown in the right part of the right part of Figure~\ref{fig-ESintersection}, we can draw a curve $\gamma$ connecting two points $x \in F$ and $x_0 \in F_0$ which crosses the boundaries of the disks $D_u$ for $u \in I^*$ at most twice.
We go along $\gamma$ from $x$ to $x_0$, and consider the \textit{first} time we cross the boundary of a disk $D_w$ for some $w \in S^* \backslash S$ (which must exist since we have to cross the boundary of $D_{w_0}$ at some point).
At that time, we reach a face in $E_{S \cup \{w\}}$ and we have crossed the face boundaries in $E_S$ at most 3 times (twice for the boundaries of the disks $D_u$ for $u \in I^*$ and once for the boundary of $D_w$).
Thus, we find a path from $F$ to a face in $E_{S \cup \{w\}}$ of length at most 3.
Based on this observation, if we ``expand'' all $A[E_{S \cup \{w\}}]$ a little bit, they will cover all vertices in $A[E_S]$.
Specifically, let $A^+[E_{S \cup \{w\}}]$ be the induced subgraph of $A[E_S]$ consisting of vertices within distance at most 3 from $A^+[E_{S \cup \{w\}}]$ for all $w \in S^* \backslash S$ with $E_S \cap D_w \neq \emptyset$.
Then the radius of each $A^+[E_{S \cup \{w\}}]$ is at most $r+3$ and all $A^+[E_{S \cup \{w\}}]$ cover the vertices in $A[E_S]$.
A simple argument shows that if a connected graph $H$ can be covered by $k$ induced subgraphs of radii at most $\rho \geq 1$, then the radius of $H$ is at most $O(k\rho)$.
Here $k = O_\varepsilon(1)$ and $\rho = r+3$.
The connectedness of $A[E_S]$ can be easily verified.
So the radius of $A[E_S]$ is bounded by a function of $\varepsilon$ and $r$.

Finally, we see that the part of the arrangement graph contained in $D_v$ is of $O_\varepsilon(1)$ radius.
Therefore, $G_3$ is of $O_\varepsilon(1)$ local radius.
This completes the overview for our reduction.

\subsection{Handling disk graphs of bounded local radius and applications}
\label{overview:applications}

We present two meta-theorems that extend the two main approaches to design (E)PTASes on planar graphs to disk graphs whose local-radius is bounded by some constant (that can depend on $\varepsilon$). In particular, let us remind that the class of planar graphs is the class of disk graphs of local-radius $1$. Then, combined with the main technical contribution of this paper, being our local-radius reduction procedure, we derive our applications for general disk graphs. Below, let us elaborate on the two meta-theorems.

\paragraph{Meta-theorem based on the subquadratic grid minor (SQGM) property.} The main concept behind this approach is that of the SQGM property of a graph class, defined as follows. Formally, a graph class $\cal G$ has the SQGM property if there exist constant $\alpha>0$ and $1\leq c<2$ such that, for any $t>0$, every graph in $\cal G$ that excludes the $t\times t$ grid as a minor has treewidth at most $\alpha\cdot t^c$ (being, in particular, subquadratic in $t$). Based on the work of Fomin et al.~\cite{DBLP:journals/jacm/FominLS18}, we know of a large class of problems that admit EPTASes on any hereditary graph class that satisfies the SQGM property. Accordingly, our main efforts are in proving that the class of disk graphs of bounded local-radius admits the SQGM property.

The main ideas in our proof are as follows. First, based on {\em (i)} a known relation between the treewidth of a disk graph and the treewidth of its arrangement graph, {\em (ii)} a simple bound on ply in terms of bounded local-radius, and {\em (iii)} a known relation between the treewidth of a planar graph and the maximum size of a grid it contains as a minor, we observe that, to assert that the class of disk graphs of bounded local-radius admits the SQGM property, it suffices to prove the following. Consider some disk graph $G$ with a realization $\cal D$ such that the corresponding arrangement graph, denoted by $A_{\cal D}$,  contains the $t'\times t'$ grid, denoted by $H$, as a minor. Then, $G$ contains the $t\times t$ grid as a minor for $t=\Omega(t'/r)$, where $r$ is the local-radius of $G$. 

To prove the statement above, we work with a grid $\widehat{H}$ that lies in the ``middle''  of $H$, particularly since it has the property that the distance between any two vertices in it is at most twice the distance between any two vertices in the sets they represent (according to a minor model function) in $G$. Then, we proceed in two steps. In the first step, we identify particular sets of vertices within $\widehat{H}$ that are far away from each other (where the definition of far away depends on the local-radius $r$ of $G$), and which will later be extended to attain the sets that will correspond to a minor model of the $t\times t$ grid in $G$. Specifically, we take vertices that are far away from each other in $\widehat{H}$, and for each such vertex create a set by taking the vertex sets of the four paths that form a ``cross'' around it. With each cross we associate a set of vertices in $G$, which consists of every vertex representing a disk such that at least one vertex in the cross represents a face contained inside that disk. While it will be easy to see that each associated set induces a connected subgraph in $G$, because we work with $\widehat{H}$ (rather than $H$), when we consider two different crosses, we will also get that the sets associated with them are disjoint. In the second step, we consider paths between the aforementioned crosses, and use these paths to decide which vertices to add to the sets associated with the crosses so that they will still induce vertex-disjoint connected subgraphs in $G$, but will also have certain edges between them so as to serve as sets for a minor model of the $t\times t$ grid.

Having proved that the class of disk graphs of bounded local-radius admits the SQGM property, we can derive (based on ~\cite{DBLP:journals/jacm/FominLS18}) that {\sc Finite-Type Component Deletion} (which encompasses, for example, {\sc $\ell$-Component Order Connnectivity} for any fixed $\ell\in\mathbb{N}$), {\sc Treewidth $\eta$-Deletion} for any fixed constant $\eta\in\mathbb{N}_0$ (which encompasses, for example, {\sc Vertex Cover} and {\sc Feedback Vertex}, {\sc Pseudoforest Deletion}, {\sc Cactus Deletion}, {\sc Path Deletion}, {\sc Caterpillar Deletion}, {\sc Max Leaf Spanning Tree}, {\sc Max Internal Spanning Tree}, and more, admit EPTASes on disk graphs of bounded local-radius. In particular, using our local-radius reduction procedure, we get that {\sc Finite-Type Component Deletion}, {\sc Vertex Cover}, {\sc Feedback Vertex Set} {\sc Path Deletion}, {\sc Caterpillar Deletion} and {\sc Pseudoforest Deletion} admit EPTAses on general disk graphs.

\paragraph{Meta-theorem based on Baker's method.} At the heart of Baker's method lies the concept of layering. Specifically, given some source vertex $s$ in a graph $G$, layer $i$ consists of all vertices at distance $i$ from $s$ in $G$. Corresponding to some parameters $\alpha,\beta\in\mathbb{N}$, each layer is assigned one label out of $\{0,1,\ldots,\alpha-1\}$ for some $\alpha\in\mathbb{N}$, so that, starting with layer $0$ and moving onwards to higher labels, the labeling looks like this: $0,0,\ldots,0,1,1,\ldots,1,\alpha-1,\alpha-1,\ldots,\alpha-1,0,0,\ldots,0,1,1,\ldots,1,\alpha-1,\alpha-1,\ldots,\alpha-1,$ and so on, where each label appears consecutively $\beta$ times. Then, a $q$-piece, for some $q\in\{0,1,\ldots,\alpha-1\}$ is a maximal sequence of consecutive layers that starts and ends (except for possibly the first and last pieces) with layers labels $q$ and does not have any layer labeled $q$ in its ``middle''. Essentially, our main efforts are in proving that, when $G$ is a disk graph of bounded local-radius, the subgraph induced by the set of vertices in any $q$-piece has treewidth that is bounded by a function depending only on $\alpha$ and $\beta$. Specifically, this is known in case of planar graphs.

To prove the aforementioned statement, we consider some piece $\cal P$ and a realization $\cal D$ of $G$. The main concept that we define is of a modified arrangement graph corresponding to $\cal P$. Roughly speaking, it is attained by considering the arrangement graph induced by all disks in the layers of $\cal P$ as well as all layers preceding it, and then contracting some of the vertices that represent faces that are not contained in any disk represented by a vertex in a layer of $\cal P$. We prove that {\em (i)} the radius of the modified arrangement graph is at most $O(\alpha\cdot\beta\cdot r)$, and {\em (ii)} the arrangement graph of the piece is a minor of the modified arrangement graph. Using this and the known result for planar graphs mentioned in the previous paragraph, we are able to prove the desired statement. 

Having this statement at hand, we are able to follow the standard way of applying Baker's approach in order to prove that any vertex deletion problem $\Pi$ to a graph class characterized by a finite set of forbidden (induced or not) subgraphs admits an EPTAS on disk graphs of bounded local-radius. In particular, using our local-radius reduction procedure, we get that {\sc $P_t$-Vertex Deletion} for any $1\leq t\leq 5$, {\sc $\ell$-Small Cycle Hitting} for any fixed constant $\ell\in\mathbb{N}$, $\ell\geq 3$, which encompasses {\sc Triangle Hitting}, and {\sc $d$-Bounded Degree Vertex Deletion} for any fixed $d\in\mathbb{N}_0$  admit EPTASes on general disk graphs.

\section{Preliminaries}\label{sec:prelims}

For a constant $c>0$, the notation $O_c(1)$ hides constants that depend only on $c$.

\paragraph{Graphs.} Let $G$ be a graph. We use $V(G)$ and $E(G)$ to denote the vertex set and edge set of $G$, respectively. For $U\subseteq V(G)$, $G[U]$ is the subgraph of $G$ induced on $U$ and $G-U$ is the subgraph of $G$ induced on $V(G)\setminus U$.  For $u\in V(G)$, $N_G(u)=\{v \in V(G)\backslash\{u\}~:~\{u,v\}\in E(G)\}$ and $N_G[u]=\{u\}\cup N_G(u)$.
Similarly, for $U \subseteq V(G)$, $N_G(U) = \{v \in V(G) \backslash U~:~\{u,v\}\in E(G) \text{ for some } u \in U\}$ and $N_G[U] = U \cup N_G[U]$.
If the graph is clear from the context, then we may remove the subscript $G$. When $G[U]$ is connected, we say that $U$ is a connected set. A path in $G$ is a sequence of distinct vertices $P=(v_0,v_1,\ldots,v_{\ell})$ such that for each $i\in [\ell]$, $\{v_{i-1},v_i\}\in E(G)$, and the length of the path is the length of the sequence minus one (i.e., the number of edges). Here, we say that $P$ is a path from $u$ to $v$. The contraction of an edge $\{u,v\}$ in $G$ is the operation that replaces $u$ and $v$ by a new vertex whose neighbor set is the union of the neighbor sets of $u$ and $v$. The contraction of a connected set $U$ in $G$ is the contraction of the edges of any spanning tree of $G[U]$ in $G$. We note that the choice of the spanning tree, as well as the order of the edge contractions, are immaterial---they yield the same graph (up to isomorphism).
For two vertices $u$ and $v$ in $G$, $\dist_G(u,v)$ is the length of a shortest length path from $u$ to $v$ in $G$. The {\em radius} of a graph $G$, denoted by $\mathsf{rad}(G)$, is the minimum integer $r$ such that there exists a vertex in $G$ whose distance to each of the other vertices is at most $r$. Moreover, the {\em diameter} of a graph $G$, denoted by $\mathsf{diam}(G)$, is the minimum integer $d$ such that the distance between any two vertices in $G$ is at most $d$.  We say that a graph $H$ is a {\em minor} of $G$ if $H$ can be obtained from a subgraph of $G$ by a series of edge contractions. 
In other words, there exists a function $\phi: V(H)\rightarrow 2^{V(G)}$, called minor model, such that for all $h\in V(H)$, $G[\phi(h)]$ is a connected graph, for all distinct $h,h'\in V(H)$, $\phi(h)\cap\phi(h')=\emptyset$, and for all $\{h,h'\}\in E(H)$, there exist $u\in \phi(h)$ and $v\in \phi(h')$ such that $\{u,v\}\in E(G)$. 

\begin{definition}[{\bf Tree Decomposition}]\label{def:treewidth}
A \emph{tree decomposition} of a graph $G$ is a pair $(T,\beta)$ 
where $T$ is a rooted tree 
and $\beta:V(T) \rightarrow 2^{V(G)}$, such that
\begin{itemize}
\itemsep0em 
\item $V(G)=\bigcup_{t\in V(T)} \beta(t)$, 
\item\label{item:twedge} for any edge $\{x,y\} \in E(G)$ there exists a node $v \in V(T)$ such that $x,y \in \beta(v)$, and
\item\label{item:twconnected} for any vertex $x \in V(G)$, the subgraph of $T$ induced by the set $T_x = \{v\in V(T): x\in\beta(v)\}$ is connected.
\end{itemize}
The {\em width} of $(T,\beta)$ is $\max_{v\in V(T)}\{|\beta(v)|\}-1$. The {\em treewidth} of $G$, denoted by $\tw(G)$, is the minimum width over all tree decompositions of $G$.
\end{definition}

In the above definition if $T$ is a path, then $(T,\beta)$ is a path decomposition and the pathwidth is defined analogously. 

A {\em $t\times t'$-grid} is the graph  whose vertex set is $\{v_{i,j}: i\in[t],j\in[t']\}$, and the edge set  is  $\{\{v_{i,j},v_{i',j'}\}: i,i'\in[t], j,j'\in[t'],|i-i'|+|j-j'|=1\}$.
A graph $G$ is {\em planar} if there is a mapping from every vertex $v$ in $V(G)$ to a point on the plane, and from every edge $e\in E(G)$ to a curve on the plane where the extreme points of the curve are the points mapped to the endpoints of $e$, and all curves are disjoint except on their extreme points. 

\paragraph{Closed and Finite-Basis Graph Classes.} In this part of the preliminaries, we consider graph classes of forms that will be relevant to our applications.
\begin{definition}
A graph class $\mathcal{G}$ is \textbf{induced-subgraph-closed} if for any $G \in \mathcal{G}$, all induced subgraphs of $G$ are also in $\mathcal{G}$.
A graph class $\mathcal{G}$ is \textbf{disjoint-union-closed} if for any $G,G' \in \mathcal{G}$, the disjoint union of $G$ and $G'$ is also in $\mathcal{G}$.
\end{definition}

We note that an induced-subgraph-closed graph class is equivalently said to be a {\em hereditary} graph class.

\begin{fact} \label{fact-closed}
Let $\mathcal{G}$ be a nonempty graph class that is induced-subgraph-closed and disjoint-union-closed.
If $G \in \mathcal{G}$, then any graph obtained from $G$ by adding new vertices with no edges is also in $\mathcal{G}$.
\end{fact}
\begin{proof}
Since $\mathcal{G}$ is induced-subgraph-closed (and nonempty), the graph with a single vertex is in $\mathcal{G}$.
As $\mathcal{G}$ is also disjoint-union-closed, any graph with no edges is in $\mathcal{G}$.
Let $G \in \mathcal{G}$ and $G'$ be a graph obtained from $G$ by adding new vertices with no edges.
Then $G'$ is the disjoint union of $G$ and a graph with no edges.
Thus, $G' \in \mathcal{G}$.
\end{proof}

In a graph $G$, we say $u \in V(G)$ is a \textit{false twin} of $v \in V(G)$ if $N_G(u) = N_G(v)$.
We use $\mathsf{FT}_G(v)$ to denote the set of all false twins of $v$ in $G$.
Then we define a \textsf{clone} operation, which adds to a graph a new vertex that is the false twin of an original vertex.
Formally, for a graph $G$ and a vertex $v \in V(G)$, define $\mathsf{clone}(G,v)$ as the graph obtained from $G$ by adding a new vertex $u$ and new edges $(u,w)$ for all $w \in N_G(v)$.

\begin{definition}[{\bf Clone-closed}]
A graph class $\mathcal{G}$ is \textbf{clone-closed} if for any $G \in \mathcal{G}$ and any vertex $v \in V(G)$, $\mathsf{clone}(G,v) \in \mathcal{G}$.
More generally, $\mathcal{G}$ is $k$-\textbf{clone-closed} for some integer $k \geq 1$ if for any $G \in \mathcal{G}$ and any vertex $v \in V(G)$ such that $|\mathsf{FT}_G(v)| \geq k$, $\mathsf{clone}(G,v) \in \mathcal{G}$.
\end{definition}

For convenience, we say a graph class $\mathcal{G}$ is \textit{closed} if $\mathcal{G}$ is induced-subgraph-closed, disjoint-union-closed, and $k$-clone-closed for some $k \geq 1$.
In fact, almost all vertex-deletion problems that have been previously studied have closed target sets.

A \textit{triangle bundle} of size $k$, denoted by $T_k$, is the graph obtained by ``gluing'' $k$ copies of triangles (i.e., $K_3$) at one vertex, i.e., picking one vertex in each triangle and identifying the $k$ picked vertices.
We say a graph $G$ is \textit{$T_k$-free} if $G$ does not contain $T_k$ as a subgraph, and a graph class $\mathcal{G}$ is \textit{$T_k$-free} if every $G \in \mathcal{G}$ is $T_k$-free.
A graph class is \textit{triangle-bundle-free} if it is $T_k$-free for some $k \geq 1$.

\begin{definition}[{\bf Finite Basis}]
Let $\mathcal{B}$ be a class of connected graphs.
The \textbf{disjoint-union closure} of $\mathcal{B}$ is the graph class $\mathcal{G}$ consisting of all graphs whose every connected component is isomorphic to some graph in $\mathcal{B}$.
A graph class $\mathcal{G}$ is \textbf{of finite basis} if there exists a finite class $\mathcal{B}$ of connected graphs such that $\mathcal{G}$ is the disjoint-union closure of $\mathcal{B}$.
\end{definition}

Note that a graph class $\mathcal{G}$ of finite basis, being the disjoint-union closure of some class $\cal B$ of connected graphs, must be disjoint-union-closed, $k$-clone-closed for some $k \geq 1$, and triangle-bundle-free. Indeed, the maximum size of a connected component in a graph $G$ in $\cal G$ is bounded by some constant $c$, being the maximum size of a graph in $\cal B$, which in particular implies that $G$ cannot have more than $c$ false twins unless they are all isolated vertices and the graph consisting of an isolated vertex belongs to $\cal B$, and it cannot have a triangle bundle of size $c$.
However, $\cal G$ may not necessarily be induced-subgraph-closed.

\paragraph{Problem Definitions.} For all the problems considered here, the input is a graph $G$ and the output is a minimum size vertex subset $S$ satisfying some property (based on the problem) which is mentioned below. 
\begin{itemize}
\setlength{\itemsep}{-2pt}
\item {\sc Vertex Cover}: $G-S$ is an edgeless graph. 
\item {\sc Feedback Vertex Set}: $G-S$ is a forest. 
\item {\sc Pseudoforest Deletion}: $G-S$ is pseudoforest. That is, each connected component in $G-S$ has at most one cycle. 
\item {\sc Caterpillar Deletion}: $G-S$ is a disjoint union of caterpillars.
\item {\sc Treewidth $\eta$-Deletion}: Treewidth of $G-S$ is at most $\eta$. 
\item {\sc Pathwidth $\eta$-Deletion}: Pathwidth of $G-S$ is at most $\eta$. {\sc Pathwidth $1$-Deletion} is equivalent to {\sc Caterpillar Deletion}.
\item {\sc Triangle Hitting}: $G-S$ is a triangle-free graph
\item {\sc $\ell$-Small Cycle Deletion}: $G-S$ does not have a cycle on at most $\ell$ vertices. {\sc $3$-Small Component Deletion} is equivalent to {\sc Triangle Hitting}.
\item {\sc $P_t$-Vertex Deletion}: $G-S$ does not contain a path of length $t-1$. 
\item {\sc $d$-Bounded Degree Deletion}: The degree of any vertex in $G-S$ is at most $d$. 
\item {\sc Cactus Deletion}: $G-S$ is a cactus graph. That is, any 2-connected component in $G-S$ is a cycle. 
\item {\sc $\ell$-Component Order Connectivity}: The number of vertices in each connected component of $G-S$ is at most $\ell$. 
\item {\sc Finite-Type Component Deletion}: $G-S \in \cal G$ for a graph class $\cal G$ of finite basis. This problem encompasses {\sc $\ell$-Component Order Connectivity} and \textsc{Vertex Cover}.
\item {\sc Path Deletion}: $G-S$ is a disjoint union of paths.
\end{itemize}

\paragraph{Geometric Intersection Graphs and Disk Graphs.}
Let $\cal O$ be a collection of geometric objects. 
Then, the {\em geometric intersection graph} of $\cal O$ is the graph $G$ with vertex set $V(G)=\{v_O: O\in{\cal O}\}$ (i.e., there is an implicit bijection between the objects and the vertices), where two vertices are adjacent if and only if their corresponding objects intersect. 
The set ${\cal O}$ is called a {\em realization} of $G$.
When $\cal O$ is a set of geometric objects on the Euclidean plane, the \textit{arrangement} of $\mathcal{O}$ refers to the subdivision of the plane formed by the boundaries of the objects in $\mathcal{O}$, which partitions the plane into \textit{faces}, interior-disjoint regions bounded by the boundaries of the objects in $\mathcal{O}$.
For convenience, sometimes we call these faces {\em arrangement faces of $\cal O$} (or simply {\em faces of $\cal O$} for short).
If ${\cal O}$ is a set of closed disks on Euclidean plane, then $G$ is called a {\em disk graph}. 
When we deal with a disk graph, we use the term realization to refer only to a realization where all objects are disks. 
We say that a collection of disks $\cal D$ is in {\em general position} if {\em (i)} no two disks in $\cal D$ are identical or tangent to each other (i.e., with boundaries intersecting at exactly one point) and {\em (ii)} no point in the plane is on the boundaries of three disks in $\cal D$.
By slightly perturbing the radii of disks in a realization of a disk graph $G$, we have the following. 

\begin{proposition}[Folklore]\label{prop:generalPosition}
Every disk graph $G$ has a realization where the disks are in general position.
\end{proposition}

So, throughout the paper and without loss of generalization, whenever we consider a (hypothetical) realization of a disk graph, we implicitly assume that it is in general position.

\begin{definition}[{\bf Ply}]
The \textbf{ply} of a disk set $\mathcal{D}$ is the largest integer $p$ such that there exists a point in the plane lying in at least $p$ disks in $\mathcal{D}$.
The \textbf{ply} of a disk graph $G$ is the minimum ply of a disk set that realizes $G$.
\end{definition}

\begin{fact} \label{fact-sparse}
Let $\mathcal{D}$ be a disk set of ply $p$.
Then any square $Q$ in the plane of side-length $\ell$ intersects at most $16p$ disks in $\mathcal{D}$ with diameter at least $\ell$.
\end{fact}
\begin{proof}
Scale $Q$ up to a square $Q'$ by a factor of $2$. So, $Q$ and $Q'$ have the same centre, and $Q'$ has side-length $2\ell$. For each disk $D$ in $\cal D$ that intersects $Q$, consider a disk of diameter exactly $\ell$ that intersects $Q$ and is contained in $D$. Let ${\cal D}'$ denote the set of these disks.  Then, the centre of each disk in ${\cal D}'$ belongs to $Q'$. Split $Q'$ into $4\times 4$ equal-size squares. Notice that, since the ply of $\cal D$ is $p$, each of these squares can only contain $p$ centres of disks from ${\cal D}'$. This yields that $|{\cal D}'|\leq 16p$.
\end{proof}

\begin{fact} \label{fact-plyedge}
For a graph $G$ of ply $p$, the number of edges in $G$ is $O(p|V(G)|)$.
\end{fact}
\begin{proof}
Note that in a disk graph $G$ with a realization $\cal D$ of ply at most $p$, every disk $D$ can intersect only $O(p)$ disks that are at least as large as $D$. This means that the number of edges in $G$ drawn by D is $O(p|V(G)|)$.
\end{proof}

For a set of objects $\mathcal{D}$, the \textit{arrangement graph} of $\cal D$, denoted by $A_{\mathcal{D}}$, is defined as follows. Every face of $\cal D$ that is contained in a disk in $\cal D$ is represented by a vertex in $A_{\cal D}$, and two vertices in $A_{\cal D}$ are \textit{adjacent} if and only if the faces that they represent are adjacent, that is, have a common boundary arc (that is not a single point).
Note that $A_{\cal D}$ is necessarily a planar graph (folklore; see, e.g., \cite{lokshtanov2022subexponential}).
For a region $R \subseteq \mathbb{R}^2$, we denote by $A_{\mathcal{D}}[R]$ the induced subgraph of $A_{\mathcal{D}}$ consisting of the arrangement faces of $\mathcal{D}$ that are contained in $R$.
In particular, for a disk $D \in \mathcal{D}$, $A_{\mathcal{D}}[D]$ is the induced subgraph of $A_{\mathcal{D}}$ consisting of the arrangement faces of $\mathcal{D}$ contained in $D$.

\begin{definition}[{\bf Local Radius and Local Diameter}]
The \textbf{local radius} (resp., {\em local diameter}) of a disk set $\mathcal{D}$ is defined as $\max_{D \in \mathcal{D}} \mathsf{rad}(A_{\mathcal{D}}[D])$ (resp., $\max_{D \in \mathcal{D}} \mathsf{diam}(A_{\mathcal{D}}[D])$).
The \textbf{local radius} (resp., {\em local diameter}) of a disk graph $G$ is the minimum local radius (resp., local diameter) of a disk set that realizes $G$.
\end{definition}

\begin{fact} \label{fact-planarlocalradius}
A graph is planar if and only if it is a disk graph of local radius at most 1.
\end{fact}

\begin{proof}
In one direction, suppose that we are given a planar graph $G$. By the Cycle Packing Theorem, we know that every planar graph admits a realization by a set of disks where every two disks are either disjoint or intersect in exactly one point)~\cite{thurston1979geometry}. Consider such a realization $\cal D$ for $G$. Observe that $\cal D$ is not in general position. However, we can scale up (i.e., keep the centre the same but increase the radius) each disk in $\cal D$ by a small enough factor $\varepsilon$ so that the realization will be in general position and the following property will hold. Each disk $D$ in $\cal D$ will consist exactly of (1) one face that is contained only in $D$ among the disks in $\cal D$; (2) for each disk $D'$ that intersects $D$, one face contained only in $D$ and $D'$ among the disks in $\cal D$, so that the face of type $1$ shares a boundary with each face of type (2).  However, this means that for the new realization $\cal D$,  $A_{\mathcal{D}}[D]$ is a star for every disk $D\in{\cal D}$, which in particular means that the local radius is $1$.

For the second direction, suppose that we are given a disk graph $G'$ of local radius at most $1$. Note that for any planar graph $H$, performing any of the two following operations keeps the graph $H$ planar: (1) adding a new vertex and making it adjacent to exactly one vertex in $H$; (2) adding a new vertex and making it adjacent to exactly two vertices in $H$, which are neighbors. So, because the class of disk graphs of local radius at most $1$ is clearly hereditary, for the sake of proving that $G'$ is planar, we may repeatedly remove from $G'$ every vertex that has either degree $1$ or degree $2$ and its two neighbors are adjacent, and prove that the resulting graph, denoted by $G$, is planar. So, in what follows, we focus on $G$, and we let $\cal D$ be some realization (in general position) of $G$ whose local radius is at most $1$.

To proceed, we will use the following claim.

\begin{claim}\label{claim:radius1a}
Let $A,B,C$ be three disks such that the boundaries of $A,B$ intersect in exactly two points, $p,q$, and $C$ contains $p$. Then, the local radius of $\{A,B,C\}$ is at least $2$.
\end{claim}
\begin{proof}
Observe that $A$ and $B$ divide the plane into four connected regions: $A\cap B$, $A\setminus B$, $B\setminus A$, and the fourth region, to which we will refer to as $\overline{A\cup B}$, which contains all points that belong to neither $A$ nor $B$. Now, observe that each one of these four regions contains a nontrivial subregion (i.e., that contains a point that is on none of the boundaries of $A,B,C$) that is contained in $C$ (see Fig.~\ref{fig:prelims}). Further, the face that represents the vertex of shortest distance to all other vertices in $A_{\{A,B,C\}}[C]$ must belong to one of the following regions: $A\cap B\cap C$, $(A\setminus B)\cap C$, $(B\setminus A)\cap C$ or $\overline{A\cup B}\cap C$. If it is $A\cap B\cap C$, then its distance to a vertex that represents a face in $\overline{A\cup B}\cap C$ is at least $2$, since these two faces do not share a boundary (that is not just a single point). So, if it is $A\cap B\cap C$, and symmetrically, if it belongs to $\overline{A\cup B}\cap C$, then the radius of $\mathsf{rad}(A_{\{A,B,C\}}[C])$ and hence the local radius of $\{A,B,C\}$ is at least $2$. If it is $(A\setminus B)\cap C$, then its distance to the vertex that represents the face $(B\setminus A)\cap C$ is at least $2$, since these two faces do not share a boundary (that is not just a single point). So, if it is $(A\setminus B)\cap C$, and symmetrically, if it is $(B\setminus A)\cap C$, then the local radius of $\{A,B,C\}$ is at least $2$.
\end{proof}

\begin{figure}[t]
\begin{center}
\includegraphics[scale=0.8]{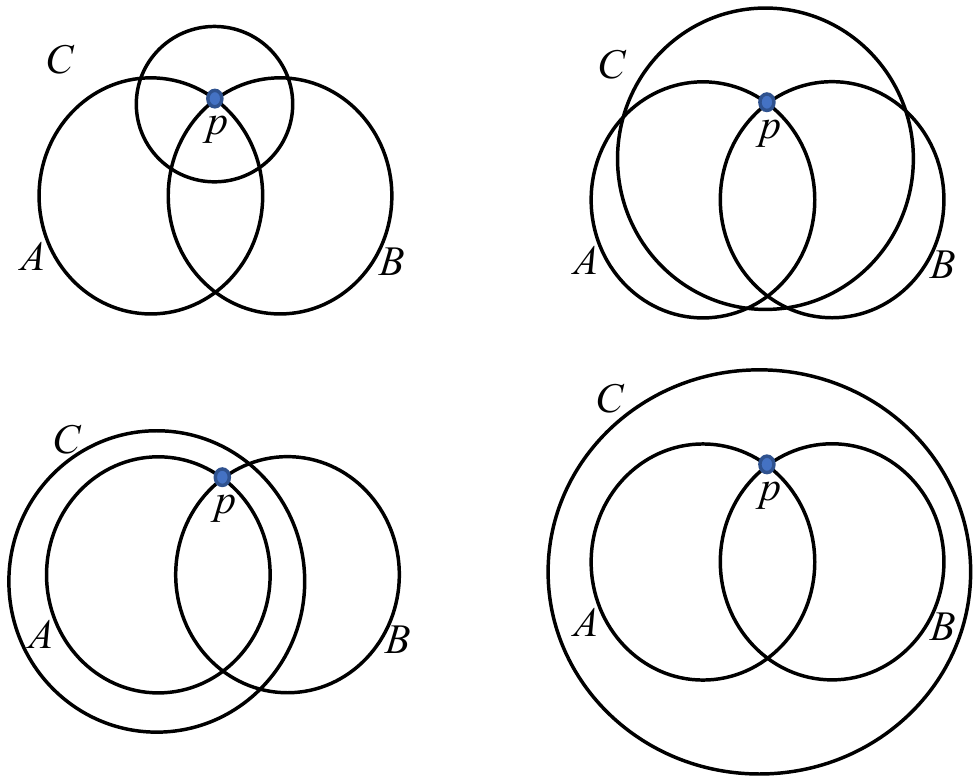}
\end{center}
\caption{An illustration for the proof of Fact \ref{fact-planarlocalradius}.}\label{fig:prelims}
\end{figure}

We now return to the proof of the fact. From Claim \ref{claim:radius1a}, we derive that $\cal D$ satisfies the following property, which we call  Property (*): There does not exist an intersection point of two disks in $\cal D$ that is contained in another disk in $\cal D$. We continue by proving another claim.

\begin{claim}\label{claim:radius1b}
There do not exist two disks $A,B\in{\cal D}$ such that $B$ contains $A$.
\end{claim}
\begin{proof}
Targeting a contradiction, suppose that there exist two disks in $\cal D$ such that one contains another. Let $A,B\in{\cal D}$ be two disks such that $B$ contains $A$ and there do not exists two disks $A',B'\in{\cal D}$ such that $B'$ contains $A'$ and $A'$ is smaller than $A$. Because $\cal D$ satisfies Property (*) and since $\cal D$ is in general position, all of the disks that intersect $A$ must contain it and are larger than it. To proceed, we consider three cases. In the first case, suppose that $A$ is contained in exactly two disks. However, this means that the vertex represented by $A$ in $G$ has exactly two neighbors in $G$ (represented by the two disks that contain $A$) and these two neighbors are adjacent. This contradicts the way we have defined $G$ based on $G'$. In the second case, suppose that $A$ is contained in exactly one disk. However, this means that the vertex represented by $A$ in $G$ has exactly one neighbor in $G$ (represented by the disk that contains $A$), and again this contradicts the way we have defined $G$ based on $G'$. In the third case, suppose that $A$ is contained in at least three disks. In particular, this implies that the ply of $\cal D$ is at least $4$. However, by Observation \ref{obs:boundedPly}, this implies that the local radius of $G$ must be at least $2$, which is a contradiction. Since every case yielded a contradiction, we conclude that the claim is true.
\end{proof}

Again, we return to the proof of the fact, and, to complete it, we note that Kedem et al.~\cite{kedem1986union} proved that any graph with a realization by disks (or, more generally, pseudodisks) that satisfies Property (*) and the property in Claim \ref{claim:radius1b} is a planar graph. Thus, we conclude that $G$ is a planar graph.
\end{proof}

\paragraph{Problems and Optimization: Other Definitions.} Given an optimization problem $\Pi$ and an instance $I$ of $\Pi$, we use $\mathsf{opt}_{\Pi}(I)$ to denote the optimal value of a solution to $I$. When $\Pi$ is the vertex deletion to a graph class $\cal G$, we denote $\mathsf{opt}_{\cal G}(I)=\mathsf{opt}_{\Pi}(I)$. When $I$ is clear from context, we drop it. For a minimization problem $\Pi$ and an instance $I$ of $\Pi$, a $c$-approximate solution is a solution whose value is at most $c\cdot\mathsf{opt}_{\Pi}(I)$.
A minimization problem $\Pi$ admits a {\em polynomial-time approximation scheme (PTAS)} if, for any fixed constant $\varepsilon>0$, given an instance $I$ of $\Pi$, a $(1+\varepsilon)$-approximate solution can be compute in time $O(n^{f(\frac{1}{\varepsilon}})$ for some function $f$ that depends only on $\varepsilon$. When the running time is further restricted to be of the form $f(\frac{1}{\varepsilon})\cdot n^{O(1)}$, where the exponent of $n$ is independent of $\varepsilon$, we say that the problem admits an {\em efficient PTAS (EPTAS)}.

Consider a problem $\Pi$ that takes a geometric graph as input. We say that an (exact or approximation) algorithm for $\Pi$ is {\em non-robust} if it supposes to be given a realization of the geometric graph as part of its input. Otherwise, that is, if the algorithm supposes to be given only the graph (i.e., its vertex set and edge set) without a realization, then it is {\em robust}.

In what follows, we define three additional properties, being separability, minor-bidimensionality, and CMSO-expressibility, which we will essentially use when invoking other results as black boxes. So, the reader may choose to skip them for the first reading of the paper.

\begin{definition}[{\bf Separability}] Let $f: \mathbb{N} \rightarrow \mathbb{N}$ be a function. A vertex  deletion problem $\Pi$ is {\em $f$-separable} if, for any graph $G$, subset $L \subseteq V(G)$, and and optimal solution $S$, it holds that
$$|S\cap L|-f(t) \leq \mathsf{opt}_{\Pi}(G[L]) \leq|S\cap L| + f(t),$$
where $t$ is the number of vertices in $L$ that are adjacent in $G$ to vertices outside $L$. $\Pi$ is {\em separable} if there exists a function $f$ such that $\Pi$ is $f$-separable. $\Pi$ is {\em linear separable} if there exists a constant $c$ such that $\Pi$ is $c\cdot t$ separable.
\end{definition}

A vertex deletion problem $\Pi$ is {\em minor closed} if given two graphs $G$ and $H$ such that $H$ is a minor of $G$, $\mathsf{opt}_{\Pi}(H)\leq \mathsf{opt}_{\Pi}(G)$.

\begin{definition}[{\bf Minor-Bidimensional}]
A vertex deletion problem $\Pi$ is {\em minor-bidimensional} if it is minor closed, and there exists a constant $\beta>0$ such that for any $k\in\mathbb{N}$, $\mathsf{opt}_{\Pi}(G)\geq\beta k^2$ where $G$ is the $k\times k$ grid.
\end{definition}

 The syntax of Monadic Second Order Logic (MSO) of graphs includes the logical connectives $\wedge, \vee, \neg, \Rightarrow, \Leftrightarrow$, variables for vertices, edges, sets of vertices, the quantifers $\forall,\exists$ that can be applied to these variables, and the following five binary relations:
 \begin{enumerate}
\item $u \in U $, where $u$ is a vertex variable and $U$ is a vertex set variable;
\item $d \in D$, where $d$ is an edge variable and $D$ is an edge set variable;
\item {\bf inc}$(d,u)$, where $d$ is an edge variable, $u$ is a vertex variable, and the interpretation is that the edge $d$ is incident with the vertex $u$;
\item {\bf adj}$(u,v)$, where $u$ and $v$ are vertex variables and the interpretation is that $u$ and $v$ are adjacent;
\item equality of variables representing vertices, edges, sets of vertices, and sets of edges.
\end{enumerate}

In addition to the usual features of MSO, if we have atomic sentences testing whether the cardinality of a set is equal to $q$ modulo $r$, where $q$ and $r$ are integers such that $0 \leq q < r$ and $r \geq 2$, then this extension of MSO is called counting monadic second-order logic (CMSO). Thus, CMSO is MSO enriched with the following atomic sentence for a set $S$: {\bf card}$_{q,r}(S)$ = {\bf true} if and only if $|S| = q$ modulo $r$. 

Min-CMSO and Max-CMSO problems are graph optimization problems for which the objective is to find a maximum- or minimum-sized vertex or edge set satisfying a CMSO-expressible property. For more information, we refer the reader to \cite{CyganFKLMPPS15}.

\section{Reduction to bounded local radius}
\label{sec:redBLR}

This section is dedicated to proving the following theorem, the formal version of Theorem~\ref{thm-reduction-informal}.

\begin{theorem} \label{thm-reduction}
Let $\mathcal{G}$ be a graph class that is closed and triangle-bundle-free.
Given an approximation factor $\varepsilon > 0$ and a graph $G$ of $n$ vertices, one can compute in $n^{O(1)}$ time an induced subgraph $G'$ of $G$ and two auxiliary sets $Y_1,Y_2 \subseteq V(G)$ such that the following conditions hold.
\begin{itemize}
    \item If $G$ is a disk graph, then $G'$ is a disk graph of local radius $(\frac{1}{\varepsilon})^{O(1)}$.
    \item Given a $(1+\frac{\varepsilon}{4})$-approximation solution for $\mathbf{P}_\mathcal{G}$ on $G'$ and the auxiliary sets $Y_1,Y_2$, one can compute in $n^{O(1)}$ time a $(1+\varepsilon)$-approximation solution for $\mathbf{P}_\mathcal{G}$ on $G$.
\end{itemize}
Furthermore, the above also holds for graph classes $\mathcal{G}$ of finite basis.
\end{theorem}

We mainly consider the case where the graph class $\mathcal{G}$ is closed and triangle-bundle-free.
For graph classes of finite basis, the reduction is exactly the same, with a slightly different analysis, which is discussed in Section~\ref{sec-finitebasis}.
Also, for convenience of exposition, we discuss the proof of Theorem~\ref{thm-reduction} in the case where the graph $G$ is a disk graph.
The same reduction applies when $G$ is not a disk graph, while the resulting graph $G'$ is not necessarily a disk graph of bounded local radius (or even not a disk graph); we shall briefly discuss this in Section~\ref{sec-nondisk}.

Let $\mathcal{G}$ be a graph class that is closed and triangle-bundle-free.
Fix a sufficiently large constant $k$ such that $\mathcal{G}$ is $k$-clone-closed and $T_k$-free.
Our reduction consists of three steps, which are given in Section~\ref{sec-step1}, \ref{sec-step2}, and \ref{sec-step3}, respectively.
Each step produces an induced subgraph of the graph obtained in the previous step.
The resulting graph of each step becomes more and more structured (and finally of bounded local-radius).
All three steps themselves are simple and easily implementable.
The interesting part lies in the analyses of the second and third steps.

\subsection{First step: reducing to bounded ply} \label{sec-step1}

The first step of our reduction is standard, which goes from a (general) disk graph to a disk graph of bounded ply.
Formally, we prove the following lemma.

\begin{lemma} \label{lem-reduction1}
Given $\varepsilon > 0$ and a disk graph $G$ of $n$ vertices, one can compute in $n^{O(1)}$ time an induced subgraph $G_1$ of $G$ such that the following conditions hold.
\begin{itemize}
    \item The ply of $G_1$ is $(\frac{1}{\varepsilon})^{O(1)}$.
    \item For any $(1+\varepsilon)$-approximation solution $S_1$ for $\mathbf{P}_\mathcal{G}$ on $G_1$, $(V(G) \backslash V(G_1)) \cup S_1$ is a $(1+\varepsilon)$-approximation solution for $\mathbf{P}_\mathcal{G}$ on $G$.
\end{itemize}
\end{lemma}

Let $\varepsilon>0$ and $G$ be as in the lemma.
The algorithm for generating $G_1$ is shown in Algorithm~\ref{alg-reduction1}.
Here $\textsc{ApprxMaxClique}$ is a constant-approximation algorithm for \textsc{Maximum Clique} on disk graphs: $\textsc{ApprxMaxClique}(G_1)$ returns a clique in $G_1$ of size $\Omega(m)$ where $m$ is the size of a maximum clique in $G_1$.
There exists such an algorithm which runs polynomial time; indeed, \textsc{Maximum Clique} on disk graphs even admits an EPTAS \cite{bonamy2021eptas}.
What Algorithm~\ref{alg-reduction1} does is quite simple.
It begins with $G_1 = G$, and keep deleting cliques of size at least $c(k,\varepsilon)$ from $G_1$ until the algorithm $\textsc{ApprxMaxClique}$ cannot find a clique in $G_1$ of size at least $c(k,\varepsilon)$.
Here $c(k,\varepsilon)$ is a parameter depending on $k$ and $\varepsilon$ to be determined shortly, and we shall guarantee that $c(k,\varepsilon)$ is polynomial in $\frac{1}{\varepsilon}$.
To bound the ply of the resulting $G_1$, we observe the following simple fact, which states that the ply of a disk graph differs from its maximum-clique size by only a constant factor.
\begin{fact} \label{fact-plyclique}
For a disk graph $H$ with maximum-clique size $h$, the ply of $H$ is in the range $[\frac{h}{4},h]$.
\end{fact}
\begin{proof}
Clearly, any realization of $H$ has ply at most $h$.
Indeed, if there is a realization of $H$ with ply larger than $h$, then there exists a point in the plane which lies in at least $h+1$ disks in the realization; these disks pairwise intersect and thus form a clique in $H$ of size at least $h+1$.
On the other hand, any realization of $H$ has ply at least $\frac{h}{4}$.
To see this, recall the well-known result by Danzer \cite{danzer1986losung} stating that a set of pairwise intersecting disks in the plane can be stabbed by $4$ points.
An $h$-clique in $H$ corresponds to $h$ pairwise intersecting disks in the realization.
Danzer's result implies the existence of $4$ points in the plane that stab all these $h$ disks.
Then one of these $4$ points must stab at least $\frac{h}{4}$ disks.
Thus, the ply of the realization is at least $\frac{h}{4}$.
\end{proof}

\begin{algorithm}[h]
    \caption{\textsc{Reduce1}$(G)$}
	\begin{algorithmic}[1]
	    \State $G_1 \leftarrow G$
	    \State $K \leftarrow \textsc{ApprxMaxClique}(G_1)$
	    \While{$|K| \geq c(k,\varepsilon)$}
	        \State $G_1 \leftarrow G_1 - K$
	        \State $K \leftarrow \textsc{ApprxMaxClique}(G_1)$
	    \EndWhile
        \State \textbf{return} $G_1$
	\end{algorithmic}
	\label{alg-reduction1}
\end{algorithm}

The graph $G_1$ returned by Algorithm~\ref{alg-reduction1} has maximum-clique size $O(c(k,\varepsilon))$ since the algorithm $\textsc{ApprxMaxClique}$ fails to find a clique in $G_1$ of size at least $c(k,\varepsilon)$.
By the above fact, the ply of $G_1$ is $O(c(k,\varepsilon))$, which is $(\frac{1}{\varepsilon})^{O(1)}$ if we choose $c(k,\varepsilon)$ polynomial in $\frac{1}{\varepsilon}$.
Now it suffices to show the second condition in Lemma~\ref{lem-reduction1}.
Consider a $(1+\varepsilon)$-approximation solution $S_1$ for $\mathbf{P}_\mathcal{G}$ on $G_1$.
Define $S_1^+ = (V(G) \backslash V(G_1)) \cup S_1$.
Clearly, $S_1^+$ is a feasible solution for $\mathbf{P}_\mathcal{G}$ on $G$, because $G- S_1^+ = G_1 - S_1 \in \mathcal{G}$.
We show that $|S_1^+| \leq (1+\varepsilon) \cdot \mathsf{opt}_\mathcal{G}(G)$ if we properly choose the parameter $c(k,\varepsilon)$.
Let $K_1,\dots,K_r$ be the cliques deleted from $G$ in Algorithm~\ref{alg-reduction1}.
Then $\{K_1,\dots,K_r,V(G_1)\}$ is a partition of $V(G)$.
Consider an optimal solution $S_\mathsf{opt} \subseteq V(G)$ for $\mathbf{P}_\mathcal{G}$ on $G$.
Note that $S_\mathsf{opt} \cap V(G_1)$ is a feasible solution for $\mathbf{P}_\mathcal{G}$ on $G_1$.
Indeed, $G_1 - (S_\mathsf{opt} \cap V(G_1))$ is an induced subgraph of $G - S_\mathsf{opt}$.
Since $G - S_\mathsf{opt} \in \mathcal{G}$ and $\mathcal{G}$ is induced-subgraph-closed, $G_1 - (S_\mathsf{opt} \cap V(G_1)) \in \mathcal{G}$.
Therefore, $|S_\mathsf{opt} \cap V(G_1)| \geq \mathsf{opt}_\mathcal{G}(G_1)$ and thus
\begin{equation} \label{eq-SoptS1}
    (1+\varepsilon) \cdot |S_\mathsf{opt} \cap V(G_1)| \geq (1+\varepsilon) \cdot \mathsf{opt}_\mathcal{G}(G_1) \geq |S_1|.
\end{equation}
Next, we observe that $(1+\varepsilon) \cdot |S_\mathsf{opt} \cap K_i| \geq |K_i|$ for all $i \in [r]$.
First notice that $|K_i| \geq c(k,\varepsilon)$ by Algorithm~\ref{alg-reduction1}.
Recall that $\mathcal{G}$ is $T_k$-free, so $G-S_\mathsf{opt}$ does not contain $T_k$ as a subgraph.
Therefore, $S_\mathsf{opt}$ contains at least $|K_i| - 3k$ vertices in $K_i$, for otherwise there is a $3k$-clique in $G-S_\mathsf{opt}$ which contains $T_k$ as a subgraph.
By setting $c(k,\varepsilon) \geq 3k \cdot (1+\frac{1}{\varepsilon})$, we have $|K_i| - 3k \geq |K_i|/(1+\varepsilon)$.
It follows that $|S_\mathsf{opt} \cap K_i| \geq |K_i|/(1+\varepsilon)$ and $(1+\varepsilon) \cdot |S_\mathsf{opt} \cap K_i| \geq |K_i|$ for all $i \in [r]$.
Combining this inequality with \eqref{eq-SoptS1}, we have
\begin{equation*}
    (1+\varepsilon) \cdot |S_\mathsf{opt}| = (1+\varepsilon) \cdot \left(|S_\mathsf{opt} \cap V(G_1)| + \sum_{i=1}^r |S_\mathsf{opt} \cap K_i|\right) \geq |S_1| + \sum_{i=1}^r |K_i| = |S_1^+|.
\end{equation*}
As a result, $S_1^+$ is a $(1+\varepsilon)$-approximation solution for $\mathbf{P}_\mathcal{G}$ on $G$.

\subsection{Second step: making neighbors almost independent} \label{sec-step2}

In this section, we discuss the second step of our reduction, which goes from a disk graph of bounded ply to a disk graph in which the neighbors of each vertex are almost an independent set.
For a graph $H$, we say a set $I \subseteq V(H)$ of vertices are \textit{$\ell$-far from independent} in $H$ if we can remove at most $\ell$ vertices from $I$ such that the remaining vertices form an independent set in $H$.
Formally, we prove the following lemma.

\begin{lemma} \label{lem-reduction2}
Given $\varepsilon > 0$ and a disk graph $G_1$ of $n$ vertices with ply $(\frac{1}{\varepsilon})^{O(1)}$, one can compute in $n^{O(1)}$ time an induced subgraph $G_2$ of $G_1$ such that the following conditions hold.
\begin{itemize}
    \item For each vertex $v \in V(G_2)$, $N_{G_2}(v)$ is $(\frac{1}{\varepsilon})^{O(1)}$-far from independent in $G_2$.
    \item For any $(1+\frac{\varepsilon}{2})$-approximation solution $S_2$ for $\mathbf{P}_\mathcal{G}$ on $G_2$, $(V(G_1) \backslash V(G_2)) \cup S_2$ is a $(1+\varepsilon)$-approximation solution for $\mathbf{P}_\mathcal{G}$ on $G_1$.
\end{itemize}
\end{lemma}

The basic idea of this reduction has already been presented in Section~\ref{overview:secBLR}.
However, there we sorted the vertices in $G_1$ by the sizes of the disks representing them.
This can be done only when a realization of $G_1$ is given.
In order to make the reduction robust, we propose a method to order the vertices of $G_1$ in a way that ``mimics'' the order induced by the disk sizes.

The ordering procedure is given in Algorithm~\ref{alg-order}, which works as follows.
In line~2, the algorithm $\textsc{ApprxMaxClique}(G_1)$ returns a $2$-approximation maximum clique in $G_1$ (we have seen the existence of such an algorithm in Section~\ref{sec-step1}), and $\delta$ stores the size of this clique.
By Fact~\ref{fact-plyclique}, the ply of $G_1$ is in the range $[\frac{\delta}{4},2\delta]$, and thus $\delta = (\frac{1}{\varepsilon})^{O(1)}$.
Then in the for-loop (line~3-7), we construct the sequence $(v_1,\dots,v_n)$ in the reversed order, i.e., we determine $v_n,\dots,v_1$ iteratively.
Here, we maintain a set $R$ of ``remaining'' vertices in $V(G_1)$; initially $R = V(G_1)$, and once a vertex $v_i$ is determined, we remove it from $R$ (line~7).
Therefore, we have $R = V(G_1) \backslash \{v_{i+1},\dots,v_n\}$ at the beginning of each iteration.
To determine $v_i$, we first compute the subset $R' \subseteq R$ consisting of the vertices which have at most $32 \delta$ neighbors in $R$ (line~4); we will see later that $R' \neq \emptyset$.
For each $v \in R'$, we define a quantity $\sigma_v = |N_{G_1}(N_{G_1}(v) \cap (V(G_1) \backslash R)) \cap R|$.
Finally, $v_i$ is chosen to be the vertex $v \in R'$ with the minimum $\sigma_v$.

\begin{algorithm}[h]
    \caption{\textsc{Order}$(G_1)$}
	\begin{algorithmic}[1]
	    \State $n \leftarrow |V(G_1)|$ and $R \leftarrow V(G_1)$
	    \State $\delta \leftarrow |\textsc{ApprxMaxClique}(G_1)|$
	    \For{$i = n,\dots,1$}
	        \State $R' \leftarrow \{v \in R: |N_{G_1}(v) \cap R| \leq 32 \delta \}$
	        \State $\sigma_v \leftarrow |N_{G_1}(N_{G_1}(v) \cap (V(G_1) \backslash R)) \cap R|$ for all $v \in R'$
	        \State $v_i \leftarrow \arg\min_{v \in R'} \sigma_v$
	        \State $R \leftarrow R \backslash \{v_i\}$
	    \EndFor
        \State \textbf{return} $(v_1,\dots,v_n)$
	\end{algorithmic}
	\label{alg-order}
\end{algorithm}

We establish certain properties of the ordering $(v_1,\dots,v_n)$ returned by Algorithm~\ref{alg-order}.
Consider a realization $\mathcal{D}$ of $G_1$ with ply $(\frac{1}{\varepsilon})^{O(1)}$.
Note that this realization is only used for analysis and is not needed by the algorithm.
For each $v \in V(G_1)$, denote by $D_v \in \mathcal{D}$ the disk representing $v$.
We first observe that the set $R'$ computed in line~4 is always nonempty (see Observation \ref{obs-nonemptyR'}), and thus the vertex $v_i$ computed in line~6 is well-defined and only neighboring to a few vertices in $\{v_1,\dots,v_i\}$.

\begin{observation} \label{obs-nonemptyR'}
We have $R' \neq \emptyset$ in each iteration of the for-loop of Algorithm~\ref{alg-order}.
Therefore, for all $i \in [n]$, $|N_{G_1}(v_i) \cap \{v_1,\dots,v_i\}| \leq 32 \delta = (\frac{1}{\varepsilon})^{O(1)}$.
\end{observation}
\begin{proof}
Consider the iteration for determining $v_i$.
We have $R = V(G_1) \backslash \{v_{i+1},\dots,v_n\}$ at the beginning of this iteration.
Let $v \in R$ be the vertex whose corresponding disk $D_v$ is the smallest.
We show that $|N_{G_1}(v) \cap R| \leq 32 \delta$ and thus $R' \neq \emptyset$.
Let $Q$ be the minimal bounding square of the disk $D_v$.
The side-length of $Q$ is equal to the diameter of $D_v$, which we denote by $l$.
Note that $D_u \cap D_v \neq \emptyset$ for all $u \in N_{G_1}(v)$ and hence $Q$ intersects $D_u$ for all $u \in N_{G_1}(v)$.
Since the ply of $\mathcal{D}$ is at most $2\delta$, by Fact~\ref{fact-sparse}, $Q$ intersects at most $32 \delta$ disks in $\mathcal{D}$ with diameter at least $l$.
In particular, $D_v$ intersects at most $32 \delta$ disks in $\mathcal{D}$ whose sizes are larger than or equal to $D_v$.
As $D_u \geq D_v$ for all $u \in R$, $D_v$ intersects at most $32 \delta$ disks in $\{D_u: u \in R\}$, which implies $|N_{G_1}(v) \cap R| \leq 32 \delta$.
Thus, $v \in R'$ and $R' \neq \emptyset$.
Furthermore, as the vertex $v_i$ is selected from $R'$ in line~7, we have $|N_{G_1}(v_i) \cap R| \leq 32 \delta$, i.e., $|N_{G_1}(v_i) \cap \{v_1,\dots,v_i\}| \leq 32 \delta = (\frac{1}{\varepsilon})^{O(1)}$.
\end{proof}

Next, we observe that the set $R'$ computed in line~4 always contains some vertex $v$ with $\sigma_v = (\frac{1}{\varepsilon})^{O(1)}$, and thus for all $i \in [n]$ the vertices in $N_{G_1}(v_i) \cap (V(G_1) \backslash \{v_{i+1},\dots,v_n\})$ are only neighboring to a few vertices in $\{v_1,\dots,v_i\}$.

\begin{observation} \label{obs-1024288}
In each iteration of the for-loop of Algorithm~\ref{alg-order}, there exists $v \in R'$ such that $\sigma_v \leq 1024 \delta^2 + 288 \delta$.
Thus, for all $i \in [n]$, $|N_{G_1}(N_{G_1}(v_i) \cap \{v_{i+1},\dots,v_n\}) \cap \{v_1,\dots,v_i\}| = (\frac{1}{\varepsilon})^{O(1)}$.
\end{observation}
\begin{proof}
Consider the iteration for determining $v_i$.
We have $R = \{v_1,\dots,v_i\}$ at the beginning of this iteration.
Again, let $v \in R$ be the vertex whose corresponding disk $D_v$ is the smallest.
In the proof of Observation~\ref{obs-nonemptyR'}, we have seen that $v \in R'$.
We shall show $\sigma_v \leq 1024 \delta^2 + 288 \delta$.
For convenience, we define $W = N_{G_1}(v) \cap (V_{G_1} \backslash R)$.
Then what we want to prove is $|N_{G_1}(W) \cap R| \leq 1024 \delta^2 + 288 \delta$.
We partition $W$ into two subsets, $W^+ = \{w \in W: D_w \geq D_v\}$ and $W^- = \{w \in W: D_w < D_v\}$.
Clearly, $|N_{G_1}(W) \cap R| \leq |N_{G_1}(W^+) \cap R| + |N_{G_1}(W^-) \cap R|$.
So in what follows, we will bound $|N_{G_1}(W^+) \cap R|$ and $|N_{G_1}(W^-) \cap R|$ separately.

We first consider $N_{G_1}(W^+) \cap R$.
We have seen in the proof of Observation~\ref{obs-nonemptyR'} that $D_v$ intersects at most $32 \delta$ disks in $\mathcal{D}$ whose sizes are larger than or equal to $D_v$.
Since $D_w \geq D_v$ and $D_w \cap D_v \neq \emptyset$ for all $w \in W^+$, we have $|W^+| \leq 32 \delta$.
By the second statement in Observation~\ref{obs-nonemptyR'}, for any $j \in [n] \backslash [i]$, we have $|N_{G_1}(v_j) \cap R| \leq |N_{G_1}(v_j) \cap \{v_1,\dots,v_j\}| \leq 32 \delta$.
This implies $|N_{G_1}(w) \cap R| \leq 32 \delta$ for all $w \in W^+$, since $W^+ \subseteq V_{G_1} \backslash R \subseteq \{v_{i+1},\dots,v_n\}$.
Therefore,
\begin{equation*}
    |N_{G_1}(W^+) \cap R| \leq \sum_{w \in W^+} |N_{G_1}(w) \cap R| \leq |W^+| \cdot 32 \delta \leq 1024 \delta^2.
\end{equation*}

Next, we consider $N_{G_1}(W^-) \cap R$.
Let $Q$ be the square in the plane centered at the center of $D_v$ with side-length $3l$, where $l$ is the the diameter of $D_v$.
We first observe that $D_w \subseteq Q$ for all $w \in W^-$.
Consider a vertex $w \in W^-$.
We have $D_w < D_v$ by the definition of $W^-$.
Also, as $W^- \subseteq N_{G_1}(v)$, we have $D_v \cap D_w \neq \emptyset$.
Therefore, any point in the disk $D_w$ must be within distance $\frac{3}{2} l$ from the center of $D_v$ and thus be contained in $Q$, which implies $D_w \subseteq Q$.
It follows that $Q$ intersects $D_u$ for all $u \in N_{G_1}(W^-)$, simply because $D_u$ intersects $D_w$ for some $w \in W^-$.
In particular, $Q$ intersects $D_u$ for all $u \in N_{G_1}(W^-) \cap R$.
Now we partition $Q$ into $9$ smaller squares $Q_1,\dots,Q_9$ with side-length $l$.
Then each disk $D_u$ for $u \in N_{G_1}(W^-) \cap R$ must intersect $Q_t$ for some $t \in [9]$.
However, the diameter of $D_u$ is at least $l$ for all $u \in N_{G_1}(W) \cap R$, because $v$ has the smallest disk $D_v$ among all vertices in $R$.
Since the ply of $\mathcal{D}$ is $2 \delta$, by Fact~\ref{fact-sparse}, each $Q_t$ for $t \in [9]$ can only intersect $32 \delta$ disks $D_u$ for $u \in N_{G_1}(W^-) \cap R$.
Therefore, $|N_{G_1}(W^-) \cap R| \leq 288 \delta$.

By the above arguments, we have $|N_{G_1}(W) \cap R| \leq 1024 \delta^2 + 288 \delta$, i.e., $\sigma_v \leq 1024 \delta^2 + 288 \delta$.
To see the second statement of the observation, note that $v_i$ is the vertex in $R'$ with the minimum $\sigma$-value, which implies $\sigma_{v_i} \leq 1024 \delta^2 + 288 \delta = (\frac{1}{\varepsilon})^{O(1)}$.
As $R = \{v_1,\dots,v_i\}$ in the iteration for determining $v_i$, we have $|N_{G_1}(N_{G_1}(v_i) \cap \{v_{i+1},\dots,v_n\}) \cap \{v_1,\dots,v_i\}| = \sigma_{v_i} = (\frac{1}{\varepsilon})^{O(1)}$.
\end{proof}


With the \textsc{Order} procedure in hand, we are ready to give the reduction for Lemma~\ref{lem-reduction2}.
The reduction procedure is described in Algorithm~\ref{alg-reduction2}, which follows the ideas presented in Section~\ref{overview:secBLR}.
In line~2, we apply the function $\textsc{Order}(G_1)$ in Algorithm~\ref{alg-order} to obtain an ordering $(v_1,\dots,v_n)$ of the $n$ vertices of $G_1$.
In the for loop (line~4-9), we consider the vertices $v_1,\dots,v_n$ iteratively.
We maintain two sets $Z$ and $U$ of vertices of $G_1$.
The set $Z$ finally contains the vertices which we want to delete from $G_1$ to obtain $G_2$, and the meaning of $U$ will be clear shortly.
Initially, $Z = \emptyset$ and $U = V(G_1)$.
In the $i$-th iteration, we first find a \textit{maximal matching} $M$ of the induced subgraph $G_1[N_{G_1}(v_i) \cap U]$, i.e., a maximal set of disjoint edges in $G_1[N_{G_1}(v_i) \cap U]$ (line~5).
If $M$ contains at least $c(k,\varepsilon)$ edges, we add $v_i$ to $Z$ (line~6); again, here $c(k,\varepsilon)$ is a parameter depending on $k$ and $\varepsilon$ to be determined shortly.
In line~7, we take an arbitrary subset $M'$ of $M$ of size $c(k,\varepsilon)$; set $M' = M$ if $|M|<c(k,\varepsilon)$.
Let $R_i$ be the set of endpoints of the edges in $M'$ (line~8).
We then remove the vertices in $R_i$ from $U$ (line~9).
After all vertices $v_1,\dots,v_n$ are considered, the graph $G_2$ is obtained by removing $Z$ from $G_1$ (line~10).

\begin{algorithm}[h]
    \caption{\textsc{Reduce2}$(G_1)$}
	\begin{algorithmic}[1]
	    \State $n \leftarrow |V(G_1)|$
	    \State $(v_1,\dots,v_n) \leftarrow \textsc{Order}(G_1)$
	    \State $Z \leftarrow \emptyset$ and $U \leftarrow V(G_1)$
	    \For{$i = 1,\dots,n$}
	        \State $M \leftarrow$ a maximal matching of $G_1[N_{G_1}(v_i) \cap U]$
	        \If{$|M| \geq c(k,\varepsilon)$}{ $Z \leftarrow Z \cup \{v_i\}$}
	        \EndIf
	        \State $M' \leftarrow$ an arbitrary subset of $M$ of size $\min\{c(k,\varepsilon),|M|\}$
	        \State $R_i \leftarrow \mathsf{endpoint}(M')$
	        \State $U \leftarrow U \backslash R_i$
	    \EndFor
	    \State $G_2 \leftarrow G_1 - Z$
        \State \textbf{return} $G_2$
	\end{algorithmic}
	\label{alg-reduction2}
\end{algorithm}

We first prove the second condition in Lemma~\ref{lem-reduction2}.
Let $S_2 \subseteq V(G_2)$ be a $(1+\frac{\varepsilon}{2})$-approximation solution for $\mathbf{P}_\mathcal{G}$ on ${G}_2$.
We want to show $(V(G_1) \backslash V(G_2)) \cup S_2$ (or equivalently $Z \cup S_2$) is a $(1+\varepsilon)$-approximation solution for $\mathbf{P}_\mathcal{G}$ on ${G}_1$.
Clearly, $Z \cup S_2$ is a feasible solution for $\mathbf{P}_\mathcal{G}$ on ${G}_1$ because $G_1 - (Z \cup S_2) = G_2 - S_2 \in \mathcal{G}$.
So it suffices to show that $|Z \cup S_2| \leq (1+\varepsilon) \cdot \mathsf{opt}_\mathcal{G}(G_1)$ if we properly choose the parameter $c(k,\varepsilon)$.
Consider an optimal solution $S_\mathsf{opt} \subseteq V(G_1)$ for $\mathbf{P}_\mathcal{G}$ on ${G}_1$.
We first observe $|(S_\mathsf{opt} \cap Z) \cup S_2| \leq (1+\frac{\varepsilon}{2}) \cdot \mathsf{opt}_\mathcal{G}(G_1)$.
Indeed, $G_2 - S_\mathsf{opt} \cap V(G_2) \in \mathcal{G}$, since $\mathcal{G}$ is closed (in particular, induced-subgraph-closed) and $G_2 - (S_\mathsf{opt} \cap V(G_2))$ is an induced subgraph of $G_1 - S_\mathsf{opt} \in \mathcal{G}$.
It follows that
\begin{equation*}
    |S_2| \leq \left(1+\frac{\varepsilon}{2}\right) \cdot \mathsf{opt}_\mathcal{G}(G_2) \leq \left(1+\frac{\varepsilon}{2}\right) \cdot |S_\mathsf{opt} \cap V(G_2)|.
\end{equation*}
Also, since $|S_\mathsf{opt}| = |S_\mathsf{opt} \cap Z| + |S_\mathsf{opt} \cap V(G_2)|$, we further have
\begin{equation} \label{eq-SoptZS2}
    |(S_\mathsf{opt} \cap Z) \cup S_2| = |S_\mathsf{opt} \cap Z| + |S_2| \leq \left(1+\frac{\varepsilon}{2}\right) \cdot |S_\mathsf{opt}| = \left(1+\frac{\varepsilon}{2}\right) \cdot \mathsf{opt}_\mathcal{G}(G_1).
\end{equation}

With the above inequality in hand, the only thing we need now is that $|Z \backslash S_\mathsf{opt}| \leq \frac{\varepsilon}{2} \cdot |S_\mathsf{opt}|$.
Intuitively speaking, this means $Z$ does not contain too many ``wrong'' vertices (i.e., vertices not in $S_\mathsf{opt}$).
Let us first briefly discuss the main intuition about our argument.
We observe that, if a vertex $v_i$ is not in $S_\mathsf{opt}$ but we add it to $Z$ in line~6 of Algorithm~\ref{alg-reduction2}, then the set $R_i$ constructed in line~8 must contain ``many'' vertices in $S_\mathsf{opt}$.
We then charge $v_i$ to the vertices in $S_\mathsf{opt} \cap R_i$.
Since $R_i$ is removed from $U$ at the end of each iteration (line~9), the set $U$ always consists of vertices that do not get charged yet, which prevents a vertex from getting charged multiple times.
Finally, each vertex $v_i \in Z \backslash S_\mathsf{opt}$ is charged to ``many'' vertices in $S_\mathsf{opt}$ and no vertex in $S_\mathsf{opt}$ gets charged more than once, which implies $|Z \backslash S_\mathsf{opt}|$ is only a small fraction of $|S_\mathsf{opt}|$.
Next, we give the formal proof.
Towards that, we make the following two simple observations.
\begin{observation} \label{obs-disjointRi}
The sets $R_1,\dots,R_n$ constructed in Algorithm~\ref{alg-reduction2} are disjoint.
\end{observation}
\begin{proof}
For $i \in [n]$, let $U_i$ denote the set $U$ at the beginning of the $i$-th iteration of the for-loop in Algorithm~\ref{alg-reduction2}.
Note that $R_i \subseteq U_i$ because the matching $M$ in the $i$-th iteration is computed in $G_1[N_{G_1}(v_i) \cap U_i]$ and $M' \subseteq M$.
Consider two indices $i,j \in [n]$ where $i>j$.
We have $U_i \cap R_j = \emptyset$ since $R_j$ is removed from $U$ at the end of the $j$-th iteration.
Thus, $R_i \cap R_j = \emptyset$.
It follows that $R_1,\dots,R_n$ are disjoint.
\end{proof}

\begin{observation} \label{obs-containmany}
For each vertex $v_i \in Z \backslash S_\mathsf{opt}$, we have $|S_\mathsf{opt} \cap R_i| \geq c(k,\varepsilon)-k$.
\end{observation}
\begin{proof}
Let $v_i \in Z \backslash S_\mathsf{opt}$, and consider the $i$-th iteration of the for-loop in Algorithm~\ref{alg-reduction2}.
The reason for why $v_i$ is added to $Z$ (in line~6) is that the matching $M$ is of size at least $c(k,\varepsilon)$.
In this case, $|M'| = c(k,\varepsilon)$.
Note that $M$ is a matching in $G_1[N_{G_1}(v_i) \cap U]$, so the endpoints of the edges in $M$ (and in particular $M'$) are all neighbors of $v_i$.
Now the edges in $M'$ are disjoint and their endpoints are all neighbors of $v_i$.
Therefore, $v_i$ and the endpoints of the edges in $M'$ (i.e., the vertices in $R_i$) form a triangle bundle $T_{|M'|}$ in $G_1$ (in which $v_i$ is the center).
Recall that $\mathcal{G}$ is $T_k$-free and thus $G_1 - S_\mathsf{opt} \in \mathcal{G}$ is $T_k$-tree.
As $v_i \notin S_\mathsf{opt}$, at least $|M'| - k+1$ edges in $M'$ must be hit by $S_\mathsf{opt}$ in order to guarantee the $T_k$-freeness of $G_1 - S_\mathsf{opt}$ (for otherwise $v_i$ and the endpoints of the edges in $M'$ not hit by $S_\mathsf{opt}$ can form a triangle bundle $T_k$ in $G_1 - S_\mathsf{opt}$).
But the edges in $M'$ are disjoint.
Thus, to hit at least $|M'| - k+1$ edges in $M'$, $S_\mathsf{opt}$ must contain at least $|M'| - k+1$ vertices in $R_i$.
It follows that $|S_\mathsf{opt} \cap R_i| \geq |M'| - k+1 = c(k,\varepsilon)-k+1 > c(k,\varepsilon)-k$.
\end{proof}

Let $L = \{i \in [n]: v_i \in Z \backslash S_\mathsf{opt}\}$.
By Observation~\ref{obs-disjointRi}, we have $|S_\mathsf{opt}| \geq \sum_{i \in L} |S_\mathsf{opt} \cap R_i|$.
On the other hand, by Observation~\ref{obs-containmany}, we have $\sum_{i \in L} |S_\mathsf{opt} \cap R_i| \geq (c(k,\varepsilon)-k) \cdot |L|$.
Therefore, if we set $c(k,\varepsilon) \geq \frac{2}{\varepsilon} + k$, then $c(k,\varepsilon)-k \geq \frac{2}{\varepsilon}$ and hence $|Z \backslash S_\mathsf{opt}| = |L| \leq \frac{\varepsilon}{2} \cdot |S_\mathsf{opt}|$.
Combining this with Equation~\ref{eq-SoptZS2}, we get $|Z \cup S_2| \leq (1+\varepsilon) \cdot \mathsf{opt}_\mathcal{G}(G_1)$.

Next, we prove the first condition in Lemma~\ref{lem-reduction2}, i.e., for every  $v \in V(G_2)$, $N_{G_2}(v)$ is $(\frac{1}{\varepsilon})^{O(1)}$-far from independent in $G_2$.
For all $i \in [n]$, let $U_i$ denote the set $U$ at the beginning of the $i$-th iteration of the for-loop in Algorithm~\ref{alg-reduction2}.
Consider an index $i \in [n]$ such that $v_i \in V(G_2)$, and our goal is to show $N_{G_2}(v_i)$ is $(\frac{1}{\varepsilon})^{O(1)}$-far from independent in $G_2$.
We partition $N_{G_2}(v_i)$ into two subsets in the following simple way.
Define $S(v_i) = N_{G_2}(v_i) \cap (\bigcup_{j=1}^i R_j)$ and $I(v_i) = N_{G_2}(v_i) \backslash (\bigcup_{j=1}^i R_j)$.
Clearly, $\{S(v_i),I(v_i)\}$ is a partition of $N_{G_2}(v_i)$.
We first observe that $I(v_i)$ is an independent set.
\begin{observation}
$I(v_i)$ is an independent set in $G_2$.
\end{observation}
\begin{proof}
Consider the $i$-th iteration of the for-loop of Algorithm~\ref{alg-reduction2}.
Line~5 computes a maximal matching $M$ in $G_1[N_{G_1}(v_i) \cap U_i]$.
Note that $U_i = V(G_1) \backslash (\bigcup_{j=1}^{i-1} R_j)$, so $M$ is a maximal matching in $G_1[N_{G_1}(v_i) \backslash (\bigcup_{j=1}^{i-1} R_j)]$.
Since $v_i \in V(G_2)$, we do not add $v_i$ to $Z$ in line~6, and thus $|M| < c(k,\varepsilon)$.
Therefore, the subset $M' \subseteq M$ computed in line~7 is of size $|M|$, i.e., $M' = M$.
Now $M'$ is a maximal matching in $G_1[N_{G_1}(v_i) \backslash (\bigcup_{j=1}^{i-1} R_j)]$, and $R_i$ is the set of endpoints of edges in $M'$.
It follows that $N_{G_1}(v_i) \backslash (\bigcup_{j=1}^i R_j)$ is an independent set in $G_1$, for otherwise there is an edge of $G_1$ with endpoints in $N_{G_1}(v_i) \backslash (\bigcup_{j=1}^i R_j)$, contradicting with the maximality of $M'$.
Finally, because $I(v_i) = N_{G_2}(v_i) \backslash (\bigcup_{j=1}^i R_j) \subseteq N_{G_1}(v_i) \backslash (\bigcup_{j=1}^i R_j)$ and $G_2$ is an induced subgraph of $G_1$, we know that $I(v_i)$ is an independent set in $G_2$.
\end{proof}

Based on the above observation, to show $N_{G_2}(v_i)$ is $(\frac{1}{\varepsilon})^{O(1)}$-far from independent, it suffices to show $|S(v_i)| = (\frac{1}{\varepsilon})^{O(1)}$.
We first split $S(v_i)$ into two parts, $S^+(v_i) = S(v_i) \cap \{v_1,\dots,v_i\}$ and $S^-(v_i) = S(v_i) \cap \{v_{i+1},\dots,v_n\}$.
In other words, $S^+(v_i)$ (resp., $S^-(v_i)$) contains the vertices in $S(v_i)$ whose disks are larger (resp., smaller) than or equal to $D_{v_i}$.
To bound $|S^+(v_i)|$, we can directly apply Observation~\ref{obs-nonemptyR'}.
Since $S^+(v_i) \subseteq N_{G_2}(v_i) \subseteq N_{G_1}(v_i)$ and $S^+(v_i) \subseteq \{v_1,\dots,v_i\}$, Observation~\ref{obs-nonemptyR'} immediately implies $|S^+(v_i)| = (\frac{1}{\varepsilon})^{O(1)}$.

Next, we try to bound $|S^-(v_i)|$ using Observation~\ref{obs-1024288}.
We observe that for any $j \in [i]$, if $S^-(v_i) \cap R_j \neq \emptyset$, then $v_j \in N_{G_1}(N_{G_1}(v_i) \cap \{v_{i+1},\dots,v_n\})$.
To see this, assume $S^-(v_i) \cap R_j \neq \emptyset$ and consider a vertex $v \in S^-(v_i) \cap R_j$.
Then we have $v \in N_{G_1}(v_i) \cap \{v_{i+1},\dots,v_n\}$, because $S^-(v_i) \subseteq N_{G_1}(v_i) \cap \{v_{i+1},\dots,v_n\}$.
Also, as $v \in R_j \subseteq N_{G_1}(v_j)$, we have $v_j \in N_{G_1}(v)$, which implies $v_j \in N_{G_1}(N_{G_1}(v_i) \cap \{v_{i+1},\dots,v_n\})$.
By Observation~\ref{obs-1024288}, there are only $(\frac{1}{\varepsilon})^{O(1)}$ indices $j \in [i]$ such that $v_j \in N_{G_1}(N_{G_1}(v_i) \cap \{v_{i+1},\dots,v_n\})$.
Therefore, there are only $(\frac{1}{\varepsilon})^{O(1)}$ indices $j \in [i]$ such that $S^-(v_i) \cap R_j \neq \emptyset$, i.e., $|J| = (\frac{1}{\varepsilon})^{O(1)}$ for $J = \{j \in [i]: S^-(v_i) \cap R_j \neq \emptyset\}$.
Note that $S^-(v_i) \subseteq S(v_i) \subseteq \bigcup_{j=1}^i R_j$ and $R_j$'s are disjoint by Observation~\ref{obs-disjointRi}.
So we have
\begin{equation*}
    |S^-(v_i)| = \sum_{j=1}^i |S^-(v_i) \cap R_j| = \sum_{j \in J} |S^-(v_i) \cap R_j| \leq \sum_{j \in J} |R_j|.
\end{equation*}
We have $|R_j| = (\frac{1}{\varepsilon})^{O(1)}$ for all $j \in [n]$ because the set $M'$ computed in line~7 of Algorithm~\ref{alg-reduction2} is of size at most $c(k,\varepsilon) = (\frac{1}{\varepsilon})^{O(1)}$.
Combining this with the fact $|J| = (\frac{1}{\varepsilon})^{O(1)}$, we finally deduce that $\sum_{j \in J} |R_j| = (\frac{1}{\varepsilon})^{O(1)}$ and hence $|S^-(v_i)| = (\frac{1}{\varepsilon})^{O(1)}$.
As a result, $|S(v_i)| = (\frac{1}{\varepsilon})^{O(1)}$ and $N_{G_2}(v_i)$ is $(\frac{1}{\varepsilon})^{O(1)}$-far from independent.
This completes the proof the first condition in Lemma~\ref{lem-reduction2}.

\subsection{Final step: getting bounded local radius} \label{sec-step3}

In this section, we give the final step of our reduction, which goes from a disk graph in which the neighbors of each vertex are almost independent to a disk graph of bounded local radius.
Formally, we prove the following lemma.

\begin{lemma} \label{lem-reduction3}
Given $\varepsilon > 0$ and a disk graph $G_2$ of $n$ vertices satisfying that $N_{G_2}(v)$ is $(\frac{1}{\varepsilon})^{O(1)}$-far from independent in $G_2$ for every vertex $v \in V(G_2)$, one can compute in $n^{O(1)}$ time an induced subgraph $G_3$ of $G_2$ such that the following conditions hold.
\begin{itemize}
    \item The local radius of $G_3$ is $(\frac{1}{\varepsilon})^{O(1)}$.
    \item Given any $(1+\frac{\varepsilon}{2})$-approximation solution for $\mathbf{P}_\mathcal{G}$ on $G_3$, one can compute in $n^{O(1)}$ time a $(1+\frac{\varepsilon}{4})$-approximation solution for $\mathbf{P}_\mathcal{G}$ on $G_2$.
\end{itemize}
\end{lemma}

Let $\varepsilon > 0$ and $G_2$ be as in the lemma.
The algorithm for generating $G_3$ is shown in Algorithm~\ref{alg-reduction3}.
In line~1, we cluster the false twins in $G_2$; specifically, we let $\mathcal{X}$ be the partition of $V(G_2)$ formed by the equivalence classes of the false-twin relation.
For each $X \in \mathcal{X}$, we let $d_X$ be the number of neighbors of $X$ in $G_2$, which is equal to the degree of every vertex in $X$ (as the vertices in $X$ are false twins in $G_2$).
Next, the algorithm arbitrarily picks a subset $V_X$ of $X$ of size $c(k,\varepsilon) \cdot d_X$ (set $V_X = X$ if $|X|<c(k,\varepsilon) \cdot d_X$) for each $X \in \mathcal{X}$ (line~4).
Again, here $c(k,\varepsilon)$ is a parameter depending on $k$ and $\varepsilon$ to be determined shortly.
Finally, $G_3$ is simply defined as the subgraph of $G_2$ induced by the vertices in all $V_X$ (line~5).

\begin{algorithm}[h]
    \caption{\textsc{Reduce3}$(G_2)$}
	\begin{algorithmic}[1]
	    \State $\mathcal{X} \leftarrow \{\mathsf{FT}_{G_2}(v): v \in V(G_2)\}$
	    \For{every $X \in \mathcal{X}$}
	        \State $d_X \leftarrow |N_{G_2}(X)|$
	        \State $V_X \leftarrow$ an arbitrary subset of $X$ of size $\min\left\{c(k,\varepsilon) \cdot d_X,|X|\right\}$
	    \EndFor
	    \State $G_3 \leftarrow G_2[\bigcup_{X \in \mathcal{X}} V_X]$
        \State \textbf{return} $G_3$
	\end{algorithmic}
	\label{alg-reduction3}
\end{algorithm}

We first show the second condition in Lemma~\ref{lem-reduction3}, which follows easily from the fact that $\mathcal{G}$ is closed.
Let $S \subseteq V(G_3)$ be a $(1+\frac{\varepsilon}{4})$-approximation solution for $\mathbf{P}_\mathcal{G}$ on $G_3$.
We create a set $S' \subseteq V(G_2)$ as follows.
For each $X \in \mathcal{X}$, let $U_X = X \cap S$ if $|X \cap S| \leq (c(k,\varepsilon)-k) \cdot d_X$ and let $U_X = (X \cap S) \cup N_{G_2}(X)$ if $|X \cap S| > (c(k,\varepsilon)-k) \cdot d_X$.
Define $S' = \bigcup_{X \in \mathcal{X}} U_X$.
We claim that $S'$ is a $(1+\frac{\varepsilon}{2})$-approximation solution for $\mathbf{P}_\mathcal{G}$ on $G_2$, if we properly choose the parameter $c(k,\varepsilon)$.
Note that for all $X \in \mathcal{X}$ we have $|U_X| \leq (1+\frac{1}{c(k,\varepsilon)-k}) \cdot |X \cap S|$ by our construction, which implies
\begin{equation*}
    |S'| \leq \sum_{X \in \mathcal{X}} |U_X| \leq \left( 1+\frac{1}{c(k,\varepsilon)-k} \right) \cdot \sum_{X \in \mathcal{X}} |X \cap S| = \left( 1+\frac{1}{c(k,\varepsilon)-k} \right) \cdot |S|.
\end{equation*}
Since $\mathcal{G}$ is closed (in particular, induced-subgraph-closed), we have $\mathsf{opt}_\mathcal{G}(G_3) \leq \mathsf{opt}_\mathcal{G}(G_2)$, and thus $|S| \leq (1+\frac{\varepsilon}{4}) \cdot \mathsf{opt}_\mathcal{G}(G_3) \leq (1+\frac{\varepsilon}{4}) \cdot \mathsf{opt}_\mathcal{G}(G_2)$.
Combining this with the above inequality, if we set $c(k,\varepsilon) \geq \frac{4+\varepsilon}{\varepsilon} +k$, then $|S'| \leq (1+\frac{\varepsilon}{4+\varepsilon}) \cdot |S| \leq (1+\frac{\varepsilon}{4+\varepsilon})(1+\frac{\varepsilon}{4}) \cdot \mathsf{opt}_\mathcal{G}(G_2) = (1+\frac{\varepsilon}{2}) \cdot \mathsf{opt}_\mathcal{G}(G_2)$.
Now it suffices to show $S'$ is a \textit{feasible} solution for $\mathbf{P}_\mathcal{G}$ on $G_2$, which is the following observation.
\begin{observation}
$G_2 - S' \in \mathcal{G}$.
\end{observation}
\begin{proof}
We have $G_3 - S \in \mathcal{G}$, which implies $G_3 - S' \in \mathcal{G}$ as $S \subseteq S'$ and $\mathcal{G}$ is induced-subgraph-closed.
Let $W = V(G_3) \backslash S'$, so $G_2[W] = G_3 - S' \in \mathcal{G}$.
Define $W^* \subseteq V(G_2) \backslash S'$ as a maximal subset such that $W \subseteq W^*$ and $G_2[W^*] \in \mathcal{G}$.
We show that $W^* = V(G_2) \backslash S'$, which directly implies $G_2 - S' = G_2[W^*] \in \mathcal{G}$ and completes the proof.

Assume $W^* \subsetneq V(G_2) \backslash S'$, then there exists a vertex $v \in V(G_2) \backslash S'$ such that $v \notin W^*$.
We shall show that $G_2[W^* \cup \{v\}] \in \mathcal{G}$, which contradicts with the maximality of $W^*$.
If $N_{G_2}(v) \cap W^* = \emptyset$, then $G_2[W^* \cup \{v\}]$ can be viewed as a graph obtained from $G_2[W^*]$ by adding a vertex with no edges.
By Fact~\ref{fact-closed}, we have $G_2[W^* \cup \{v\}] \in \mathcal{G}$, because $G_2[W^*] \in \mathcal{G}$.
It suffices to consider the case that $N_{G_2}(v) \cap W^* \neq \emptyset$.
Let $X \in \mathcal{X}$ such that $v \in X$.
Observe that $v \notin V_X$, for otherwise $v \in V(G_3) \backslash S' = W \subseteq W^*$.
This implies $V_X \neq X$ and thus $|V_X| = c(k,\varepsilon) \cdot d_X$.
On the other hand, we observe that $|X \cap S| \leq (c(k,\varepsilon)-k) \cdot d_X$; indeed, if $|X \cap S| > (c(k,\varepsilon)-k) \cdot d_X$, then $N_{G_2}(X) \subseteq U_X \subseteq S'$ and thus $N_{G_2}(v) = N_{G_2}(X) \subseteq S'$, which implies $N_{G_2}(v) \cap W^* = \emptyset$ as $W^* \subseteq V(G_2) \backslash S'$. This contradicts our assumption that $N_{G_2}(v) \cap W^* \neq \emptyset$. Thus, we have that $|X \cap S| \leq (c(k,\varepsilon)-k) \cdot d_X$. This implies that $U_X = X \cap S$. 

\begin{claim}
\label{claimXprime}
$X \cap U_{X'} = \emptyset$ for all $X' \in \mathcal{X}$ with $X' \neq X$.
\end{claim}
\begin{proof}
If $N_{G_2}(X') \cap X = \emptyset$, then clearly $X \cap U_{X'} = \emptyset$ because $U_{X'} \subseteq X' \cup N_{G_2}(X')$ by our construction.
If $N_{G_2}(X') \cap X \neq \emptyset$, then $X \subseteq N_{G_2}(X')$ because the vertices in $X$ are false twins in $G_2$.
Since $v \notin S'$, we have $v \notin U_{X'}$.
But $v \in X \subseteq N_{G_2}(X')$, which implies $N_{G_2}(X') \nsubseteq U_{X'}$.
Therefore, $U_{X'} = X' \cap S$. 
This implies that $U_{X'}\subseteq X'$ and 
hence $X \cap U_{X'} = \emptyset$, because $X$ and $X'$ are pairwise disjoint.  
\end{proof}

By Claim \ref{claimXprime} and that fact that $S'=\bigcup_{Y\in {\cal X}} U_Y$, we obtain that $X \cap S'=X \cap U_X=X \cap S$. Here, the last equality follows from the fact that $U_X = X \cap S$. Hence, 
$|X \cap S'|=|X \cap S|\leq (c(k,\varepsilon)-k) \cdot d_X$. This along with  the fact $|V_X| = c(k,\varepsilon) \cdot d_X$ obtained above, we get $|V_X \backslash S'| \geq k \cdot d_X$. By the assumption $N_{G_2}(v) \cap W^* \neq \emptyset$, we have $d_X = |N_{G_2}(v)| \geq 1$. Let $v' \in V_X \backslash S' \subseteq W^*$ be an arbitrary vertex. Such a vertex $v'$ exists because $|V_X \backslash S'| \geq k \cdot d_X \geq k$.
Since $|V_X \backslash S'| \geq k$, we have $|\mathsf{FT}_{G_2[W]}(v')| \geq k$ and thus $|\mathsf{FT}_{G_2[W^*]}(v')| \geq k$.
Note that $G_2[W^* \cup \{v\}]$ is isomorphic to $\mathsf{clone}(G_2[W^*],v')$.
As $\mathcal{G}$ is $k$-clone-closed, we finally have $G_2[W^* \cup \{v\}] \in \mathcal{G}$. This competes the proof of the observation. 
\end{proof}

The rest of this section is dedicated to showing that the local radius of $G_3$ is $(\frac{1}{\varepsilon})^{O(1)}$, which completes the proof of Lemma~\ref{lem-reduction3}.
This part is technical, and we begin with the following simple observation, which bounds the size of the sets $V_X$ we constructed in Algorithm~\ref{alg-reduction3}.
\begin{observation} \label{obs-smallVX}
$|V_X| = (\frac{1}{\varepsilon})^{O(1)}$ for every $X \in \mathcal{X}$.
\end{observation}
\begin{proof}
Since $N_{G_2}(v)$ is $(\frac{1}{\varepsilon})^{O(1)}$-far from independent in $G_2$ for all $v \in V(G_2)$, a maximum clique in $G_2$ is of size $(\frac{1}{\varepsilon})^{O(1)}$.
By Fact~\ref{fact-plyclique}, the ply of $G_2$ is $(\frac{1}{\varepsilon})^{O(1)}$.
Thus, the ply of $H$ is $(\frac{1}{\varepsilon})^{O(1)}$ for any induced subgraph $H$ of $G_2$.
Consider a class $X \in \mathcal{X}$.
Let $H = G_2[X \cup N_{G_2}(X)]$.
Because the vertices in $X$ are false twins in $G_2$, every vertex in $X$ is adjacent to every vertex in $N_{G_2}(X)$, which implies that $H$ has at least $|X| \cdot d_X$ edges.
One the other hand, as the ply of $H$ is $(\frac{1}{\varepsilon})^{O(1)}$, by Fact~\ref{fact-plyedge}, the number of edges in $H$ is at most $(\frac{1}{\varepsilon})^{O(1)} \cdot |V(H)|$, which is $(\frac{1}{\varepsilon})^{O(1)} \cdot (|X|+d_X)$.
As a result, we have $|X| \cdot d_X = (\frac{1}{\varepsilon})^{O(1)} \cdot (|X|+d_X)$, which implies $\min\{|X|, d_X\} = (\frac{1}{\varepsilon})^{O(1)}$.
If $|X| = (\frac{1}{\varepsilon})^{O(1)}$, then $|V_X| = (\frac{1}{\varepsilon})^{O(1)}$ as $V_X \subseteq X$.
Otherwise, $d_X = (\frac{1}{\varepsilon})^{O(1)}$, and thus we have $|V_X| = (\frac{1}{\varepsilon})^{O(1)}$ because $|V_X| \leq c(k,\varepsilon) \cdot d_X$ and we chose $c(k,\varepsilon) = (\frac{1}{\varepsilon})^{O(1)}$.
\end{proof}

\begin{corollary} \label{cor-fewFT}
$|\mathsf{FT}_{G_3}(v)| = (\frac{1}{\varepsilon})^{O(1)}$ for every $v \in V(G_3)$.
\end{corollary}
\begin{proof}
Let $v \in V(G_3)$ and $X \in \mathcal{X}$ be the class that contains $v$.
We show that $\mathsf{FT}_{G_3}(v) = V_X$, and hence $|\mathsf{FT}_{G_3}(v)| = (\frac{1}{\varepsilon})^{O(1)}$ by Observation~\ref{obs-smallVX}.
First observe that $V_X \subseteq \mathsf{FT}_{G_3}(v)$.
Indeed, if two vertices of $G_3$ are false twins in $G_2$, then they are also false twins in $G_3$, since $G_3$ is an induced subgraph of $G_2$.
It suffices to show $\mathsf{FT}_{G_3}(v) \subseteq V_X$.
Consider a vertex $v' \in \mathsf{FT}_{G_3}(v)$.
We show by contradiction that $v' \in X$, which implies $v' \in V_X$ as $V(G_3) \cap X = V_X$.
Assume $v' \notin X$.
Then $v$ and $v'$ are not false twins in $G_2$, so there exists a vertex $u \in V(G_2)$ that is neighboring to exactly one of $v$ and $v'$ in $G_2$.
Let $Y \in \mathcal{X}$ be the class that contains $u$.
Note that $V_Y \neq \emptyset$.
Every vertex in $V_Y$ is neighboring to exactly one of $v$ and $v'$ in $G_2$.
Since $V_Y \subseteq V(G_3)$, the vertices in $V_Y$ witness that $v$ and $v'$ are not false twins in $G_3$, i.e., contradicting with the fact $v' \in \mathsf{FT}_{G_3}(v)$.
\end{proof}

Since $G_3$ is an induced subgraph of $G_2$, $N_{G_3}(v)$ is $(\frac{1}{\varepsilon})^{O(1)}$-far from independent in $G_3$ for every vertex $v \in V(G_3)$.
Therefore, for each $v \in V(G_3)$, we can partition $N_{G_3}(v)$ into two sets $S(v)$ and $I(v)$, where $|S(v)| = (\frac{1}{\varepsilon})^{O(1)}$ and $I(v)$ is an independent set in $G_3$; for convenience, we call the vertices in $S(v)$ \textit{singular neighbors} of $v$ and the vertices in $I(v)$ \textit{independent neighbors} of $v$.

Now we start to bound the local radius of $G_3$.
In fact, we shall prove a stronger statement: any realization of $G_3$ has $(\frac{1}{\varepsilon})^{O(1)}$ local radius.
Let $\mathcal{D}$ be a realization of $G_3$, and we denote by $D_v \in \mathcal{D}$ the disk representing the vertex $v \in V(G_3)$.
To show the local radius of $\mathcal{D}$ is $(\frac{1}{\varepsilon})^{O(1)}$, it suffices to have $\textsf{rad}(A_\mathcal{D}[D_v]) = (\frac{1}{\varepsilon})^{O(1)}$ for every vertex $v \in V(G_3)$.
Fix a vertex $v \in V(G_3)$.
First, we partition $N_{G_3}(v)$ into two sets $S^*$ and $I^*$ where $S^* \supseteq S(v)$ and $I^* \subseteq I(v)$ as follows.
For convenience, we write $S^2(v) = \bigcup_{w \in S(v)} S(w)$, i.e., the set of singular neighbors of singular neighbors of $v$.
A vertex $u \in N_{G_3}(v)$ is included in $S^*$ if (at least) one of the following conditions holds.
\begin{itemize}
    \item $u \in S(v)$, i.e., $u$ is a singular neighbor of $v$.
    \item $u \in S^2(v)$, i.e., $u$ is a singular neighbor of a singular neighbor of $v$.
    \item $N_{G_3}(u) \subseteq \{v\} \cup S(v) \cup S^2(v)$, i.e., every neighbor of $u$ except $v$ is either a singular neighbor of $v$ or a singular neighbor of a singular neighbor of $v$.
\end{itemize}
Then simply define $I^* = N_{G_3}(v) \backslash S^*$.
We observe that the partition $\{S^*,I^*\}$ of $N_{G_3}(v)$ has a similar property as the partition $\{S(v),I(v)\}$, i.e., $S^*$ is small and $I^*$ is an independent set.
\begin{observation} \label{obs-S*I*pro1}
$|S^*| = (\frac{1}{\varepsilon})^{O(1)}$ and $I^*$ is an independent set in $G_3$.
\end{observation}
\begin{proof}
It is clear that $I^*$ is an independent set as $I^* \subseteq I(v)$ and $I(v)$ is an independent set.
To see $|S^*| = (\frac{1}{\varepsilon})^{O(1)}$, we use fact that every vertex of $G_3$ has only $(\frac{1}{\varepsilon})^{O(1)}$ singular neighbors.
Therefore, the number of singular neighbors of singular neighbors of a vertex is also bounded by $(\frac{1}{\varepsilon})^{O(1)}$.
In particular, $|S^2(v)| = (\frac{1}{\varepsilon})^{O(1)}$ and $|\{v\} \cup S(v) \cup S^2(v)| = (\frac{1}{\varepsilon})^{O(1)}$.
Now it suffices to bound the number of vertices $u \in N_{G_3}(v)$ satisfying $N_{G_3}(u) \subseteq \{v\} \cup S(v) \cup S^2(v)$.
For each subset $Z \subseteq \{v\} \cup S(v) \cup S^2(v)$, the number of vertices $u \in N_{G_3}(v)$ satisfying $N_{G_3}(u) = Z$ is $(\frac{1}{\varepsilon})^{O(1)}$ by Corollary~\ref{cor-fewFT}.
Next, we observe that, although the number of subsets of $\{v\} \cup S(v) \cup S^2(v)$ is exponential in $|\{v\} \cup S(v) \cup S^2(v)|$, only polynomially many subsets $Z \subseteq \{v\} \cup S(v) \cup S^2(v)$ satisfying $N_{G_3}(u) = Z$ for some $u \in N_{G_3}(v)$.
The reason is that the VC-dimension of the neighbor sets of a disk graph (as a set system) is constant \cite{aronov2021pseudo,bonamy2021eptas}, and thus the number of distinct neighbor sets on $\{v\} \cup S(v) \cup S^2(v)$ is polynomial in the size of $\{v\} \cup S(v) \cup S^2(v)$ by Sauer–Shelah lemma.
It follows that there are $(\frac{1}{\varepsilon})^{O(1)}$ vertices $u \in N_{G_3}(v)$ satisfying $N_{G_3}(u) \subseteq \{v\} \cup S(v) \cup S^2(v)$ and hence $|S^*| = (\frac{1}{\varepsilon})^{O(1)}$.
\end{proof}

Furthermore, $S^*$ and $I^*$ satisfy an additional good property: the disk representing a vertex in $I^*$ is not contained in the union of the disks representing the vertices in $\{v\} \cup S^*$.
This property is crucial for the rest of our proof.

\begin{observation} \label{obs-S*I*pro2}
$D_u \nsubseteq \bigcup_{w \in \{v\} \cup S^*} D_w$ for all $u \in I^*$.
\end{observation}
\begin{proof}
Consider a vertex $u \in I^*$, and we show by contradiction that $D_u \nsubseteq \bigcup_{w \in \{v\} \cup S^*} D_w$.
Assume that $D_u \subseteq \bigcup_{w \in \{v\} \cup S^*} D_w$.
We claim that $N_{G_3}(u) \subseteq \{v\} \cup S(v) \cup S^2(v)$, which implies that $u \in S^*$ and thus contradicts with the assumption $u \in I^*$.
Let $z \in N_{G_3}(u) \backslash \{v\}$.
We have to show that $z \in S(v) \cup S^2(v) =  \bigcup_{w \in \{v\} \cup S(v)} S(w)$, or equivalently, $z \in S(w)$ for some $w \in \{v\} \cup S(v)$.
Note that $D_z \cap D_u \neq \emptyset$, as $z \in N_{G_3}(u)$.
Since $D_u \subseteq \bigcup_{w \in \{v\} \cup S^*} D_w$, we must have $D_z \cap D_u \cap D_w \neq \emptyset$ for some $w \in \{v\} \cup S^* \subseteq N_{G_3}[v]$.
Observe that $w \in \{v\} \cup S(v)$; indeed, if $w \notin \{v\} \cup S(v)$, then $w \in I(v)$, which implies $D_u \cap D_w = \emptyset$ because $u \in I^* \subseteq I(v)$ and $I(v)$ is an independent set.
Now it suffices to show $z \in S(w)$.
Clearly, $z,u \in N_{G_3}(w)$ since $D_z \cap D_u \cap D_w \neq \emptyset$.
However, at most one of $z$ and $u$ can be in $I(w)$, because $D_z \cap D_u \neq \emptyset$ and $I(w)$ is an independent set. Next we claim that 
$u \in I(w)$ and hence $z \in S(w)$ which will complete the proof. 
Since $w \in \{v\} \cup S(v)$, we cannot have $u \in S(w)$ for otherwise $u \in S^*$ (contradicting with the assumption  $u \in I^*$).
Thus, $z \notin I(w)$ and $z \in S(w)$, which completes the proof.
\end{proof}

In what follows, we exploit the properties of $S^*$ and $I^*$ in Observation~\ref{obs-S*I*pro1} and~\ref{obs-S*I*pro2} to show $\textsf{rad}(A_\mathcal{D}[D_v]) = (\frac{1}{\varepsilon})^{O(1)}$.
To this end, we need the following geometric result in \cite{lokshtanov2022subexponential} (Lemma~5.7).
For completeness, we give a short proof here.
\begin{lemma} \label{lem-inttwice}
Let $D_1,\dots,D_m$ be disks such that their intersection $E = \bigcap_{i=1}^m D_i$ is nonempty.
If a disk $D$ satisfies $D \nsubseteq \bigcup_{i=1}^m D_i$ and $D,D_1,\dots,D_m$ are in general position, then the boundary of $D$ intersects the boundary of $E$ at most twice.
\end{lemma}
\begin{proof}
Let $D$ be a disk such that $D \nsubseteq \bigcup_{i=1}^m D_i$.
We first observe that $\partial D \nsubseteq \bigcup_{i=1}^m D_i$ where $\partial D$ is the boundary of $D$.
Assume $\partial D \subseteq \bigcup_{i=1}^m D_i$ for a contradiction.
Pick a point $x \in E = \bigcap_{i=1}^m D_i$.
For every point $y \in \partial D$, note that the segment $\overline{xy}$ must be contained in $\bigcup_{i=1}^m D_i$, i.e., $\overline{xy} \subseteq \bigcup_{i=1}^m D_i$.
Indeed, by our assumption, $y \in D_i$ for some $i \in [m]$.
Then we have $x,y \in D_i$ and thus $\overline{xy}$ is contained in $D_i$ by the convexity of the disk $D_i$.
It is easy to see that $D \subseteq \bigcup_{y \in \partial D} \overline{xy}$ (as for any point $z \in D$ one can find a point $y \in \partial D$ such that $z \in \overline{xy}$).
It follows that $D \subseteq \bigcup_{y \in \partial D} \overline{xy} \subseteq \bigcup_{i=1}^m D_i$, contradicting with the fact $D \nsubseteq \bigcup_{i=1}^m D_i$.

Because $\partial D \nsubseteq \bigcup_{i=1}^m D_i$, there exists a point $x \in \partial D \backslash (\bigcup_{i=1}^m D_i)$.
Now $\partial D \backslash \{x\}$ is homeomorphic to the real line $\mathbb{R}$, and we just view it as the real line.
For each $i \in [m]$, $\partial D \cap D_i$ is either empty or a (closed) connected portion of $\partial D$ that does not contain $x$, and is thus a closed interval on the real line $\partial D \backslash \{x\}$.
Thus, $\partial D \cap E = \bigcap_{i=1}^m (\partial D \cap D_i)$ is the intersection of a set of closed intervals, which is either empty or also a closed interval on $\partial D \backslash \{x\}$.
If $\partial D \cap E = \emptyset$, then $\partial D$ does not intersect $E$ (and hence the boundary of $E$).
If $\partial D \cap E$ is a closed interval on $\partial D \backslash \{x\}$, then $\partial D$ intersects the boundary of $E$ only at the two endpoints of $\partial D \cap E$ (under the general-position assumption).
\end{proof}


Let $\mathcal{D}' = \{D_u: u \in N_{G_3}[v]\}$.
Clearly, $A_{\mathcal{D}'}[D_v] = A_\mathcal{D}[D_v]$ because the faces of the arrangement of $\mathcal{D}$ that are contained in $D_v$ are determined by only the disks that intersect $D_v$.
For each $S \subseteq S^*$, we write $\mathcal{D}_S = \{D_u: u \in \{v\} \cup S\}$, $E_S = \bigcap_{D \in \mathcal{D}_S} D$, and $\mathcal{D}_S^+ = \{D_u: u \in \{v\} \cup S \cup I^*\}$.
Then we have $\mathcal{D}' = \mathcal{D}_{S^*}^+$.
Recall that a \textit{star} is a tree in which there is a node (called the \textit{center} of the star) adjacent to every other node of the tree.
The following observation follows from Lemma~\ref{lem-inttwice}.
\begin{figure}
    \centering
    \includegraphics[height=5.5cm]{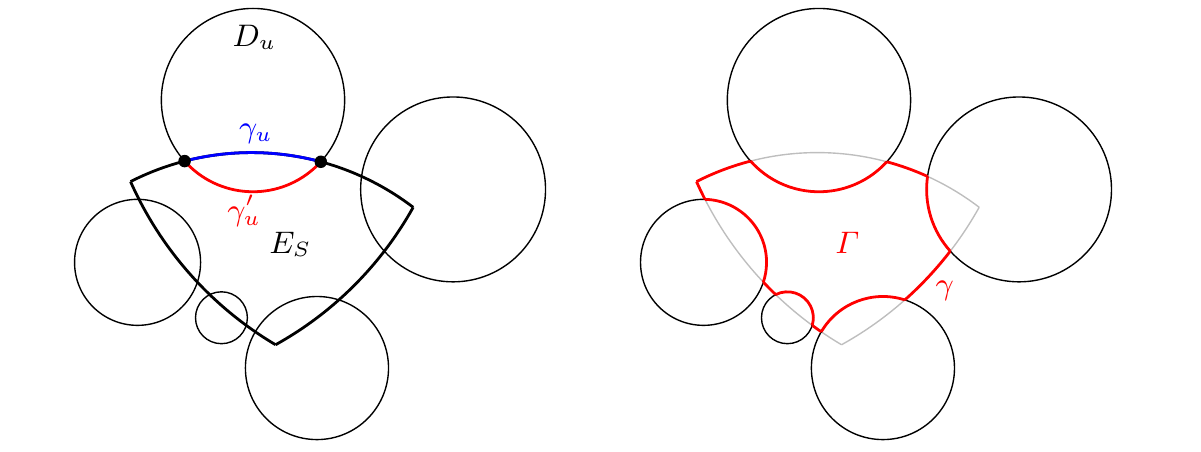}
    \caption{Illustration of the proof of Observation~\ref{obs-star}. Left part illustrates the curves $\gamma_u$ and $\gamma'_u$. Right part illustrates the construction of the closed curve $\gamma$ and the region $\varGamma$ it encloses. The five disks intersecting $E_S$ are the disks $D_u$ for $u \in I^*$, which are disjoint.}
    \label{fig-star}
\end{figure}
\begin{observation} \label{obs-star}
For every $S \subseteq S^*$ such that $E_S \neq \emptyset$, $A_{\mathcal{D}_S^+}[E_S]$ is a star.
\end{observation}
\begin{proof}
Since $I^*$ is an independent set in $G_3$, the disks in $\mathcal{D}_S^+ \backslash \mathcal{D}_S = \{D_u: u \in I^*\}$ are disjoint.
If $E_S \subseteq D_u$ for some $u \in I^*$, then $E_S \cap D_{u'}$ for all $u' \in I^*$ with $u' \neq u$.
In this case, $E_S$ itself is a face of the arrangement of $\mathcal{D}_S^+$ and thus $A_{\mathcal{D}_S^+}[E_S]$ is a star with a single vertex.
So assume no disk in $\{D_u: u \in I^*\}$ contains $E_S$.
We investigate what are the arrangement faces of $\mathcal{D}_S^+$ contained in $E_S$.
Because of the disjointness of the disks in $\{D_u: u \in I^*\}$, each disk $D_u$ for $u \in I^*$ with $D_u \cap E_S \neq \emptyset$ corresponds to a face $F_u = D_u \cap E_S$ in $E_S$.
In addition to these faces, we show that $E_S$ contains exactly one more face.
Since $E_S$ is the intersection of the disks in $\mathcal{D}_S$, it is topologically homeomorphic to a disk.
By Observation~\ref{obs-S*I*pro2}, each disk $D_u$ for $u \in I^*$ is not contained in $\bigcup_{w \in \{v\} \cup S^*} D_w$ and in particular is not contained in $\bigcup_{w \in \{v\} \cup \mathcal{D}_S} D_w$.
Therefore, the boundary $\partial D_u$ of $D_u$ intersects the boundary $\partial E_S$ of $E_S$ at most twice by Lemma~\ref{lem-inttwice}.
By the general position assumption, either $\partial D_u$ does not intersect $\partial E_S$ or $\partial D_u$ intersects $\partial E_S$ twice.
Note that $D_u \nsubseteq E_S$ since $D_u \nsubseteq \bigcup_{w \in \{v\} \cup S^*} D_w$.
Therefore, if $\partial D_u$ does not intersect $\partial E_S$, then $D_u \cap E_S = \emptyset$.
So we only need to consider the disks $D_u$ for $u \in I^*$ such that $\partial D_u$ intersects $\partial E_S$ twice.
Without loss of generality, assume $\partial D_u$ intersects $\partial E_S$ twice for all $u \in I^*$.
For each $u \in I^*$, let $\gamma_u = \partial E_S \cap D_u$ (resp., $\gamma_u' = \partial D_u \cap E_S$) be the part of $\partial E_S$ (resp., $\partial D_u$) contained in $D_u$ (resp., $E_S$).
See the left part of Figure~\ref{fig-star} for an illustration.
Since $D_u$ and $E_S$ are both (topological) disks whose boundaries intersect twice, $\gamma_u$ (resp., $\gamma'_u$) is a connected portion of $\partial E_S$ (resp., $\partial D_u$) and thus is a plane curve.
Also, $\gamma_u$ and $\gamma'_u$ share the same endpoints, which are the two intersection points of $\partial D_u$ and $\partial E_S$.
As the disks in $\{D_u: u \in I^*\}$ are disjoint, the curves $\gamma_u$ (resp., $\gamma_u'$) for all $u \in I^*$ are disjoint.
Now we obtain a curve $\gamma$ from $\partial E_S$ by replacing each $\gamma_u$ with $\gamma'_u$ (we can do this as $\gamma_u$'s are disjoint curves lying on $E_S$ and each $\gamma_u$ shares the same endpoints with $\gamma'_u$).
Note that $\gamma$ is still a simple closed curve, since the curves $\gamma_u'$ are disjoint and each $\gamma'_u$ only intersects $\partial E_S$ on its two endpoints.
By Jordan curve theorem, $\gamma$ bounds a region $\varGamma$ in $\mathbb{R}^2$ that is homeomorphic to a disk.
See the right part of Figure~\ref{fig-star} for an illustration.
We have $\varGamma \cap F_u = \gamma'_u$ for all $u \in I^*$ and $\varGamma \cup (\bigcup_{u \in I^*} F_u) = E_S$.
Observe that $\varGamma$ is a single face of the arrangement of $\mathcal{D}_S^+$ as the boundary of any disk in $\mathcal{D}_S^+$ does not intersect the interior of $\varGamma$.
Therefore, the arrangement faces of $\mathcal{D}_S^+$ contained in $E_S$ are exactly $\varGamma$ and $F_u$'s.
The faces $F_u$ are non-adjacent to each other as the disks $D_u$ are disjoint.
On the other hand, the face $\varGamma$ is adjacent to every $F_u$ as $\varGamma \cap F_u = \gamma'_u$.
As a result, $A_{\mathcal{D}_S^+}[E_S]$ is a star whose center is the vertex corresponding to $\varGamma$. 
\end{proof}

Let $\mathcal{S} = \{S \subseteq S^*: E_S \neq \emptyset\}$.
For each vertex $F$ of $A_{\mathcal{D}'}[D_v]$ (viewed as an arrangement face of $\mathcal{D}'$ contained in $D_v$), define $S_F = \{u \in S^*: F \subseteq D_u\}$.
Note that $S_F \in \mathcal{S}$ for all vertices $F$ of $A_{\mathcal{D}'}[D_v]$, because the existence of $F$ witnesses the non-emptiness of $E_{S_F}$.
An element $S \in \mathcal{S}$ is \textit{maximal} if $S \nsubseteq S'$ for all $S' \in \mathcal{S}$ with $S' \neq S$.
The following observation shows that for a face $F$ either $S_F$ is maximal or $F$ is ``close to" another face $F'$ such that $S_F$ is a proper subset of $S_{F'}$.

\begin{figure}
    \centering
    \includegraphics[height=5cm]{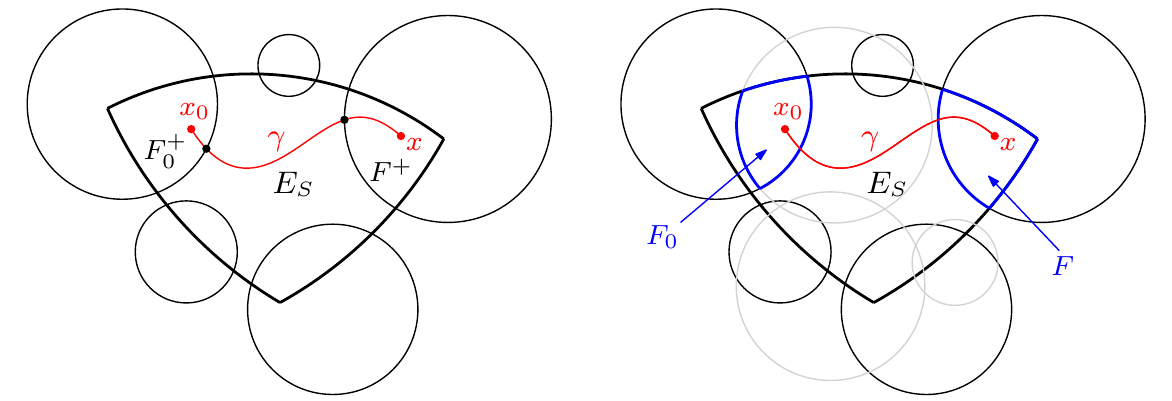}
    \caption{Illustration of the proof of Observation~\ref{obs-maxorinc}. The black disks are those corresponding to $I^*$. The grey disks (right part) are those in $\mathcal{D}' \backslash \mathcal{D}_S^+$. Left part only shows the disks in $\mathcal{D}_S^+$ for illustrating $F^+$ and $F_0^+$. The path $\gamma$ only intersects the boundaries of the black disks twice.}
    \label{fig-faces}
\end{figure}

\begin{observation} \label{obs-maxorinc}
For every vertex $F$ of $A_{\mathcal{D}'}[D_v]$, if $S_F$ is not maximal in $\mathcal{S}$, then there exists another vertex $F'$ of $A_{\mathcal{D}'}[D_v]$ such that $S_F \subsetneq S_{F'}$ and the shortest-path distance between $F$ and $F'$ in $A_{\mathcal{D}'}[D_v]$ is at most $3$.
\end{observation}
\begin{proof}
Let $F$ be a vertex of $A_{\mathcal{D}'}[D_v]$, viewed as an arrangement face of $\mathcal{D}'$ contained in $D_v$.
For convenience, we write $S = S_F$.
Clearly, $F \subseteq E_S$.
Assume $S$ is not maximal in $\mathcal{S}$.
Then there exists $u \in S^* \backslash S$ such that $S \cup \{u\} \in \mathcal{S}$ and $D_u \cap E_S \neq \emptyset$.
Let $F_0$ be a face of the arrangement of $\mathcal{D}'$ contained in $D_u \cap E_S$.
Then $S \subsetneq S_{F_0}$.
Now consider the set $\mathcal{F}$ of arrangement faces of $\mathcal{D}_S^+$ contained in $E_S$.
Since $\mathcal{D}_S^+ \subseteq \mathcal{D}'$, every face of the arrangement of $\mathcal{D}'$ in $E_S$ must be contained in some face in $\mathcal{F}$.
In particular, $F$ (resp., $F_0$) are contained in some face $F^+$ (resp., $F_0^+$) in $\mathcal{F}$.
By Observation~\ref{obs-star}, $A_{\mathcal{D}_S^+}[E_S]$ is a star.
So there exists a path $\pi$ from $F^+$ to $F_0^+$ in $A_{\mathcal{D}_S^+}[E_S]$ of length at most 2.
It follows that for any two points $x \in F^+$ and $x_0 \in F_0^+$, we can draw a plane curve $\gamma$ in $E_S$ connecting $x$ and $x_0$ that intersects at most 3 faces in $\mathcal{F}$ (which are the faces on the path $\pi$) and hence crosses the boundaries of the disks in $\mathcal{D}_S^+$ at most twice (when leaving from one face to the next face).
Let us pick $x$ (resp., $x_0$) in the interior of $F$ (resp., $F_0$), and draw the aforementioned curve $\gamma$.
See Figure~\ref{fig-faces} for an illustration (the left part only shows the arrangement of $\mathcal{D}_S^+$ while the right part shows the arrangement of the entire $\mathcal{D}'$).
Now we go along $\gamma$ from $x$ to $x_0$.
During this procedure, we may cross the boundaries of the disks in $\mathcal{D}'$ (when leaving from one face to another face of the arrangement of $\mathcal{D}'$).
If we cross the boundaries of the disks in $\mathcal{D}'$ at most 3 times, then we visit at most 4 faces of the arrangement of $\mathcal{D}'$, which give us a path in $A_{\mathcal{D}'}[E_S]$ from $F$ to $F_0$ of length at most 3.
In this case, we can simply set $F' = F_0$ to complete the proof.
If we cross the boundaries of the disks in $\mathcal{D}'$ more than 3 times, then in one of the first 3 times we must cross the boundary of a disk $D_w$ for $w \in S^* \backslash S$, because $\gamma$ only crosses the boundaries of the disks in $\mathcal{D}_S^+$ at most twice.
This implies that among the first 4 faces (of the arrangement of $\mathcal{D}'$) we visit, there is one face $F'$ contained in $D_w$.
Now we have a path from $F$ to $F'$ in $A_{\mathcal{D}'}[E_S]$ of length at most 3, and $S \subsetneq S_{F'}$ because $w \in S_{F'} \backslash S$.
\end{proof}

Based on Observation~\ref{obs-maxorinc}, a simple induction argument shows that for every vertex $F$ of $A_{\mathcal{D}'}[D_v]$, there exists a vertex $F^*$ of $A_{\mathcal{D}'}[D_v]$ such that $S_{F^*}$ is maximal in $\mathcal{S}$ and the shortest-path distance between $F$ and $F^*$ in $A_{\mathcal{D}'}[D_v]$ is at most $3 \delta_F$, where $\delta_F = |S^* \backslash S_F|$.
When $\delta_F = 0$, the statement holds because $S_F = S^*$ and thus $S_F$ itself is maximal in $\mathcal{S}$ (so we can simply set $F^* = F$).
Suppose the statement holds for all $\delta_F \leq d-1$ and consider a face $F$ with $\delta_F = d$.
If $S_F$ is maximal in $\mathcal{S}$, we are done as we can simply set $F^* = F$.
Otherwise, by Observation~\ref{obs-maxorinc}, there exists a vertex $F'$ of $A_{\mathcal{D}'}[D_v]$ within distance 3 from $F$ such that $S_F \subsetneq S_{F'}$ and thus $\delta_{F'} \leq \delta_F-1 = d-1$.
By our induction hypothesis, there exists a vertex $F^*$ of $A_{\mathcal{D}'}[D_v]$ within distance $3 (d-1)$ from $F'$ such that $S_{F^*}$ is maximal in $\mathcal{S}$.
So $F^*$ is within distance $3 d$ from $F$.
The statements holds for $\delta_F = d$.
We conclude that every vertex of $A_{\mathcal{D}'}[D_v]$ is within distance $3|S^*|$ from a vertex $F^*$ of $A_{\mathcal{D}'}[D_v]$ such that $S_{F^*}$ is maximal in $\mathcal{S}$.

Now consider a maximal $S \in \mathcal{S}$.
We notice that $A_{\mathcal{D}'}[E_S] = A_{\mathcal{D}_S^+}[E_S]$, because no disk $D_u$ for $u \in S^* \backslash S$ intersects $E_S$, due to the maximality of $S$.
By Observation~\ref{obs-star}, this implies $A_{\mathcal{D}'}[E_S]$ is a star.
We denote by $\mathsf{ctr}(S)$ the center of the star $A_{\mathcal{D}'}[E_S]$.
We claim that every vertex $F$ of $A_{\mathcal{D}'}[D_v]$ is within distance $3|S^*|+1$ from $\mathsf{ctr}(S)$ for some maximal $S \in \mathcal{S}$.
As argued above, $F$ is within distance $3|S^*|$ from a vertex $F^*$ of $A_{\mathcal{D}'}[D_v]$ such that $S_{F^*}$ is maximal in $\mathcal{S}$.
We have $F^* \subseteq E_{S_{F^*}}$, and thus $F^*$ is within distance 1 from $\mathsf{ctr}(S_{F^*})$.
So $F$ is within distance $3|S^*|+1$ from $\mathsf{ctr}(S_{F^*})$.
For every maximal $S \in \mathcal{S}$, let $B_S$ be the ball in $A_{\mathcal{D}'}[D_v]$ centered at $\mathsf{ctr}(S)$ with radius $3|S^*|+1$, that is, the subgraph of $A_{\mathcal{D}'}[D_v]$ induced by all vertices within distance $3|S^*|+1$ from $\mathsf{ctr}(S)$.
Then the set of balls $B_S$ cover all vertices of $A_{\mathcal{D}'}[D_v]$.
We observe the following simple fact.
\begin{fact} \label{fact-smallcover}
If a connected graph $H$ can be covered by $p$ induced subgraphs $H_1,\dots,H_p$, i.e., $V(H) = \bigcup_{i=1}^p V(H_i)$, then we have $\mathsf{rad}(H) \leq \sum_{i=1}^p (2\mathsf{rad}(H_i)+1)$.
\end{fact}
\begin{proof}
We show that the shortest path between any two vertices $s,t \in V(H)$ in $H$ is of length at most $\sum_{i=1}^p (2\mathsf{rad}(H_i)+1)$, which implies $\mathsf{rad}(H) \leq \sum_{i=1}^p (2\mathsf{rad}(H_i)+1)$.
Suppose the shortest path between $s$ and $t$ in $H$ is $\pi = (v_0,v_1,\dots,v_r)$ where $v_0 = s$ and $v_r = t$.
We claim that $|\{v_0,v_1\dots,v_r\} \cap V(H_i)| \leq 2 \mathsf{rad}(H_i)+1$ for all $i \in [p]$.
Assume $|\{v_0,v_1\dots,v_r\} \cap V(H_i)| > 2 \mathsf{rad}(H_i)+1$.
Then there exist $j,j' \in [r] \cup \{0\}$ with $j'-j > 2 \mathsf{rad}(H_i)$ such that $v_j,v_{j'} \in V(H_i)$.
But there exists a path between $v_j$ and $v_{j'}$ in $H_i$ (and thus in $H$) of length at most $2 \mathsf{rad}(H_i)$.
This implies the subpath $(v_j,\dots,v_{j'})$ of $\pi$ is not a shortest path in $H$, which contradicts the fact that $\pi$ is a shortest path.
Therefore, $|\{v_0,v_1\dots,v_r\} \cap V(H_i)| \leq 2 \mathsf{rad}(H_i)+1$ for all $i \in [p]$.
Because $V(H) = \bigcup_{i=1}^p V(H_i)$, we have $r = |\{v_0,v_1\dots,v_r\}| \leq \sum_{i=1}^p |\{v_0,v_1\dots,v_r\} \cap V(H_i)| \leq \sum_{i=1}^p (2\mathsf{rad}(H_i)+1)$.
\end{proof}

It is easy to see that $A_{\mathcal{D}'}[D_v]$ is connected\footnote{To see this, consider two arrangement faces $F$ and $F'$ of $\mathcal{D}'$ contained in $D_v$. One can carefully select one point $x \in F$ and one point $x' \in F'$ such that the segment $\overline{xx'}$ connecting $x$ and $x'$ does not intersect any vertices of the arrangement (i.e., intersection points of the disk boundaries). Note that $\overline{xx'}$ is contained in $D_v$. So if we go along $\overline{xx'}$ from $x$ to $x'$, the faces we visit gives us a path from $F$ to $F'$ in $A_{\mathcal{D}'}[D_v]$.}.
Each ball $B_S$ is of radius $3|S^*|+1$.
We then observe that the number of maximal $S \in \mathcal{S}$ is bounded by $O((|S^*|+1)^2)$, and thus the number of the balls $B_S$ is $O((|S^*|+1)^2)$.
Indeed, if $S \in \mathcal{S}$ and $S' \in \mathcal{S}$ are both maximal and $S \neq S'$, then $E_S \cap E_{S'} = \emptyset$, for otherwise any vertex $u \in S \backslash S'$ satisfies $D_u \cap E_{S'} \neq \emptyset$ (and $S \backslash S' \neq \emptyset$ by the maximality of $S$), contradicting with the maximality of $S'$.
So the regions $E_S$ for maximal $S \in \mathcal{S}$ are disjoint.
Now consider the arrangement of the disks in $\{D_u: u \in \{v\} \cup S^*\}$.
The total number of faces in this arrangement is $O((|S^*|+1)^2)$.
Each region $E_S$ for a maximal $S \in \mathcal{S}$ contains at least one face of this arrangement.
By the disjointness of the regions $E_S$ for maximal $S \in \mathcal{S}$, there can be at most $O((|S^*|+1)^2)$ maximal $S \in \mathcal{S}$.
We have $O((|S^*|+1)^2)$ balls $B_S$ covering $A_{\mathcal{D}'}[D_v]$, each of which is of radius $3|S^*|+1$.
Fact~\ref{fact-smallcover} then shows that $\mathsf{rad}(A_{\mathcal{D}'}[D_v]) = O((|S^*|+1)^3)$.
By Observation~\ref{obs-S*I*pro1}, $|S^*| = (\frac{1}{\varepsilon})^{O(1)}$.
So we finally have $\mathsf{rad}(A_{\mathcal{D}'}[D_v]) = (\frac{1}{\varepsilon})^{O(1)}$, which in turn implies that $\mathsf{rad}(A_\mathcal{D}[D_v]) = (\frac{1}{\varepsilon})^{O(1)}$ and the local radius of $G_3$ is $(\frac{1}{\varepsilon})^{O(1)}$.


\subsection{Putting everything together}

The proof of Theorem~\ref{thm-reduction} (for a closed and triangle-bundle-free graph class $\mathcal{G}$, and for a disk graph $G$) follows immediately from Lemma~\ref{lem-reduction1}, \ref{lem-reduction2}, and \ref{lem-reduction3}.
Specifically, given a disk graph $G$, we apply Lemma~\ref{lem-reduction1} to obtain a graph $G_1$, then apply Lemma~\ref{lem-reduction2} on $G_1$ to obtain the graph $G_2$, and finally apply Lemma~\ref{lem-reduction3} to obtain the graph $G_3$.
The first property of $G_1$ (resp., $G_2$) in Lemma~\ref{lem-reduction1} (resp., Lemma~\ref{lem-reduction2}) guarantees that Lemma~\ref{lem-reduction2} (resp., Lemma~\ref{lem-reduction3}) is applicable to $G_1$ (resp., $G_2$).
Each $G_i$ is an induced subgraph of $G_{i-1}$ and $G_1$ is an induced subgraph of $G$.
So $G_1,G_2,G_3$ are all induced subgraphs of $G$.
Computing $G_1,G_2,G_3$ all take $n^{O(1)}$ time by Lemmata~\ref{lem-reduction1}, \ref{lem-reduction2}, and \ref{lem-reduction3}.

We then set the graph $G'$ in Theorem~\ref{thm-reduction} to be $G_3$, and set the two auxiliary sets $Y_1,Y_2 \subseteq V(G)$ to be the vertex sets of $G_1,G_2$, respectively.
The local radius of $G'$ is $(\frac{1}{\varepsilon})^{O(1)}$ by Lemma~\ref{lem-reduction3}.
Furthermore, given a $(1+\frac{\varepsilon}{4})$-approximation solution for $\textbf{P}_\mathcal{G}$ on $G'$, we can iteratively apply the second conditions in Lemmata~\ref{lem-reduction3}, \ref{lem-reduction2}, and \ref{lem-reduction1} to recover a $(1+\varepsilon)$-approximation solution for $\textbf{P}_\mathcal{G}$ on $G$ in $n^{O(1)}$ time (using the auxiliary sets $Y_1,Y_2$ to identify the intermediate graphs $G_1,G_2$).


\subsection{Analysis for the finite-basis case} \label{sec-finitebasis}
In this section, we give the proof of Theorem~\ref{thm-reduction} for a graph class $\mathcal{G}$ of finite basis (i.e., the last statement in Theorem~\ref{thm-reduction}), assuming $G$ is a disk graph.
The reduction procedure is totally the same as the one for closed and triangle-bundle-free graph classes.
Let $\mathcal{B}$ be the finite class of connected graphs such that $\mathcal{G}$ is the disjoint-union closure of $\mathcal{B}$, and define $k = \max_{H \in \mathcal{B}} |V(H)|$.
Given a disk graph $G$, we iteratively apply Algorithm~\ref{alg-reduction1}, \ref{alg-reduction2}, and~\ref{alg-reduction3} to obtain the graphs $G_1$, $G_2$, and $G_3$, and then set $G' = G_3$.
Since we use exactly the same reduction algorithms (which are independent of the graph class $\mathcal{G}$), the first conditions in Lemma~\ref{lem-reduction1}, \ref{lem-reduction2}, and \ref{lem-reduction3} still hold.
In particular, $G'$ is of $(\frac{1}{\varepsilon})^{O(1)}$ local radius.
So it suffices to prove the second conditions in Lemma~\ref{lem-reduction1}, \ref{lem-reduction2}, and \ref{lem-reduction3}.

\paragraph{Transferring solutions from $G_1$ to $G$.}
We first consider Lemma~\ref{lem-reduction1}.
We want to show that if $S_1$ is a $(1+\varepsilon)$-approximation solution for $\mathbf{P}_\mathcal{G}$ on $G_1$, then $(V(G) \backslash V(G_1)) \cup S_1$ is a $(1+\varepsilon)$-approximation solution for $\mathbf{P}_\mathcal{G}$ on $G$, if we choose the parameter $c(k,\varepsilon)$ properly.
Set $S_1^+ = (V(G) \backslash V(G_1)) \cup S_1$.
Clearly, $S_1^+$ is a feasible solution for $\mathbf{P}_\mathcal{G}$ on $G$, because $G - S_1^+ = G_1 - S_1 \in \mathcal{G}$.
It suffices to show $|S_1^+| \leq (1+\varepsilon) \cdot \mathsf{opt}_\mathcal{G}(G)$.
Let $K_1,\dots,K_r$ be the cliques deleted from $G$ in Algorithm~\ref{alg-reduction1}.
Then $\{K_1,\dots,K_r,V(G_1)\}$ is a partition of $V(G)$.
Consider an optimal solution $S_\mathsf{opt} \subseteq V(G)$ for $\mathbf{P}_\mathcal{G}$ on $G$.
As $G - S_\mathsf{opt} \in \mathcal{G}$, every connected component of $G - S_\mathsf{opt}$ is an element in $\mathcal{B}$ and thus contains at most $k$ vertices.
We have the following simple observation.
\begin{observation} \label{obs-SoptKi}
$|S_\mathsf{opt} \cap (\bigcup_{i=1}^r K_i)| \geq (1 - \frac{k}{c(k,\varepsilon)}) \cdot \sum_{i=1}^r |K_i|$.
\end{observation}
\begin{proof}
It suffices to show $|S_\mathsf{opt} \cap K_i| \geq (1 - \frac{k}{c(k,\varepsilon)}) \cdot |K_i|$ for every $i \in [r]$.
Note that $S_\mathsf{opt}$ must contain all but at most $k$ vertices in $K_i$, for otherwise there exists a clique of size larger than $k$ in $G-S_\mathsf{opt}$, contradicting with the fact that every connected component of $G-S_\mathsf{opt}$ contains at most $k$ vertices.
According to Algorithm~\ref{alg-reduction1}, we have $|K_i| \geq c(k,\varepsilon)$.
Thus, $|K_i| - k \geq (1 - \frac{k}{c(k,\varepsilon)}) \cdot |K_i|$ and hence $|S_\mathsf{opt} \cap K_i| \geq (1 - \frac{k}{c(k,\varepsilon)}) \cdot |K_i|$.
\end{proof}

Let $S_\mathsf{opt}' \supseteq S_\mathsf{opt}$ be the set consisting of all vertices in $S_\mathsf{opt}$ and the vertices of all connected components of $G - S_\mathsf{opt}$ which are not entirely contained in $G_1$.
We have $G - S_\mathsf{opt}' \in \mathcal{G}$ since each connected component of $G - S_\mathsf{opt}'$ is also a connected component of $G - S_\mathsf{opt}$.
Furthermore, $V(G) \backslash V(G_1) \subseteq S_\mathsf{opt}'$, which implies $G_1 - (S_\mathsf{opt}' \cap V(G_1)) = G - S_\mathsf{opt}' \in \mathcal{G}$.
Therefore, $S_\mathsf{opt}' \cap V(G_1)$ is a feasible solution for $\mathbf{P}_\mathcal{G}$ on $G_1$ and thus $|S_\mathsf{opt}' \cap V(G_1)| \geq \mathsf{opt}_\mathcal{G}(G_1) \geq |S_1|/(1+\varepsilon)$.
Then we have
\begin{equation} \label{eq-S1+Sopt'}
    |S_1^+| - |S_\mathsf{opt}'| = |S_1| - |S_\mathsf{opt}' \cap V(G_1)| \leq \varepsilon |S_\mathsf{opt}' \cap V(G_1)| \leq \varepsilon |S_\mathsf{opt} \cap V(G_1)| + \varepsilon |S_\mathsf{opt}' \backslash S_\mathsf{opt}|.
\end{equation}
Next, we try to make $|S_\mathsf{opt}' \backslash S_\mathsf{opt}| \leq \frac{\varepsilon}{1+\varepsilon} \cdot |S_\mathsf{opt} \cap (\bigcup_{i=1}^r K_i)|$ by choosing $c(k,\varepsilon)$ carefully, and combine this with the above inequality to obtain $|S_1^+| \leq (1+\varepsilon) \cdot |S_\mathsf{opt}|$.
We know that $S_\mathsf{opt}'$ consists of $S_\mathsf{opt}$ and the connected components of $G - S_\mathsf{opt}$ which contain at least one vertex in $\bigcup_{i=1}^r K_i$.
The number of these connected components can be at most $|(\bigcup_{i=1}^r K_i) \backslash S_\mathsf{opt}|$, because each of them contains at least one vertex in $(\bigcup_{i=1}^r K_i) \backslash S_\mathsf{opt}$.
The sizes of these connected components are at most $k$ as they all belong to $\mathcal{B}$.
Therefore, we have $|S_\mathsf{opt}' \backslash S_\mathsf{opt}| \leq k |(\bigcup_{i=1}^r K_i) \backslash S_\mathsf{opt}|$.
Note that $|(\bigcup_{i=1}^r K_i) \backslash S_\mathsf{opt}| + |S_\mathsf{opt} \cap (\bigcup_{i=1}^r K_i)| = \sum_{i=1}^r |K_i|$.
By Observation~\ref{obs-SoptKi}, this further implies that $|(\bigcup_{i=1}^r K_i) \backslash S_\mathsf{opt}| \leq \frac{k}{c(k,\varepsilon)}/(1 - \frac{k}{c(k,\varepsilon)}) \cdot |S_\mathsf{opt} \cap (\bigcup_{i=1}^r K_i)|$.
By choosing $c(k,\varepsilon)$ sufficiently large to guarantee $\frac{k^2}{c(k,\varepsilon)}/(1 - \frac{k}{c(k,\varepsilon)}) \leq \frac{\varepsilon}{1+\varepsilon}$, we have $|S_\mathsf{opt}' \backslash S_\mathsf{opt}| \leq k |(\bigcup_{i=1}^r K_i) \backslash S_\mathsf{opt}| \leq \frac{\varepsilon}{1+\varepsilon} \cdot |S_\mathsf{opt} \cap (\bigcup_{i=1}^r K_i)|$.
Combining this with Equation~\ref{eq-S1+Sopt'}, it follows that
    \begin{align*}
        |S_1^+| - |S_\mathsf{opt}| & = (|S_1^+| - |S_\mathsf{opt}'|) + (|S_\mathsf{opt}'| - |S_\mathsf{opt}|) \\
        & \leq \varepsilon |S_\mathsf{opt} \cap V(G_1)| + \varepsilon |S_\mathsf{opt}' \backslash S_\mathsf{opt}| + |S_\mathsf{opt}' \backslash S_\mathsf{opt}| \\
        & = \varepsilon |S_\mathsf{opt} \cap V(G_1)| + (1+\varepsilon) \cdot |S_\mathsf{opt}' \backslash S_\mathsf{opt}| \\
        & \leq \varepsilon |S_\mathsf{opt} \cap V(G_1)| + \varepsilon |S_\mathsf{opt} \cap (\bigcup\nolimits_{i=1}^r K_i)| \\
        & = \varepsilon |S_\mathsf{opt}|.
    \end{align*}
Therefore, we have $|S_1^+| \leq (1+\varepsilon) \cdot |S_\mathsf{opt}|$.

\paragraph{Transferring solutions from $G_2$ to $G_1$.}
Next, we consider Lemma~\ref{lem-reduction2}.
We want to show that if $S_2$ is a $(1+\frac{\varepsilon}{2})$-approximation solution for $\mathbf{P}_\mathcal{G}$ on $G_2$, then $(V(G_1) \backslash V(G_2)) \cup S_2$ is a $(1+\varepsilon)$-approximation solution for $\mathbf{P}_\mathcal{G}$ on $G_1$, if we choose the parameter $c(k,\varepsilon)$ properly.
Let $Z$ be as in Algorithm~\ref{alg-reduction2}.
Then $(V(G_1) \backslash V(G_2)) \cup S_2 = Z \cup S_2$.
Clearly, $Z \cup S_2$ is a feasible solution for $\mathbf{P}_\mathcal{G}$ on $G_1$, because $G_1 - (Z \cup S_2) = G_2 - S_2 \in \mathcal{G}$.
It suffices to show $|Z \cup S_2| \leq (1+\varepsilon) \cdot \mathsf{opt}_\mathcal{G}(G_1)$.

Consider an optimal solution $S_\mathsf{opt} \subseteq V(G_1)$ for $\mathbf{P}_\mathcal{G}$ on $G_1$.
We first bound the size of $Z \backslash S_\mathsf{opt}$.
This part is similar to that in Section~\ref{sec-step2}.
We notice that Observation~\ref{obs-disjointRi} and~\ref{obs-containmany} still hold.
Indeed, Observation~\ref{obs-disjointRi} is independent of the graph class $\mathcal{G}$.
Observation~\ref{obs-containmany} relies on the $T_k$-freeness of $\mathcal{G}$, which is satisfied by our graph class $\mathcal{G}$ here because every connected component of a graph in $\mathcal{G}$ is of size at most $k$.
Then we can use the same argument as the one after Observation~\ref{obs-disjointRi} and~\ref{obs-containmany} to deduce $|Z \backslash S_\mathsf{opt}| \leq |S_\mathsf{opt}|/(c(k,\varepsilon)-k)$.
Now we can apply the reasoning above for transferring solutions from $G_1$ to $G$.
Let $S_\mathsf{opt}' \supseteq S_\mathsf{opt}$ be the set consisting of all vertices in $S_\mathsf{opt}$ and the vertices of all connected components of $G_1 - S_\mathsf{opt}$ which are not entirely contained in $G_2$.
We have $G_1 - S_\mathsf{opt}' \in \mathcal{G}$ since each connected component of $G_1 - S_\mathsf{opt}'$ is also a connected component of $G_1 - S_\mathsf{opt}$.
Furthermore, we have $Z \subseteq S_\mathsf{opt}'$, which implies $G_2 - (S_\mathsf{opt}' \cap V(G_2)) = G_1 - S_\mathsf{opt}' \in \mathcal{G}$.
Therefore, $S_\mathsf{opt}' \cap V(G_2)$ is a feasible solution for $\mathbf{P}_\mathcal{G}$ on $G_2$ and thus $|S_\mathsf{opt}' \cap V(G_2)| \geq \mathsf{opt}_\mathcal{G}(G_2) \geq |S_2|/(1+\frac{\varepsilon}{2})$.
Then,
\begin{align} \label{eq-ZS2Sopt'}
    |Z \cup S_2| - |S_\mathsf{opt}'| &= |S_2| - |S_\mathsf{opt}' \cap V(G_2)| \nonumber \\
    & \leq \frac{\varepsilon}{2} \cdot |S_\mathsf{opt}' \cap V(G_2)| \nonumber \\
    & \leq \frac{\varepsilon}{2} \cdot |S_\mathsf{opt} \cap V(G_2)| + \frac{\varepsilon}{2} \cdot |S_\mathsf{opt}' \backslash S_\mathsf{opt}|.
\end{align}
We know that $S_\mathsf{opt}'$ consists of $S_\mathsf{opt}$ and the connected components of $G_1 - S_\mathsf{opt}$ which contain at least one vertex in $Z$.
The number of these connected components can be at most $|Z \backslash S_\mathsf{opt}|$, because each of them contains at least one vertex in $Z \backslash S_\mathsf{opt}$.
The sizes of these connected components are at most $k$ as they all belong to $\mathcal{B}$.
We have shown that $|Z \backslash S_\mathsf{opt}| \leq |S_\mathsf{opt}|/(c(k,\varepsilon)-k)$.
Therefore, we have $|S_\mathsf{opt}' \backslash S_\mathsf{opt}| \leq k |Z \backslash S_\mathsf{opt}| \leq \frac{k}{c(k,\varepsilon)-k} \cdot |S_\mathsf{opt}|$.
Now choose $c(k,\varepsilon)$ sufficiently large to guarantee that $\frac{k}{c(k,\varepsilon)-k} \leq \frac{\varepsilon}{2+\varepsilon}$.
So $|S_\mathsf{opt}' \backslash S_\mathsf{opt}| \leq \frac{\varepsilon}{2+\varepsilon} \cdot |S_\mathsf{opt}|$.
Combining this with Equation~\ref{eq-ZS2Sopt'}, it follows that
\begin{align*}
    |Z \cup S_2| - |S_\mathsf{opt}| & = (|Z \cup S_2| - |S_\mathsf{opt}'|) + (|S_\mathsf{opt}'| - |S_\mathsf{opt}|) \\
    & \leq \frac{\varepsilon}{2} \cdot |S_\mathsf{opt} \cap V(G_2)| + \frac{\varepsilon}{2} \cdot |S_\mathsf{opt}' \backslash S_\mathsf{opt}| + |S_\mathsf{opt}' \backslash S_\mathsf{opt}| \\
    & = \frac{\varepsilon}{2} \cdot |S_\mathsf{opt} \cap V(G_2)| + \left(1+\frac{\varepsilon}{2} \right) \cdot |S_\mathsf{opt}' \backslash S_\mathsf{opt}| \\
    & \leq \frac{\varepsilon}{2} \cdot |S_\mathsf{opt} \cap V(G_2)| + \frac{\varepsilon}{2} \cdot |S_\mathsf{opt}| \\
    & \leq \varepsilon |S_\mathsf{opt}|.
\end{align*}
Therefore, we have $|Z \cup S_2| \leq (1+\varepsilon) \cdot |S_\mathsf{opt}|$.

\paragraph{Transferring solutions from $G_3$ to $G_2$.}
Finally, we consider Lemma~\ref{lem-reduction3}.
Given a $(1+\frac{\varepsilon}{4})$-approximation solution $S \subseteq V(G_3)$ for $\mathbf{P}_\mathcal{G}$ on $G_3$, we want to compute a $(1+\frac{\varepsilon}{2})$-approximation solution for $\mathbf{P}_\mathcal{G}$ on $G_2$.
Let $\mathcal{X}$, $V_X$, $d_X$ be as in Algorithm~\ref{alg-reduction3}.
We distinguish two cases: $\mathcal{B}$ contains the trivial graph (i.e., the graph of a single verex), and $\mathcal{B}$ does not contain the trivial graph.

\paragraph{[Case 1] $\mathcal{B}$ contains the trivial graph.}
In this case, we can construct an approximation solution for $\mathbf{P}_\mathcal{G}$ on $G_2$ using a method similar to the one in Section~\ref{sec-step3}.
Specifically, we create a set $S' \subseteq V(G_2)$ as follows.
For each $X \in \mathcal{X}$, if $|X \cap S| \leq (c(k,\varepsilon)-k) \cdot d_X$, we let $U_X = X \cap S$.
If $|X \cap S| > (c(k,\varepsilon)-k) \cdot d_X$, we define $U_X$ as the set consisting of all vertices in $X \cap S$ and the vertices of all connected components of $G_3 - S_3$ that contain at least one vertex in $N_{G_2}(X)$.
Then we define $S' = \bigcup_{X \in \mathcal{X}} U_X$.
We first show $S'$ is a feasible solution for $\mathbf{P}_\mathcal{G}$ on $G_2$.
\begin{observation}
If $c(k,\varepsilon) \geq k$, then $G_2 - S' \in \mathcal{G}$.
\end{observation}
\begin{proof}
Suppose $c(k,\varepsilon) \geq k$.
It suffices to show that every connected component of $G_2 - S'$ belongs to $\mathcal{B}$.
We shall show that every connected component of $G_2 - S'$ is either a connected component of $G_3 - S$ or only consists of one vertex; in both cases, the component belongs to $\mathcal{B}$.
Consider a connected component $\varPhi$ of $G_2 - S'$.
We distinguish two cases: $\varPhi$ contains a vertex in $V(G_3)$, and $\varPhi$ does not contain any vertex in $V(G_3)$.

If $\varPhi$ contains a vertex $v \in V(G_3)$, then $v \in V(G_3) \backslash S$, because $S \subseteq S'$ and $V(\varPhi) \cap S' = \emptyset$.
Let $\varPsi$ be the connected component $\varPhi$ of $G_3 - S$ that contains $v$, and we claim that $\varPsi = \varPhi$.
First note that all vertices in $V(\varPsi) \subseteq V(\varPhi)$.
Indeed, by the construction of $S'$, each connected component of $G_3 - S$ is either contained in $S'$ or disjoint from $S'$.
Since $v \in V(\varPsi)$, $S'$ does not contain $\varPsi$ and thus $V(\varPsi) \cap S' = \emptyset$, i.e., $V(\varPsi) \subseteq V(G_2) \backslash S'$.
As $\varPsi$ is connected, it is contained in one connected component of $G_2 - S'$, which must be $\varPhi$.
To see $\varPsi = \varPhi$, it suffices to show $V(\varPhi) \subseteq V(\varPsi)$.
Suppose $V(\varPsi) \subsetneq V(\varPhi)$ for a contradiction.
Then we can find a vertex $u \in V(\varPhi) \backslash V(\varPsi)$ that is neighboring to some vertex $u' \in V(\varPsi)$, because $\varPhi$ is connected.
Let $X \in \mathcal{X}$ such that $u \in X$.
If $|X \cap S| > (c(k,\varepsilon)-k) \cdot d_X$, then $u' \in U_X$, because by our construction $U_X$ contains all connected components of $G_3 - S_3$ that contain at least one vertex in $N_{G_2}(X)$ and hence $U_X$ contains $\varPsi$ as $u' \in V(\varPsi)$.
However, this contradicts with the fact that $V(\varPhi) \cap S' = \emptyset$, since $u' \in V(\varPhi) \cap U_X \subseteq V(\varPhi) \cap S'$.
If $|X \cap S| < (c(k,\varepsilon)-k) \cdot d_X$, then $U_X = X \cap S$.
Note that $u \in X \backslash V_X$, because $u \notin V(G_3)$ (for otherwise $\varPsi$ cannot be a connected component of $G_3 - S$).
It follows that $V_X \subsetneq X$ and $|V_X| = c(k,\varepsilon) \cdot d_X$, according to line~4 of Algorithm~\ref{alg-reduction3}.
We have $d_X \geq 1$ since $u' \in N_{G_2}(X)$.
So $|V_X \backslash S| = |V_X| - |X \cap S| \geq k d_X \geq k$.
Now the vertices in $V_X \backslash S$ and $u'$ for a connected subgraph of $G_3 - S$ of size at least $k+1$, contradicting with the fact that $G_3 - S \in \mathcal{G}$.
This implies $\varPsi = \varPhi$.
Thus, $\varPhi$ is a connected component of $G_3 - S$ and $\varPhi \in \mathcal{B}$.

If $\varPhi$ does not contain any vertex in $V(G_3)$, then we claim that $\varPhi$ only consists of one vertex.
Consider a vertex $v \in V(\varPhi)$.
Assume there exists another vertex $v' \in V(\varPhi)$ such that $v' \in N_{G_2}(v)$.
We have $v' \notin V(G_3)$ as $\varPhi$ does not contain any vertex in $V(G_3)$.
Let $X,X' \in \mathcal{X}$ such that $v \in X$ and $v' \in X'$.
As $X$ and $X'$ consist of false twins, $X \subseteq N_{G_2}(X')$ and $X' \subseteq N_{G_2}(X)$, which implies $|X| \leq d_{X'}$ and $|X'|  $.
Since $v,v' \notin V(G_3)$, we must have $v \in X \backslash V_X$ and $v \in X' \backslash V_{X'}$.
In particular, $V_X \neq X$ and $V_{X'} \neq X'$, which implies $c(k,\varepsilon) \cdot d_X < |X|$ and $c(k,\varepsilon) \cdot d_{X'} < |X'|$.
Now we can deduce $c(k,\varepsilon) \cdot d_X < |X| \leq d_{X'}$ and $c(k,\varepsilon) \cdot d_{X'} < |X'| \leq d_X$.
These two inequalities contradict with each other because $c(k,\varepsilon) \geq k \geq 1$.
Therefore, $\varPhi$ only consists of one vertex.
\end{proof}

Next, we show that $|S'| \leq (1+\frac{\varepsilon}{2}) \cdot \mathsf{opt}_\mathcal{G}(G_2)$, if we properly choose $c(k,\varepsilon)$.
We first compare $|S'|$ with $\mathsf{opt}_\mathcal{G}(G_3)$.
It suffices to compare $|U_X|$ with $|X \cap S|$ for each $X \in \mathcal{X}$.
\begin{observation} \label{obs-compareS'optG3}
If $c(k,\varepsilon) > k$, then we have $|U_X| \leq (1+\frac{k}{c(k,\varepsilon)-k}) \cdot |X \cap S|$ for all $X \in \mathcal{X}$ and hence $|S'| \leq (1+\frac{k}{c(k,\varepsilon)-k}) \cdot |S| \leq (1+\frac{k}{c(k,\varepsilon)-k})(1+\frac{\varepsilon}{4}) \cdot \mathsf{opt}_\mathcal{G}(G_3)$.
\end{observation}
\begin{proof}
If $|X \cap S| \leq (c(k,\varepsilon)-k) \cdot d_X$, then we have $|U_X| = |X \cap S| \leq (1+\frac{k}{c(k,\varepsilon)-k}) \cdot |X \cap S|$.
If $|X \cap S| > (c(k,\varepsilon)-k) \cdot d_X$, then $U_X$ contains $X \cap S$ as well as the connected components of $G_3 - S_3$ that contain at least one vertex in $N_{G_2}(X)$.
The number of such connected components is at most $|N_{G_2}(X)| = d_X$, and each component contains at most $k$ vertices because $G_3 - S_3 \in \mathcal{G}$.
Thus, the total number of vertices in these components is at most $k \cdot d_X$, which implies that $|U_X| \leq |X \cap S| + k \cdot d_X < (1+\frac{k}{c(k,\varepsilon)-k}) \cdot |X \cap S|$, since $|X \cap S| > (c(k,\varepsilon)-k) \cdot d_X$.
Finally, we have $|S'| = \sum_{X \in  \mathcal{X}} |U_X| \leq \sum_{X \in  \mathcal{X}} (1+\frac{k}{c(k,\varepsilon)-k}) \cdot |X \cap S| = (1+\frac{k}{c(k,\varepsilon)-k}) \cdot |S|$.
\end{proof}

Now it suffices to compare $\mathsf{opt}_\mathcal{G}(G_3)$ and $\mathsf{opt}_\mathcal{G}(G_2)$.
Here we need to be careful because it does not necessarily hold that $\mathsf{opt}_\mathcal{G}(G_3) \leq  \mathsf{opt}_\mathcal{G}(G_2)$, as $\mathcal{G}$ is not necessarily induced-subgraph-closed.
Consider an optimal solution $S_\mathsf{opt} \subseteq V(G_2)$ for $\mathbf{P}_\mathcal{G}$ on $G_2$.
We define another set $S_\mathsf{opt}' \supseteq S_\mathsf{opt}$ as follows.
Besides vertices in $S_\mathsf{opt}$, for each $X \in \mathcal{X}$ such that $|X| \geq c(k,\varepsilon)$ and $|X \backslash S_\mathsf{opt}| < k$, we include in $S_\mathsf{opt}'$ the vertices of all connected components of $G_2 - S_\mathsf{opt}$ that contain at least one vertex in $X$.
We have the following observation.
\begin{observation} \label{obs-compareG3G2}
$G_3 - (S_\mathsf{opt}' \cap V(G_3)) \in \mathcal{G}$ and $|S_\mathsf{opt}'| \leq (1+\frac{k^2}{c(k,\varepsilon)-k}) \cdot |S_\mathsf{opt}|$.
\end{observation}
\begin{proof}
To see $G_3 - (S_\mathsf{opt}' \cap V(G_3)) \in \mathcal{G}$, we observe that $G_2 - S_\mathsf{opt}' \in \mathcal{G}$, simply because $G_2 - S_\mathsf{opt} \in \mathcal{G}$ and $S_\mathsf{opt}'$ consists of the vertices in $S_\mathsf{opt}$ and some connected components of $G_2 - S_\mathsf{opt}$.
We then show that each connected component of $G_3 - (S_\mathsf{opt}' \cap V(G_3))$ is a connected component of $G_2 - S_\mathsf{opt}'$ and is thus in $\mathcal{B}$.
Consider a connected component $\varPsi$ of $G_3 - (S_\mathsf{opt}' \cap V(G_3))$.
Clearly, $V(\varPsi) \subseteq V(G_2) \backslash S_\mathsf{opt}'$, so there exists a connected component $\varPhi$ that contains $\varPhi$.
We claim that $\varPsi = \varPhi$.
Assume $\varPsi \subsetneq \varPhi$ for a contradiction.
Then there exists a vertex $v \in V(\varPhi) \backslash V(\varPsi)$ that is neighboring to a vertex $v' \in V(\varPsi)$, as $V(\varPhi)$ is connected.
Now we must have $v \notin V(G_3)$, for otherwise $\varPsi$ cannot be a connected component of $G_3 - (S_\mathsf{opt}' \cap V(G_3))$ (because $v$ is neighboring to $\varPsi$).
Let $X \in \mathcal{X}$ such that $v \in X$.
It follows that $v \in X \backslash V_X$ and thus $V_X \neq X$.
Thus, $|X| > c(k,\varepsilon) \cdot d_X$.
Note that $d_X \geq 1$ because of the existence of $v'$, implying $|X| > c(k,\varepsilon)$.
If $|X \backslash S_\mathsf{opt}| < k$, then $S_\mathsf{opt}'$ contains all vertices in $X$ by our construction, contradicting with the fact that $v \notin S_\mathsf{opt}'$.
If $|X \backslash S_\mathsf{opt}| \geq k$, then the vertices in $X \backslash S_\mathsf{opt}$ and $v'$ form a connected subgraph of $G_2 - S_\mathsf{opt}$, which contradicts with the fact that $G_2 - S_\mathsf{opt} \in \mathcal{G}$.
This implies $\varPsi = \varPhi$ and thus $\varPsi \in \mathcal{B}$.

To see $|S_\mathsf{opt}'| \leq (1+\frac{k^2}{c(k,\varepsilon)-k}) \cdot |S_\mathsf{opt}|$, consider a class $X \in \mathcal{X}$ such that $|X| \geq c(k,\varepsilon)$ and $|X \backslash S_\mathsf{opt}| < k$.
The number of connected components of $G_2 - S_\mathsf{opt}$ that contain at least one vertex in $X$ is at most $k-1$, as the connected component can only contain vertices in $X \backslash S_\mathsf{opt}$ which is of size smaller than $k$.
The total number of vertices in these connected components is at most $(k-1)k < k^2$, as each connected component of $G_2 - S_\mathsf{opt}$ is in $\mathcal{B}$ and hence is of size at most $k$.
On the other hand, $|X \cap S_\mathsf{opt}| = |X| - |X \backslash S_\mathsf{opt}| > c(k,\varepsilon) - k$.
Therefore, for each class $X \in \mathcal{X}$, we include at most $\frac{k^2}{c(k,\varepsilon) - k} \cdot |X \cap S_\mathsf{opt}|$ vertices to $S_\mathsf{opt}'$.
In total, we include $\frac{k^2}{c(k,\varepsilon) - k} \cdot |S_\mathsf{opt}|$ vertices to $S_\mathsf{opt}'$, i.e., $|S_\mathsf{opt}' \backslash S_\mathsf{opt}| \leq \frac{k^2}{c(k,\varepsilon) - k} \cdot |S_\mathsf{opt}|$, which implies $|S_\mathsf{opt}'| \leq (1+\frac{k^2}{c(k,\varepsilon)-k}) \cdot |S_\mathsf{opt}|$.
\end{proof}

The first statement in Observation~\ref{obs-compareG3G2} guarantees that $|S_\mathsf{opt}'| \geq |S_\mathsf{opt}' \cap V(G_3)| \geq \mathsf{opt}_\mathcal{G}(G_3)$.
Then by the second statement in Observation~\ref{obs-compareG3G2} we have
\begin{equation*}
    \mathsf{opt}_\mathcal{G}(G_3) \leq \left(1+\frac{k^2}{c(k,\varepsilon)-k}\right) \cdot |S_\mathsf{opt}| = \left(1+\frac{k^2}{c(k,\varepsilon)-k}\right) \cdot \mathsf{opt}_\mathcal{G}(G_2).
\end{equation*}
Combining the above inequality with Observation~\ref{obs-compareS'optG3}, if $c(k,\varepsilon) > k$, we have
\begin{equation*}
    |S'| \leq \left(1+\frac{k}{c(k,\varepsilon)-k}\right) \left(1+\frac{k^2}{c(k,\varepsilon)-k}\right) \left(1+\frac{\varepsilon}{4}\right) \cdot \mathsf{opt}_\mathcal{G}(G_2).
\end{equation*}
Finally, choosing $c(k,\varepsilon) > k$ sufficiently large such that $(1+\frac{k}{c(k,\varepsilon)-k}) (1+\frac{k^2}{c(k,\varepsilon)-k}) \leq 1 + \frac{\varepsilon}{4+\varepsilon}$, we have $|S'| \leq (1+\frac{\varepsilon}{2}) \cdot \mathsf{opt}_\mathcal{G}(G_2)$.

\paragraph{[Case 2] $\mathcal{B}$ does not contain the trivial graph.}
In this case, we claim $S' = (V(G_2) \backslash V(G_3)) \cup S$ is a $(1+\frac{\varepsilon}{2})$-approximation solution for $\mathbf{P}_\mathcal{G}$ on $G_2$, if we choose the parameter $c(k,\varepsilon)$ properly.
Clearly, $S'$ is a feasible solution for $\mathbf{P}_\mathcal{G}$ on $G_2$ because $G_2 - S' = G_3 - S \in \mathcal{G}$.
Now it suffices to have $|S'| \leq (1+\frac{\varepsilon}{2}) \cdot \mathsf{opt}_\mathcal{G}(G_2)$.
Consider an optimal solution $S_\mathsf{opt}$ for $\mathbf{P}_\mathcal{G}$ on $G_2$.
\begin{observation} \label{obs-|X|-k}
$|X \cap S_\mathsf{opt}| > |X| - k$ for every $X \in \mathcal{X}$.
\end{observation}
\begin{proof}
If $N_{G_2}(X) \subseteq S_\mathsf{opt}$, then we must have $X \subseteq S_\mathsf{opt}$ and thus $|X \cap S_\mathsf{opt}| = |X| > |X| - k$, for otherwise a vertex in $X \backslash S_\mathsf{opt}$ forms a connected component in $G_2 - S_\mathsf{opt}$ that is not in $\mathcal{B}$ (as $\mathcal{B}$ does not contain the trivial graph), contradicting with the fact that $G_2 - S_\mathsf{opt} \in \mathcal{G}$.
If $N_{G_2}(X) \nsubseteq S_\mathsf{opt}$, then $S_\mathsf{opt}$ must contain all but at most $k-1$ vertices in $X$ and thus $|X \cap S_\mathsf{opt}| > |X| - k$, for otherwise $k$ vertices in $X \backslash S_\mathsf{opt}$ and a vertex in $N_{G_2}(X) \backslash S_\mathsf{opt}$ form a connected subgraph in $G_2$ of size $k+1$, contradicting with the fact that $G_2 - S_\mathsf{opt} \in \mathcal{G}$.
\end{proof}

Set $c(k,\varepsilon) \geq k$.
Using the above observation, we now construct another optimal solution $S_\mathsf{opt}' \subseteq V(G_2)$ for $\mathbf{P}_\mathcal{G}$ on $G_2$ such that $V(G_2) \backslash V(G_3) \subseteq S_\mathsf{opt}'$.
For each $X \in \mathcal{X}$, let $U_X \subseteq X$ be a subset such that $X \backslash V_X \subseteq U_X$ and $|U_X| = |X \cap S_\mathsf{opt}|$.
Note that such a subset $U_X$ always exists.
Indeed, since $c(k,\varepsilon) \geq k$, we have either $V_X = X$ (thus $|X \backslash V_X| \leq |X \cap S_\mathsf{opt}|$) or $|X \backslash V_X| \leq |X| - c(k,\varepsilon) \leq |X| - k < |X \cap S_\mathsf{opt}|$ by Observation~\ref{obs-|X|-k}.
Then define $S_\mathsf{opt}' = \bigcup_{X \in \mathcal{X}} U_X$.
Clearly, $|S_\mathsf{opt}'| = |S_\mathsf{opt}|$ as $|U_X| = |X \cap S_\mathsf{opt}|$ for all $X \in \mathcal{X}$.
Furthermore, for every $X \in \mathcal{X}$, since $|X \cap S_\mathsf{opt}'| = |U_X| = |X \cap S_\mathsf{opt}|$, we have $|X \backslash S_\mathsf{opt}'| = |X \backslash S_\mathsf{opt}|$.
This implies $G_2 - S_\mathsf{opt}'$ is isomorphic to $G_2 - S_\mathsf{opt}$, because $\mathcal{X}$ consists of false-twin classes.
Therefore, we have $G_2 - S_\mathsf{opt}' \in \mathcal{G}$.
Finally, $V(G_2) \backslash V(G_3) \subseteq S_\mathsf{opt}'$, since $X \backslash V_X \subseteq U_X$ for all $X \in \mathcal{X}$.
As $S'$ and $S_\mathsf{opt}'$ both contain $V(G_2) \backslash V(G_3)$, we have $|S'| - |S_\mathsf{opt}'| = |S| - |S_\mathsf{opt}' \cap V(G_3)|$.
Note that $G_3 - (S_\mathsf{opt}' \cap V(G_3)) = G_2 - S_\mathsf{opt}' \in \mathcal{G}$, so $S_\mathsf{opt}' \cap V(G_3)$ is a  feasible solution for $\mathbf{P}_\mathcal{G}$ on $G_3$.
Thus, $|S_\mathsf{opt}' \cap V(G_3)| \geq \mathsf{opt}_\mathcal{G}(G_3) \geq |S|/(1+\frac{\varepsilon}{4})$.
It follows that
\begin{equation*}
    |S'| - |S_\mathsf{opt}'| = |S| - |S_\mathsf{opt}' \cap V(G_3)| \leq \frac{\varepsilon}{4} \cdot |S_\mathsf{opt}' \cap V(G_3)| \leq \frac{\varepsilon}{4} \cdot |S_\mathsf{opt}'|.
\end{equation*}
So we have $|S'| \leq (1+\frac{\varepsilon}{4}) \cdot |S_\mathsf{opt}'| = (1+\frac{\varepsilon}{4}) \cdot \mathsf{opt}_\mathcal{G}(G_2) \leq (1+\frac{\varepsilon}{2}) \cdot \mathsf{opt}_\mathcal{G}(G_2)$.

\subsection{When $G$ is not a disk graph} \label{sec-nondisk}
Although we only discussed the proof of Theorem~\ref{thm-reduction} in the case where $G$ is a disk graph, it actually applies to any graph (note that when $G$ is not a disk graph, Theorem~\ref{thm-reduction} does not have any structural requirement for $G'$).
In fact, no matter whether $G$ is a disk graph or not, we can always apply the reductions in Section~\ref{sec-step1}, \ref{sec-step2}, and \ref{sec-step3} to iteratively generate the graphs $G_1$, $G_2$, and $G_3$.
Also, one can check that the procedure of transforming an approximation solution on $G_3$ to $G_2$, $G_1$, and $G$ iteratively does not rely on the fact that $G$ is a disk graph.
Therefore, our reduction works even if $G$ is not a disk graph.


\section{Handling bounded local-radius case}\label{sec:boundedLocalRadius}

In the two following subsections, we present two general theorems to handle the case where we are given a disk graph of bounded local-radius in a robust manner (that is, if the given graph is not a disk graph of bounded local-radius, then we are allowed to output this conclusion). Towards that, we will need the following immediate observation.

\begin{observation}\label{obs:radDiam}
Let $G$ be a disk graph. Then, its local-diameter is at most twice its local-radius.
\end{observation}

We also present a simple observation stating that bounded local-radius implies bounded ply.

\begin{observation}\label{obs:boundedPly}
Let $G$ be a disk graph with local-radius $r$. Then, any realization of $G$ has ply $p\leq 2r+1$.
\end{observation}

\begin{figure}[t]
\begin{center}
\includegraphics[scale=0.8]{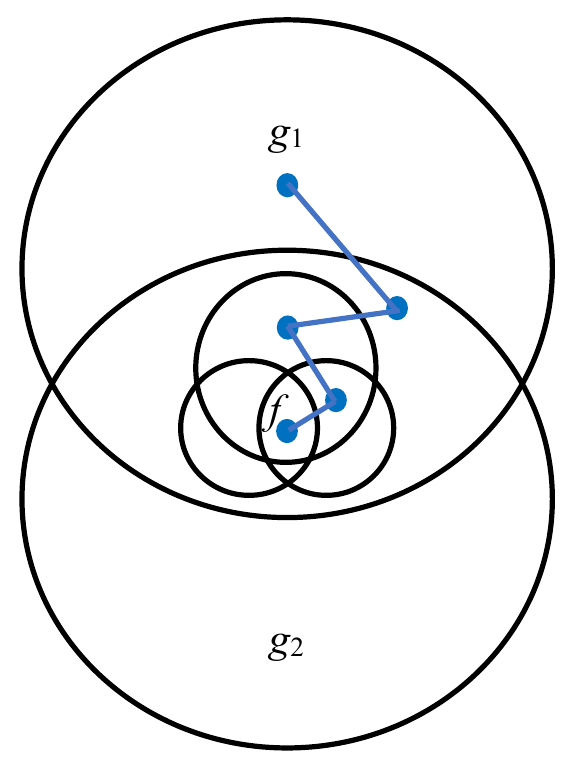}
\end{center}
\caption{An illustration for the proof of Observation \ref{obs:boundedPly}. Here, ${\cal D}'={\cal D}$. The face $f$ is contained in five disks, and its distance to either $g_1$ or $g_2$ is $4$. One possible path of length $4$ between $f$ and $g_1$ is shown in blue.}\label{fig:app1}
\end{figure}

\begin{proof}
Let $\cal D$ be some realization of $G$. Let $f$ be a face of $\cal D$ that is contained in $p$ disks. Let ${\cal D}'$ denote the set of these disks. Then, due to our general position assumption, there exists a disk $D\in{\cal D}$ that contains a face $g$ that does not belong to any other disk in ${\cal D}'$. For every disk $D'\in{\cal D}'\setminus\{D\}$, every path from $f$ to $g$ in $A_{\cal D}$ will have to traverse at least one vertex $v\neq g$ such that the face corresponding to $v$ is contained in $D'$ while the face corresponding to the successor of $v$ on the path is not contained in $D'$ (see Fig.~\ref{fig:app1} for an example). Thus, the diameter of $A_{\cal D}[D]$ is at least $p-1$. Due to Observation \ref{obs:boundedPly}, the local-radius of $G$ is at least $\frac{p-1}{2}$, and hence $p\leq 2r+1$.
\end{proof}

\subsection{Meta-theorem based on the subquadratic grid minor property}

For our first meta-theorem, we need three definitions: the SQGM property, reducibility, and treewidth $\eta$-modulated. We start with the first definition, concerning a property of a graph class.

\begin{definition}[{\bf SQGM Property}, Definition 8 in \cite{DBLP:journals/jacm/FominLS18}]
A graph class $\cal G$ has the {\em subquadratic grid minor (SQGM) property} if there exist constants $\alpha > 0$ and $1 \leq c < 2$ such that, for any $t > 0$, every graph $G \in {\cal G}$, excluding the $t\times t$-grid as a minor, has treewidth at most $\alpha\cdot t^c$. When we need to specify the parameters $\alpha$ and $c$, we say that graph class $\cal G$ has the SQGM property with parameters $\alpha$ and $c$.
\end{definition}

Next, we prove that a lemma that will help us proving that the class of disk graphs of bounded local-radius has the SQGM property. Towards that, we need the following definition and proposition.

\begin{definition}[Definition 5.15 in \cite{lokshtanov2022subexponential}]
Let $H$ be a $3t\times 3t$ grid with a corresponding vertex set $V(H)=\{v_{i,j}: i,j\in\{1,2,\ldots,3t\}\}$. Then, the {\em $H$-interior} is the grid $H[\{v_{i,j}: i,j\in\{t+1,\ldots,2t\}\}]$.
\end{definition}

\begin{proposition}[Observation 5.16 in \cite{lokshtanov2022subexponential}]\label{prop:distances}
Let $G$ be a planar graph that has a $3t\times 3t$ grid $H$ as a minor with a minor model $\varphi$. Let $\widehat{H}$ be the $H$-interior. Let $x,y\in V(\widehat{H})$, $u\in \varphi(x)$ and $v\in \varphi(y)$. Then, $\displaystyle{\frac{\dist_{\widehat{H}}(x,y)}{2}\leq\dist_G(u,v)}$.
\end{proposition}

We now state and prove the aforementioned lemma.

\begin{lemma}\label{lem:grids}
Let $G$ be a disk graph with local-radius $r$. Let $\cal D$ be some realization of $G$, and let $t'\in\mathbb{N}$. If $A_{\cal D}$ contains the grid of size $t'\times t'$ as a minor, then $G$ contains a grid of size $t\times t$ as a minor for $t=\Omega(t'/r)$.
\end{lemma}

\begin{proof}
Let $d$ be the local-diameter of $G$, and note that $d\leq 2r$ (Observation \ref{obs:radDiam}). Let $t''=\lfloor t'/3\rfloor$. Let $H$ denote the $3t''\times 3t''$ grid, and suppose that $A_{\cal D}$ contains $H$ as a minor. Let $\widehat{H}$ denote the $H$-interior, with corresponding vertex set $V(\widehat{H})=\{v_{i,j}: i,j\in\{1,2,\ldots,t''\}\}$. Let $\varphi: V(H)\rightarrow 2^{V(G)}$ denote a minor model of $H$ in $A_{\cal D}$. Notice that $\varphi|_{V(\widehat{H})}$ is a minor model of $\widehat{H}$ in $A_{\cal D}$. Let $d^\star=2d+1$. Let $t=\lfloor t''/3d^\star\rfloor$, and note that $t=\Omega(t'/r)$. We define the following vertices and sets (see Fig.~\ref{fig:app2}):
\begin{itemize}
    \item For every $x,y\in\{1,2,\ldots,t\}$, let  $c_{x,y}=v_{i,j}$ where $i=1+d^\star+(x-1)\cdot (1+3d^\star)$ and $j=1+d^\star+(y-1)\cdot (1+3d^\star)$.
    
    \item For every $x,y\in\{1,2,\ldots,t\}$, let $C_{x,y}=\{v_{i,j'}: j'\in\{j-d^\star,j-d^\star+1,\ldots,j+d^\star\}\}\cup \{v_{i',j}: i'\in\{i-d^\star,i-d^\star+1,\ldots,i+d^\star\}\}$ where $i,j$ are such that $c_{x,y}=v_{i,j}$.
    
    \item For every $x,y\in\{1,2,\ldots,t-1\}$, let $P^{\mathsf{left}}_{x,y}=\{v_{i,j'}: j'\in\{j+d^\star+1,j+d^\star+2,\ldots,j+2d^\star\}\}$ and $P^{\mathsf{bottom}}_{x,y}=\{v_{i',j}: i'\in\{i+d^\star+1,i+d^\star+2,\ldots,i+2d^\star\}\}$ where $i,j$ are such that $c_{x,y}=v_{i,j}$.
\end{itemize}

\begin{figure}[t]
\begin{center}
\includegraphics[scale=1.25]{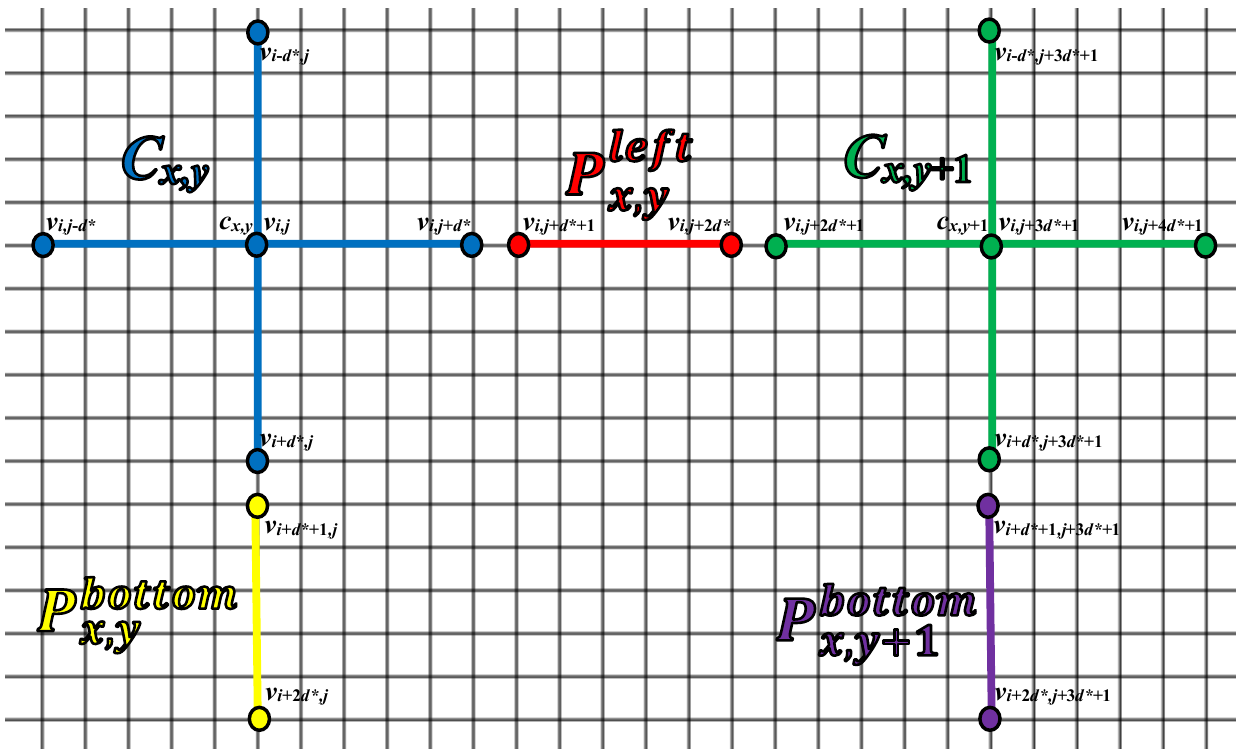}
\end{center}
\caption{An illustration for the proof of Lemma \ref{lem:grids}. Here, $d^\star=5$.}\label{fig:app2}
\end{figure}

Let ${\cal C}=\{C_{x,y}: x,y\in\{1,2,\ldots,t\}\}$, ${\cal P}^{\mathsf{left}}=\{P^{\mathsf{left}}_{x,y}: x,y\in\{1,2,\ldots,t\}\}$ and ${\cal P}^{\mathsf{bottom}}=\{P^{\mathsf{bottom}}_{x,y}: x,y\in\{1,2,\ldots,t\}\}$. Given $U\subseteq V(\widehat{H})$, let $\varphi(U)=\bigcup_{u\in U}\varphi(u)$, and given $u\in V(A_{\cal D})$, let ${\cal D}_u$ denote the set of disks in $\cal D$ that contain the face that $u$ represents, and ${\cal D}_U=\bigcup_{u\in U}{\cal D}_u$. The proof will rely on the three following claims.

\begin{claim}\label{claim:help1}
Let $u,v\in V(\widehat{H})$ such that $\dist_{\widehat{H}}(u,v)\geq d^\star$. Let $u'\in\varphi(u)$ and $v'\in\varphi(v)$. Then, ${\cal D}_{u'}\cap{\cal D}_{v'}=\emptyset$.
\end{claim}

\begin{proof}
Correctness follows from Proposition \ref{prop:distances} (which implies that $\dist_{A_{\cal D}}(u',v')\geq d^\star/2>d$), and since the local-diameter of $G$ is $d$.
\end{proof}

\begin{claim}\label{claim:help2}
Let $u,v\in V(A_{\cal D})$. Let $D_u\in{\cal D}_u$ and $D_v\in{\cal D}_v$. Then, $D_u$ and $D_v$ have non-empty intersection.
\end{claim}

\begin{proof}
Correctness is immediate from the definition of $A_{\cal D}$.
\end{proof}

\begin{claim}\label{claim:help3}
The two following items hold.
\begin{itemize}
    \item For every $x,y,x',y'\in\{1,2,\ldots,t\}$ such that $(x,y)\neq (x',y')$, ${\cal D}_{\varphi(C_{x,y})}\cap {\cal D}_{\varphi(C_{x',y'})}=\emptyset$. Moreover, for every $x,y,x',y'\in\{1,2,\ldots,t\}, \alpha,\beta\in\{\mathsf{left},\mathsf{bottom}\}$ such that $(x,y,\alpha)\neq (x',y',\beta)$, ${\cal D}_{\varphi(P^\alpha_{x,y})}\cap {\cal D}_{\varphi(P^\beta_{x',y'})}=\emptyset$. Lastly, for every $x,y\in\{1,2,\ldots,t\},\alpha\in\{\mathsf{left},\mathsf{bottom}\}$, the only $x',y'\in\{1,2,\ldots,t\}$ such that possibly ${\cal D}_{\varphi(P^\alpha_{x,y})}\cap {\cal D}_{\varphi(C_{x',y'})}\neq\emptyset$ are such that $(x',y')=(x,y)$, or $(x',y')=(x,y+1)$ if $\alpha=\mathsf{left}$, or $(x',y')=(x+1,y)$ if $\alpha=\mathsf{bottom}$.
    \item For every $x,y\in\{1,2,\ldots,t\}$, the vertices represented by the disks in $D_{\varphi(C_{x,y})}$ induce a connected subgraph in $G$; the same holds also for $P^{\mathsf{left}}_{x,y}$ and $P^{\mathsf{bottom}}_{x,y}$.
\end{itemize}
\end{claim}

\begin{proof}
The correctness of the first item directly follows from Claim \ref{claim:help1}, while the correctness of the second item directly follows from Claim \ref{claim:help2}. 
\end{proof}

We now return to the proof of the lemma. For every $x,y\in\{1,2,\ldots,t\}$, initialize $M_{x,y}={\cal D}_{\varphi(C_{x,y})}$.
Perform the following process for every $x\in\{1,2,\ldots,t\}$ and $y\in\{1,2,\ldots,t-1\}$. Let $i,j$ are such that $c_{x,y}=v_{i,j}$. Then, from the definition of a minor model, there exists a path $Q$ in $A_{\cal D}$ from a vertex $u\in \varphi(C_{x,y})$ to a vertex $v\in \varphi(C_{x,y+1})$ such that all of the internal vertices of $Q$ belong to $\varphi(P^{\mathsf{left}}_{x,y})$. Denote $Q=q_1-q_2-\cdots-q_\ell$ where $q_1=u$ and $q_\ell=v$. Let $s$ be the smallest index in $\in\{2,3,\ldots,\ell\}$ such that ${\cal D}_{q_s}\cap {\cal D}_{\varphi(C_{x,y+1})}\neq\emptyset$. Then, add all of the disks in ${\cal D}_{\{q_2,q_2,\ldots,q_{s-1}\}}$ to $M_{x,y}$. Symmetrically, perform the following process for every $x\in\{1,2,\ldots,t-1\}$ and $y\in\{1,2,\ldots,t\}$. Let $i,j$ are such that $c_{x,y}=v_{i,j}$. Then, from the definition of a minor model, there exists a path $Q$ in $A_{\cal D}$ from a vertex $u\in \varphi(C_{x,y})$ to a vertex $v\in \varphi(C_{x+1,y})$ such that all of the internal vertices of $Q$ belong to $\varphi(P^{\mathsf{bottom}}_{x,y})$. Denote $Q=q_1-q_2-\cdots-q_\ell$ where $q_1=u$ and $q_\ell=v$. Let $s$ be the smallest index in $\in\{2,3,\ldots,\ell\}$ such that ${\cal D}_{q_s}\cap {\cal D}_{\varphi(C_{x+1,y})}\neq\emptyset$. Then, add all of the disks in ${\cal D}_{\{q_2,q_2,\ldots,q_{s-1}\}}$ to $M_{x,y}$. 

Let $\widetilde{H}$ be the $t\times t$ grid, corresponding to the vertex set $\{h_{x,y}: x,y\in\{1,2,\ldots,t\}\}$. Define $\psi: V(\widetilde{H})\rightarrow 2^{V(G)}$ such that for every $h_{x,y}\in V(\widetilde{H})$, $\psi(h_{x,y})$ is the set of vertices represented by the disks in $M_{x,y}$. Due to Claims \ref{claim:help2} and \ref{claim:help3}, we derive that $\psi$ is a minor model of $\widetilde{H}$ in $G$. This concludes the proof of the lemma.
\end{proof}

In addition, we will need the following two propositions.

\begin{proposition}[Folklore; see also \cite{smith1998geometric}]\label{prop:twArrangement}
Let $G$ be a geometric (not necessarily disk) graph that has a realization of ply $p$ whose arrangement graph has treewidth $w$. Then, the treewidth of $G$ is $O(w \cdot p)$.
\end{proposition}

\begin{proposition}[\cite{gu2012improved}]\label{prop:planarTW}
Let $G$ be a planar graph of treewidth $w$. Then, $G$ contains the $\lfloor w/5\rfloor \times \lfloor w/5 \rfloor$ grid as a minor.
\end{proposition}

Now, we are ready to prove that the class of disk graphs of bounded local-radius has the SQGM property.

\begin{lemma}\label{lem:diskSQGM}
Let $r\in\mathbb{N}$. Let $\cal G$ be the class of disk graphs of local-radius at most $r$. Then, $\cal G$ has the SQGM property with parameters $\alpha$ and $c$ such that $\alpha$ depends on $r$ and $c=1$.
\end{lemma}

\begin{proof}
Let $G$ be a disk graph of local-radius at most $r$. Let $t>0$ such that $G$ excludes the $t\times t$ grid as a minor. Let $\cal D$ be some realization of $G$. Due to Observation \ref{obs:boundedPly} and Proposition \ref{prop:twArrangement}, to conclude the correctness of the lemma, it suffices to show that the treewidth of $A_{\cal D}$ is at most $\widetilde{\alpha}\cdot t$ for some $\widetilde{\alpha}$ that can depend on $r$. Towards that, let $\beta$ be the constant hidden in the $\Omega$-notation in Lemma \ref{lem:grids}, and choose $\widetilde{\alpha}=10r/\beta$.

Targeting a contradiction, suppose that the treewidth of $A_{\cal D}$ is larger than $\widetilde{\alpha}\cdot t$. By Proposition \ref{prop:planarTW}, this means that $A_{\cal D}$ contains the $t'\times t'$ grid as a minor for $t'=\lfloor\widetilde{\alpha}\cdot t/5\rfloor\geq tr/\beta$. However, by Lemma \ref{lem:grids}, this means that $G$ contains the $t\times t$ grid as a minor, which is a contradiction.
\end{proof}

We now turn to present the second definition, concerning a property of a problem. The satisfaction of this property is much simpler than it might appear in first glance. Indeed, it is often straightforward to show that a problem is reducible by simply picking $\Pi'=\Pi$, $G'=G-X$ for the first item, and $S=S'\cup X$ for the second item. 

\begin{definition}[{\bf Reducibility}, Definition 6 in \cite{DBLP:journals/jacm/FominLS18}]
A graph optimization problem $\Pi$ defined by a predicate $\phi_{\Pi}(G,S)$ is {\em reducible} if there exists a Min/Max-CMSO problem $\Pi'$ with CMSO-expressible property $\phi_{\Pi'}$, a function $f : \mathbb{N} \rightarrow \mathbb{N}$, and a constant $\rho_{\Pi}$ such that
\begin{enumerate}
    \item there is a polynomial-time algorithm that, given a graph $G$ and a set $X \subseteq V(G)$, outputs a graph $G'$ such that $\mathsf{tw}(G') \leq f(\mathsf{tw}(G-X))$ and  $|\mathsf{opt}_{\Pi'}(G')-\mathsf{opt}_{\Pi}(G)| \leq\rho_{\Pi}\cdot |X|$; and
    \item there is a polynomial-time algorithm that, given a graph $G$ and a set $X \subseteq V(G)$, a graph $G'$ and a vertex (edge) set $S' \subseteq V(G')$ ($S' \subseteq E(G')$) such that $\phi_{\Pi'}(G',S')$ holds, outputs $S \subseteq V(G)$ such that $\phi_{\Pi}(G,S)$ is true and $||S|-|S'||\leq\rho_{\Pi}\cdot |X|.$
\end{enumerate}
\end{definition}

The more restrictive property that we will require of problems to satisfy is the following. Here, often (but not necessarily), the output set $X$ is a constant-factor approximate solution to the problem.

\begin{definition}[{\bf Treewidth $\eta$-Modulated}, Definition 7 in \cite{DBLP:journals/jacm/FominLS18}]
For a nonnegative integer $\eta$, a graph optimization problem $\Pi$ is {\em treewidth $\eta$-modulated}, or simply, {\em $\eta$-modulated}, if there is a polynomial-time algorithm that, given a graph $G$, outputs a set $X$ of size $O(\mathsf{opt}_{\Pi}(G))$ such that $\mathsf{tw}(G-X) \leq\eta$.
\end{definition}

The following proposition will make the relevance of the definitions given so far in this subsection clearer.

\begin{proposition}[Theorem 1 in \cite{DBLP:journals/jacm/FominLS18}]\label{prop:meta1}
Let $\Pi$ be an $\eta$-modulated and reducible graph optimization problem. Then, $\Pi$ has an EPTAS on every hereditary graph class  with the SQGM property. More precisely, for any fixed $\varepsilon>0$, $\Pi$ admits a $(1+\varepsilon)$-approximation algorithm with runtime $f(\varepsilon)\cdot |I|^{O(1)}$ (for some computable function $f$) on every hereditary graph class with the SQGM property with parameters $\alpha$ and $c$ where $\alpha$ can depend on $\varepsilon$. 
\end{proposition}

To achieve robustness for our theorem, we need to a robust version of Proposition \ref{prop:appMeta1}. However, with the addition of a simple test, such a version directly follows from the proof of \ref{prop:appMeta1}, as we explain now. Let $G$ be the input graph. Then, the proof of Proposition \ref{prop:appMeta1} consists of the computation of a subset $M\subseteq V(G)$ such that, if $G$ belongs to the graph class under consideration, then: {\em (i)} $|M|\leq\varepsilon\cdot\mathsf{opt}$, and {\em (ii)} $G-M$ is of treewidth $O_\varepsilon(1)$. Then, the algorithm outputs  $M\cup\mathit{S}$ as the solution where $S$ is an \textit{optimal} solution for $G-M$. The computation of $S$ is efficient as long as $G-M$ indeed has treewidth $O_\varepsilon(1)$, but, then, if the computation takes longer than it is supposed to, we can stop it and output that $G$ is ``invalid''; alternatively, we can directly verify that the treewidth of $G-M$ is $O_\varepsilon(1)$ by using known efficient algorithms for this purpose (see \cite{CyganFKLMPPS15}). Note that $|S|\geq(1-\varepsilon)\cdot\mathsf{opt}$ if $G$ is ``valid''. Thus, if $|M|>\frac{\varepsilon}{1-\varepsilon}\cdot|S|$, then we can report ``invalid input''; otherwise, $M\cup\mathit{S}$ is an $(1+O(\varepsilon))$-approximation solution because $|S|\leq\mathsf{opt}$. Thus, we derive the following corollary.

\begin{corollary}\label{cor:meta1Robust}
Let $\Pi$ be an $\eta$-modulated and reducible graph optimization problem. Then, $\Pi$ has a robust EPTAS on every hereditary graph class  with the SQGM property. More precisely, for any fixed $\varepsilon>0$, $\Pi$ admits a $(1+\varepsilon)$-approximation algorithm with runtime $f(\varepsilon)\cdot |I|^{O(1)}$ (for some computable function $f$) on every hereditary graph class with the SQGM property with parameters $\alpha$ and $c$ where $\alpha$ can depend on $\varepsilon$, and where, if the given graph does not belong to the graph class under consideration, then the algorithm may output this conclusion instead of a $(1+\varepsilon)$-approximate solution. 
\end{corollary}

So, combining Lemma \ref{lem:diskSQGM} and Corollary \ref{cor:meta1Robust}, we have the following.

\begin{theorem}\label{thm:meta1}
Let $\Pi$ be an $\eta$-modulated and reducible graph optimization problem. Then, $\Pi$ has a robust EPTAS on the class of disk graphs of local-radius $O_{\varepsilon}(1)$. More precisely, for any fixed $\varepsilon>0$, $\Pi$ admits a  $(1+\varepsilon)$-approximation algorithm with runtime $f(\varepsilon)\cdot |I|^{O(1)}$ (for some computable function $f$) on the class of disk graphs of local-radius $O_{\varepsilon}(1)$, where, if the given graph is not a disk graph of bounded local-radius, then the algorithm may output this conclusion instead of a $(1+\varepsilon)$-approximate solution. 
\end{theorem}

Towards the presentation of our application, we need the following proposition.

\begin{proposition}[\cite{DBLP:journals/jacm/FominLS18}]\label{prop:appMeta1}
Restricted to hereditary graph classes with the SQGM property, every minor-bidimensional, linear-separable problem is reducible and $\eta$-modulated for some fixed constant $\eta$. For example, this includes {\sc Treewidth $\lambda$-Deletion} for any fixed constant $\lambda$ (which encompasses {\sc Vertex Cover} and {\sc Feedback Vertex Set} for $\lambda=0$ and $\lambda=1$, respectively), {\sc Path Deletion}, {\sc Pathwidth $\lambda$-Deletion} for any fixed constant $\lambda$ (which encompasses {\sc Caterpillar Deletion} for $\lambda=1$), {\sc Pseudoforest Deletion}, and {\sc  Cactus Graph Deletion}.
\end{proposition}

Additionally, we prove the following lemma.

\begin{lemma}\label{lem:appMeta1}
Let $\cal G$ be a graph class of finite basis. Let $\Pi$ be {\sc Finite-Type Component Deletion} with target class $\cal G$. Then, $\Pi$ is reducible and $\eta$-modulated where $\eta$ is the maximum size of a connected graph in $\cal G$.
\end{lemma}

\begin{proof}
For reducibility, note that $\Pi$ is a Min-CMSO problem, because we minimize the size of a vertex set $X$ so that each connected component in $G-X$ is isomorphic to some graph in $\cal G$, while connectivity and isomorphism are trivially expressible in CMSO. To see that the two items in the definition of reducibility hold, let $\Pi'=\Pi$,  let $f$ be the identity function and $\rho_\Pi=\eta$. For the first item, pick $G'=G-X$. Clearly, $\mathsf{tw}(G')=f(\mathsf{tw}(G-X))$ and $\mathsf{opt}_{\Pi}(G)\leq \mathsf{opt}_{\Pi}(G')+|X|$. Moreover, suppose that we are given an optimal solution $S$ for $G$. Then, every component of $G-S$ belongs to $\cal G$. So, in $(G-S)-(X\setminus S)$, at most $|X\setminus S|$ components do not belong to $\cal G$, and altogether they contain at most  $|X\setminus S|\cdot(\eta-1)\leq \rho_\Pi\cdot |X|$ many vertices. This implies that $\mathsf{opt}_{\Pi}(G')\leq \mathsf{opt}_{\Pi}(G)+\rho_\Pi\cdot |X|$. For the second item, pick $S=S'$; then, clearly, the condition is satisfied.

Now, we argue that $\Pi$ is $\eta$-modulated. Given a graph $G$, the algorithm works as follows. We initialize $X=\emptyset$. As long as there exists $U\subseteq V(G)$ of size $\eta+1$ such that $G[U]$ is connected, insert $U$ into $X$. Since each considered set $U$ must be hit (i.e., intersected) by any solution, we have that $|T|\leq (\eta+1)\cdot\mathsf{opt}_\Pi$. Since the maximum size of a connected component of $G-X$ is $\eta$, its treewidth is at most $\eta-1$.
\end{proof}

From Theorem \ref{thm:meta1}, Proposition \ref{prop:appMeta1} and Lemma \ref{lem:appMeta1}, we derive the following theorem.

\begin{restatable}{theorem}{metaOne}\label{cor:meta1}
Each of the following problems admits a robust EPTAS on the class of disk graphs of local-radius $O_{\varepsilon}(1)$. More precisely, for any fixed $\varepsilon>0$, each of the following problems admits   $(1+\varepsilon)$-approximation algorithm with runtime $f(\varepsilon)\cdot |I|^{O(1)}$ (for some computable function $f$) on the class of disk graphs of local-radius $O_{\varepsilon}(1)$, where, if the given graph is not a disk graph of bounded local-radius, then the algorithm may output this conclusion instead of a $(1+\varepsilon)$-approximate solution.  Every minor-bidimensional, linear-separable problem. For example, this includes {\sc Treewidth $\lambda$-Deletion} for any fixed constant $\lambda$ (which encompasses {\sc Vertex Cover} and {\sc Feedback Vertex Set} for $\lambda=0$ and $\lambda=1$, respectively), {\sc Path Deletion}, {\sc Pathwidth $\lambda$-Deletion} for any fixed constant $\lambda$ (which encompasses {\sc Caterpillar Deletion} for $\lambda=1$), {\sc Pseudoforest Deletion}, and {\sc Cactus Graph Deletion}. Additionally, each problem in the class of {\sc Finite-Type Component Deletion} problems. For example, this includes the {\sc $\ell$-Component Order Connectivity} problem for any fixed constant $\ell$.
\end{restatable}

We remark that Fomin et al.~\cite{DBLP:journals/jacm/FominLS18} proved that additional problems (sometimes after preprocessing), such as {\sc Max Leaf Spanning Tree} and {\sc Max Internal Spanning Tree}, are also reducible and $\eta$-modulated. However, we preferred to state only some of the problems for our corollary above for bounded local-radius disk graphs, since these are the problems that will yield our applications in Section \ref{sec:application} for general disk graphs. Specifically, for additional applications for bounded local-radius disk graphs, we refer the interested reader to \cite{DBLP:journals/jacm/FominLS18}.

\subsection{Meta-theorem based on Baker's method}

For our second meta-theorem, we will employ Baker's method. At the heart of this method are the following observation concerning treewidth, and the concept of {\em layering} that will be immediately defined.

\begin{proposition}[\cite{robertson1984graph}; implicit in \cite{baker1994approximation}]\label{prop:twByRadius}
A planar graph of radius $r$ has treewidth $O(r)$.
\end{proposition}

\begin{figure}[t]
\begin{center}
\includegraphics[scale=1.1]{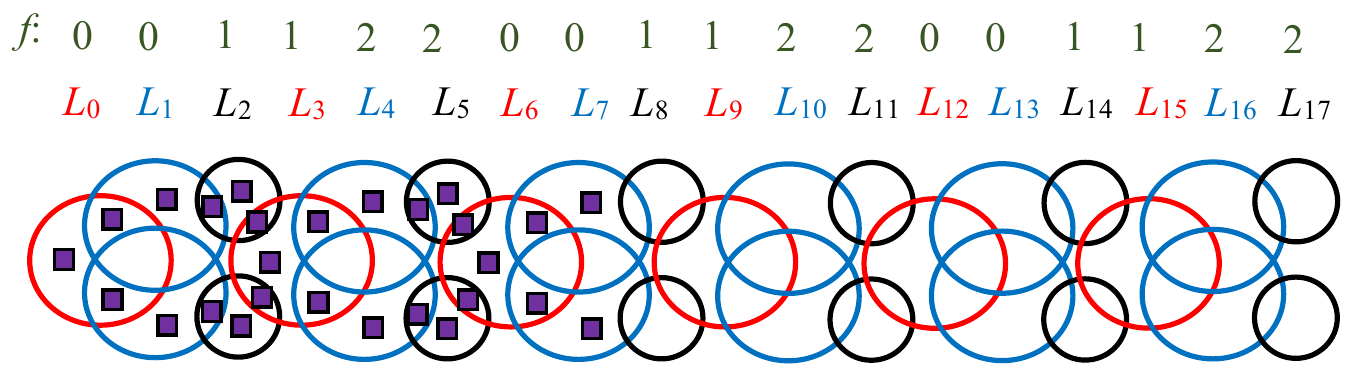}
\end{center}
\caption{An illustration for Definitions \ref{def:piece} and \ref{def:modifiedPiece}. Let $G$ be the intersection graph of the disks shown in the figure. Let $\alpha=3$ and $\beta=2$. Then, the $1$-pieces are $(L_0,L_1,L_2,L_3),(L_2,L_3,\ldots,L_9),(L_8,L_9,\ldots,L_{15})$ and $(L_{14},L_{15},L_{16},L_{17})$. For the $1$-piece ${\cal P}=(L_8,L_9,\ldots,L_{15})$, one possibility (out of several possibilities) to attain a modified arrangement graph for $\mathsf{Ext}({\cal P})$ is to contract the set of vertices corresponding to the faces marked by purple squares.}\label{fig:app3}
\end{figure}

We proceed with the definition of a layering, and then with a definition of ``pieces'''---composed of some consecutive layers---of a layering (see Fig.~\ref{fig:app3}).

\begin{definition}[{\bf Baker Layering}]
Let $G$ be a graph, $s\in V(G)$ and $\alpha,\beta\in\mathbb{N}$. An {\em $(s,\alpha,\beta)$-Baker layering} (for short, {\em $(s,\alpha,\beta)$-layering}) is a pair $({\cal L},f)$ where
${\cal L}=(L_0,L_1,\ldots,L_\ell)$ for $\ell$ being the longest distance of a vertex from $s$, such that each $L_i\in{\cal L}$ (called a {\em layer}) contains all vertices at distance $i$ from $s$, and $f:{\cal L}\rightarrow\{0,1,\ldots,\alpha-1\}$ such that for each $L_i\in{\cal L}$, $f(L_i)$ (called {\em the label of $L_i$}) equals to $ \lfloor\frac{i}{\beta}\rfloor \mod \alpha$.
\end{definition}

\begin{definition}[{\bf Piece}]\label{def:piece}
Let $G$ be a graph with an $(s,\alpha,\beta)$-layering $({\cal L},f)$. Given $q\in\{0,1,\ldots,\alpha-1\}$, a {\em $q$-piece} of $({\cal L},f)$ is any maximal\footnote{The size of each piece, besides possibly the first piece and the last piece, is exactly $(\alpha+1)\beta$.} set of consecutive layers ${\cal P}= (L_i,L_{i+1},\ldots,L_j)\subseteq {\cal L}$ such that there do not exist $x,y,z\in \{i,i+1,\ldots,j\}$ that satisfy $f(L_x)=f(L_z)\neq q$, $x<y<z$, and $f(L_x)\neq f(L_y)$. The {\em extension} of $\cal P$ is $\mathsf{Ext}({\cal P})=(L_0,L_1,\ldots,L_j)$. When $q$ is immaterial, a $q$-piece is simply called a piece.
\end{definition}

Given a disk graph $G$ with a realization $\cal D$ and ${\cal U}\subseteq 2^{V(G)}$, we will use the following abbreviation: $A_{\cal  D}[{\cal U}]=A_{\{D\in{\cal D}: D\ \mathrm{represents\ a\ vertex\ in\ } \bigcup{\cal U}\}}$. In particular, if $\cal U$ is a piece, then this notation concerns the arrangement graph of $G$ restricted to that piece. However, for our analysis of the arrangement graph corresponding to a piece, we will also need to take into account a modified version of the arrangement graph corresponding to an extension of a piece. To this end, we will use the following definition (see Fig.~\ref{fig:app3}).
 
\begin{definition}[{\bf Modified Arrangement Graph of an Extended Piece}]\label{def:modifiedPiece} Let $G$ be a disk graph with an $(s,\alpha,\beta)$-layering $({\cal L},f)$, a piece $\cal P$ and a realization $\cal D$. A {\em modified arrangement graph for $\mathsf{Ext}({\cal P})$ with respect to $\cal D$} is obtained from $A_{\cal  D}[\mathsf{Ext}({\cal P})]$ by contracting a set of vertices $U\subseteq V(A_{\cal  D}[\mathsf{Ext}({\cal P})])$ such that: {\em (i)} $(A_{\cal  D}[\mathsf{Ext}({\cal P})])[U]$ is connected; {\em (ii)} for each vertex $v\in (\bigcup\mathsf{Ext}({\cal P}))\setminus (\bigcup{\cal P})$, $U$ contains at least one vertex that represents a face that is contained in the disk represented by $v$; and {\em (iii)} $U$ does not contain any vertex that represents a face that is contained in a disk represented by a vertex from $\bigcup{\cal P}$.
\end{definition}

The following observation directly follows from the definitions of an arrangement graph and a layering.

\begin{observation}
Let $G$ be a disk graph with an $(s,\alpha,\beta)$-layering $({\cal L},f)$, a piece $\cal P$ and a realization $\cal D$. Then, there exists a modified arrangement graph for $\mathsf{Ext}({\cal P})$ with respect to $\cal D$.
\end{observation}

Due to the above observation, given $G$, $({\cal L},f)$, $\cal P$ and $\cal D$, we can denote by $M_{{\cal P},{\cal D}}$ some modified arrangement graph for $\mathsf{Ext}({\cal P})$ with respect to $\cal D$ (if there are several choices, we pick one arbitrarily). For this graph, we have the following observation.

\begin{observation}\label{obs:boundRadius}
Let $G$ be a disk graph with an $(s,\alpha,\beta)$-layering $({\cal L},f)$, a piece $\cal P$ and a realization $\cal D$. Then, the radius of $M_{{\cal P},{\cal D}}$ is bounded from above by $(|{\cal P}|+1)\cdot (d+1)\leq ((\alpha+1)\beta+1)\cdot (d+1)$ where $d$ is the local-diameter of $G$.
\end{observation}

\begin{proof}
Consider the vertex $u$ resulting from the set $U$ that was contracted to attain $M_{{\cal P},{\cal D}}$. Then, because the local-diameter is $d$,  $u$ can reach the set of all vertices in $M_{{\cal P},{\cal D}}$ that represent a face contained in a disk in $(\bigcup\mathsf{Ext}({\cal P}))\setminus(\bigcup {\cal P})$ using paths of length at most $d$.  Denote the aforementioned set by $R$. Let ${\cal P}=(L_i,L_2,\ldots,L_j)$. Due to the definition of layering and because the local-diameter is $d$, any vertex representing a face contained in a disk in $L_p$, $p\in\{i,i+1,\ldots,j\}$, can be reached from at least one vertex in $R$ by using at most $(p-i+1)\cdot (d+1)\leq |{\cal P}|\cdot (d+1)$ additional edges.
\end{proof}

Next, we reveal the main connection between $A_{\cal  D}[{\cal P}]$ and $M_{{\cal P},{\cal D}}$ that is relevant to our proof.

\begin{observation}\label{obs:minorOf}
Let $G$ be a disk graph with an $(s,\alpha,\beta)$-layering $({\cal L},f)$, a piece $\cal P$ and a realization $\cal D$. Then, $A_{\cal  D}[{\cal P}]$ is a minor of $M_{{\cal P},{\cal D}}$.
\end{observation}

\begin{proof}
For this observation, we notice two arguments. First, by contracting the edge between two vertices representing adjacent faces in an arrangement graph, we merge them into a new vertex representing the face that is the union of the two faces. Second, notice that every face of $A_{\cal  D}[{\cal P}]$ is the union of some non-empty set of faces of $M_{{\cal P},{\cal D}}$ --- in particular, this holds since the set $U$ that was contracted to attain $M_{{\cal P},{\cal D}}$ does not contain any vertex that represents a face contained in a disk corresponding to a vertex in $\bigcup{\cal P}$. So, by contracting some of the edges of $M_{{\cal P},{\cal D}}$, we attain a supergraph of $A_{\cal  D}[{\cal P}]$, and hence the observation follows.
\end{proof}

We will now combine the arguments given so far in this subsection, in order to prove the following lemma.

\begin{lemma}\label{lem:twPiece}
Let $G$ be a disk graph with an $(s,\alpha,\beta)$-layering $({\cal L},f)$ and a piece $\cal P$. Then, the treewidth of $G[\bigcup{\cal P}]$ is bounded by $O(\alpha\cdot \beta\cdot r^2)$ where $r$ is the local-radius of $G$.
\end{lemma}

\begin{proof}
Let $\cal D$ be a realization of $G$.
Due to Observation \ref{obs:boundedPly} and Proposition \ref{prop:twArrangement}, to prove that the treewidth of $G[\bigcup{\cal P}]$ is bounded by $O(\alpha\cdot \beta\cdot r^2)$, it suffices to prove that the treewidth of $A_{\cal D}[{\cal P}]$ is bounded by $O(\alpha\cdot \beta\cdot r)$. Due to Observations \ref{obs:radDiam} and \ref{obs:minorOf} (and since a minor of a graph has treewidth bounded by that of the graph), it further suffices to prove that the treewidth of $M_{{\cal P},{\cal D}}$ is bounded by $O(\alpha\cdot \beta\cdot d)$ where $d$ is the local-diameter of $G$. Since $M_{{\cal P},{\cal D}}$ is a planar graph (being a minor of an arrangement graph, which is planar), the latter claim directly follows from Proposition \ref{prop:twByRadius} and Observation \ref{obs:boundRadius}.
\end{proof}

Having small treewidth is particularly useful due to the following proposition, which is a consequence of, e.g., Courcelle's theorem \cite{courcelle1990monadic} (or, more efficiently, it can be proved using dynamic programming directly as in \cite{LokshtanovPSSZ20,DBLP:series/txcs/DowneyF13}).

\begin{proposition}[\cite{courcelle1990monadic}]\label{prop:algForSmallTW}
Let $\cal G$ be a graph class characterized by a finite set of forbidden (induced or not) subgraphs. Then, $\mathbf{P}_\mathcal{G}$ is in FPT parameterized by the treewidth $\mathsf{tw}$ of the input graph. Specifically, it is solvable in time $f(\mathsf{tw})\cdot |I|$ for some computable function $f$ of $\mathsf{tw}$ and where $|I|$ is the input size.
\end{proposition}

Besides the above lemma and proposition concerning treewidth, towards the proof of our meta-theorem, we will need two additional observations. The first will be helpful for us in proving that our algorithm necessarily returns a solution.

\begin{observation}\label{obs:unionIsSol}
Let $\cal G$ be a graph class characterized by a finite set of forbidden (induced or not) subgraphs. Let $c_{\cal G}$ be the maximum size of a graph in this forbidden set.  Let $G$ be a disk graph with an $(s,\alpha,\beta)$-layering $({\cal L},f)$, a realization $\cal D$, and $i\in\{1,2,\ldots,\alpha\}$. Let $\mathbf{P}$ be the collection of all $i$-pieces. For every ${\cal P}\in\mathbf{P}$, let $S_{\cal P}$ be a solution to $\mathbf{P}_\mathcal{G}$ on the input $G[\bigcup{\cal P}]$. If $\beta\geq c_{\cal G}$, then $S^\star=\bigcup_{{\cal P}\in\mathbf{P}}S_{\cal P}$ is a solution for the input $G$.
\end{observation}

\begin{proof}
The observation directly follows from the definition of a piece by noting that, for $\beta\geq c_{\cal G}$, every (induced or not) subgraph of $G$ that belongs to the forbidden set is fully contained in at at least one $i$-piece.
\end{proof}

The second observation, stated below, will be useful for deriving the desired approximation ratio.

\begin{observation}\label{obs:smallSol}
Let $\cal G$ be a graph class characterized by a finite set of forbidden (induced or not) subgraphs. Let $G$ be a disk graph with an $(s,\alpha,\beta)$-layering $({\cal L},f)$, and a realization $\cal D$. There exists $i\in\{0,1,\ldots,\alpha-1\}$ such that the following holds. Let $\mathbf{P}_i$ be the collection of all $i$-pieces. For every ${\cal P}\in\mathbf{P}_i$, let $S_{\cal P}$ be an optimal solution to $\mathbf{P}_\mathcal{G}$ on the input $G[\bigcup{\cal P}]$. Let $S^\star=\bigcup_{{\cal P}\in\mathbf{P}_i}S_{\cal P}$. Then, $|S^\star|\leq (1+\frac{1}{\alpha})\mathsf{opt}_{\cal G}$.
\end{observation}

\begin{proof}
Consider some optimal solution. By the pigeonhole-principle, there exists $i\in\{1,2,\ldots,\alpha\}$ such that the number of vertices from layers labeled $i$ which $S$ contains is at most $\frac{1}{\alpha}|S|$. Consider this particular $i$. Now, notice that the restriction of $S$ to any ${\cal P}\in\mathbf{P}_i$ is a solution to $\mathbf{P}_\mathcal{G}$ on the input $G[\bigcup{\cal P}]$, and hence its size is bounded from below by $|S_{\cal P}|$. Together, this implies that $|S^\star|=\sum_{{\cal P}\in\mathbf{P}_i}|S_{\cal P}|\leq \sum_{{\cal P}\in\mathbf{P}_i}|S\cap (\bigcup{\cal P})|\leq (1+\frac{1}{\alpha})|S|$, and hence the observation follows.
\end{proof}

We are now ready to prove the main result of this subsection.

\begin{restatable}{theorem}{metaTwo}\label{thm:meta2}
Let $\cal G$ be a graph class characterized by a finite set of forbidden (induced or not) subgraphs. Then, $\mathbf{P}_\mathcal{G}$ admits a robust EPTAS on the class of disk graphs of local-radius $O_{\varepsilon}(1)$. More precisely, for any fixed $\varepsilon>0$, $\mathbf{P}_\mathcal{G}$ admits   $(1+\varepsilon)$-approximation algorithm with runtime $f(\varepsilon)\cdot |I|^{O(1)}$ (for some computable function $f$) on the class of disk graphs of local-radius $O_{\varepsilon}(1)$, where, if the given graph is not a disk graph of bounded local-radius, then the algorithm may output this conclusion instead of a $(1+\varepsilon)$-approximate solution. 
\end{restatable}

\begin{proof}
Fix some $\varepsilon>0$. We fix $\alpha=\frac{1}{\varepsilon}$ and $\beta=c_{\cal G}$ where  $c_{\cal G}$ be the maximum size of a graph in the forbidden set. Given a (presumably) disk graph $G$ of local-radius $r=O_{\varepsilon}(1)$, the algorithm works as follows. Pick some $s\in V(G)$ (arbitrarily), and compute the $(s,\alpha,\beta)$-layering $({\cal L},f)$. Then, or every $i\in\{0,1,\ldots,\alpha-1\}$, it performs the following procedure:
\begin{enumerate}
\item Let ${\bf P}_i$ be the collection of $i$-pieces.
\item For every piece ${\cal P}\in{\bf P}_i$, if the treewidth of $G[\bigcup{\cal P}]$ is not bounded by $c\cdot\alpha\cdot\beta\cdot r^2$ where $c$ is the hidden constant in Lemma \ref{lem:twPiece} (this condition can be tested efficiently using known algorithms~\cite{CyganFKLMPPS15}), then output ``invalid input''; otherwise, use the algorithm in Proposition \ref{prop:algForSmallTW} to solve $\Pi$ optimally on input $G[\bigcup {\cal P}]$, and let $S_{\cal P}$ denote the result.
\item Compute $S_i=\bigcup_{{\cal P}\in{\bf P}_i} S_{\cal P}$.
\end{enumerate}
Afterwards, return the set $S$ of minimum size among the sets $S_1,S_2,\ldots,S_\alpha$. This completes the description of the algorithm.

Due to Observation \ref{obs:unionIsSol}, the algorithm returns a solution, and due to Observation \ref{obs:smallSol}, its size is at most $(1+\varepsilon)\mathsf{opt}_{\cal G}$. Further, due to Lemma \ref{lem:twPiece}, the treewidth of each graph of the form $G[{\bigcup P}]$ considered by the algorithm is bounded by a polynomial in $1/\varepsilon$ and $c_{\cal G}$, which is, in particular, a function of $\varepsilon$ only (since $c_{\cal G}$ is a constant independent of the input). Thus, due to Proposition \ref{prop:algForSmallTW}, it follows that the algorithm runs in time of the form $f(\varepsilon)\cdot |I|^{O(1)}$ where $|I|$ is the input size. 
\end{proof}

\section{Applications}\label{sec:application}

For our application for general disk graphs, we need the following observation.

\begin{observation}\label{obs:graphClasses}
Each of the following graph classes is triangle-bundle free and closed or of finite basis: any graph class of finite basis; forests; pseudoforests; disjoint unions of paths; disjoint unions of caterpillars; $P_t$-free graphs for any $2\leq t\leq 5$; graphs excluding cycles of length at most $\ell$, for any fixed constant $\ell\in\mathbb{N}$, $\ell\geq 3$, and in particular triangle-free graphs; bounded-degree graphs. Further, the last three graph classes can be characterzed by a finite set of forbidden subgraphs.
\end{observation}

\begin{proof}
The observation trivially holds for any graph class of finite basis. So, from now on, we consider only the other graph classes mentioned in the observation.
It should be clear that each of them is induced-subgraph-closed and disjoint-union-closed. Now, observe that each of the classes of forests, disjoint unions of paths, disjoint unions of caterpillars, and graphs excluding cycles of length at most $\ell$, for any fixed constant $\ell\in\mathbb{N}$, $\ell\geq 3$, are $T_1$-free. Further, pseudoforests and $P_t$-free graphs for any $2\leq t\leq 5$ are $T_2$-free (since a triangle bundle of size $2$ is a connected component with more than one cycle and it contains a path on five vertices), and graphs of maximum degree $d$ are $T_{d+1}$-free (in fact, even $T_s$-free for $s=\lceil\frac{d+1}{2}\rceil$). 

Next, consider a graph $G$ with a vertex $u_1$ that has two false twins $u_2$ and $u_3$. Notice that if $|N_G(u_i)\cap N_G(u_j)|\geq 2$ for any $i,j\in\{1,2,3\}$, $i\neq j$, then $G$ contains $K_{2,3}$ (the complete bipartite graph with two vertices on one side and three on the other) as a subgraph and in particular it contains a connected component with at least two (non-disjoint) cycles of length $4$. So, none of the classes of forests, pseudoforests, disjoint unions of paths, disjoint unions of caterpillars, $P_t$-free graphs for any $2\leq t\leq 5$, and graphs excluding cycles of length at most $\ell$, for any fixed constant $\ell\in\mathbb{N}$, $\ell\geq 4$, can contain more than two false twins. (In fact, it can be shown that none of them, expect pseudoforests, can contain even two false twins.) Now, consider a graph $G$ of any of the aforementioned graph classes. The addition of a  degree-$1$ vertex $v$ to $G$ so that $v$ is made adjacent to a vertex that already has three degree-$1$ vertices adjacent to it, keeps the graph within the same graph class. So, we conclude that all of these graph classes are $3$-clone-closed. (In fact, some of them are even $k$-clone-closed for $k\in\{1,2\}$.) We proceed to consider a triangle-free graph $G$. Notice that if the addition of a vertex $v$ to $G$ creates a triangle, then there could not have existed in $G$ any vertex whose neighborhood is the same as $v$ (since then it would have also been part of a triangle). In particular, this means that the class of triangle-free graphs is $1$-clone-closed. Lastly, consider a graph $G$ of maximum degree $d$. Then, $G$ cannot contain more than $d$ false twins, unless they are isolated vertices. In particular, this implies that the class of graphs of maximum degree $d$ is $(d+1)$-clone-closed.
\end{proof}

Combined with Theorems \ref{thm-reduction}, \ref{cor:meta1} and \ref{thm:meta2}, we derive the following.

\combined*

\begin{proof}
Consider an input $G$ to any of the problems mentioned in the theorem. We first apply the algorithm stated in Theorem \ref{thm-reduction} and attain a graph $G'$ such that {\em (i)} if $G$ is a disk graph, then $G'$ is a disk graph of local-radius $r=(\frac{1}{\varepsilon})^{O(1)}$, and {\em (ii)} so that to solve the given instance, it suffices to find a $(1+\frac{\varepsilon}{4})$-approximate solution for $G'$. By Observation \ref{obs:graphClasses}, this application is legal. To find the required $(1+\frac{\varepsilon}{4})$-approximate solution (or output ``invalid input''), we simply apply the algorithm in either Theorem \ref{cor:meta1} or Theorem \ref{thm:meta2} (with parameter $\varepsilon'=\frac{\varepsilon}{4}$).
\end{proof}

\section{Conclusion and future work}\label{sec:conclusion}
We initiated a systematic study of approximation schemes for fundamental optimization problems on disk graphs, which is a graph class that simultaneously generalizes planar graphs and unit-disk graphs.
A general framework was proposed for designing efficient polynomial-time approximation schemes (EPTASes) for vertex-deletion problems on disk graphs, resulting in robust EPTASes for a large variety of fundamental problems on disk graphs.
Our framework is based on a new invariant of disk graphs, local radius, introduced in this paper.
On the one hand, the core of our framework is a reduction for a broad class of vertex-deletion problem from (general) disk graphs to disk graphs of bounded local radius.
On the other hand, we prove that disk graphs of bounded local radius preserve certain nice properties of planar graphs such as the Excluded Grid Minor property and locally bounded treewidth.
Powerful tools for designing approximation schemes on planar graphs, bidimensionality and Baker's approach, were then extended to obtain EPTASes on disk graphs of bounded local radius, which in turn result in EPTASes on general disk graphs by our reduction.
We believe that our framework can possibly have broader applications, because each of the two parts of our framework (the reduction and the results for disk graphs of bounded local radius) can be applied to more problems than the ones we achieve EPTASes for.
For example, our reduction from disk graphs to disk graphs of bounded local radius applies to \textsc{Odd Cycle Transversal} (OCT), a central vertex-deletion problem; we fail to obtain an EPTAS for OCT simply because the question of whether OCT admits a (E)PTAS is a notorious open problem on planar graphs (and thus on disk graphs of bounded local radius).
Also, our results for disk graphs of bounded local radius apply to the \textsc{Treewidth-$\eta$ Deletion} problem, the problem of hitting any fixed forbidden subgraph, and more, but for these problems our reduction does not apply.

We pose some interesting open questions for future work.
The first one is the aforementioned: can we obtain PTASes/EPTASes for more general classes of vertex-deletion problems on disk graphs (by either modifying our framework or applying new approaches)?
One example is the problem of hitting a fixed forbidden subgraph $F$ (or \textsc{$F$-Hitting}).
Baker's technique directly gives EPTAS for \textsc{$F$-Hitting} on planar graphs for any $F$, while on disk graphs our framework only works for some special cases of $F$, i.e., when $F$ is (a subgraph of) a triangle bundle.
Another interesting question is whether one can obtain similar results on broader classes of geometric intersection graphs, such as pseudo-disk graphs, ball graphs, etc.
Finally, we would like to ask whether we can extend our framework to the \textit{weighted} version of vertex-deletion problems, where each vertex has a weight and the goal is to delete a minimum-weighted subset.


\bibliographystyle{plainurl}
\bibliography{my_bib.bib}

\end{document}

%% file: saketPreamble.tex
\usepackage{wrapfig}
\usepackage{bm}

\usepackage{graphicx,color}
\usepackage{empheq}
\usepackage{tcolorbox}
\definecolor{blueish}{rgb}{0.122, 0.435, 0.698}
\definecolor{dagstuhlyellow}{rgb}{0.99,0.78,0.07}
\definecolor{lightgray}{rgb}{0.9,0.9,0.9}
\newtcbox{\colbox}{
size=title,
  nobeforeafter,
  colframe=white,
  colback=blue!5!white,
  arc=10pt,
  tcbox raise base}

\usepackage{xspace}




%% file: main.bbl
\begin{thebibliography}{10}

\bibitem{AgarwalKS98}
Pankaj~K. Agarwal, Marc~J. van Kreveld, and Subhash Suri.
\newblock Label placement by maximum independent set in rectangles.
\newblock {\em Comput. Geom.}, 11(3-4):209--218, 1998.
\newblock \href {https://doi.org/10.1016/S0925-7721(98)00028-5}
  {\path{doi:10.1016/S0925-7721(98)00028-5}}.

\bibitem{aronov2021pseudo}
Boris Aronov, Anirudh Donakonda, Esther Ezra, and Rom Pinchasi.
\newblock On pseudo-disk hypergraphs.
\newblock {\em Computational Geometry}, 92:101687, 2021.

\bibitem{Baker94}
Brenda~S. Baker.
\newblock Approximation algorithms for np-complete problems on planar graphs.
\newblock {\em J. {ACM}}, 41(1):153--180, 1994.
\newblock URL: \url{http://doi.acm.org/10.1145/174644.174650}, \href
  {https://doi.org/10.1145/174644.174650} {\path{doi:10.1145/174644.174650}}.

\bibitem{baker1994approximation}
Brenda~S Baker.
\newblock Approximation algorithms for np-complete problems on planar graphs.
\newblock {\em Journal of the ACM (JACM)}, 41(1):153--180, 1994.

\bibitem{bonamy2021eptas}
Marthe Bonamy, {\'{E}}douard Bonnet, Nicolas Bousquet, Pierre Charbit, Panos
  Giannopoulos, Eun~Jung Kim, Pawel Rzazewski, Florian Sikora, and
  St{\'{e}}phan Thomass{\'{e}}.
\newblock {EPTAS} and subexponential algorithm for maximum clique on disk and
  unit ball graphs.
\newblock {\em J. {ACM}}, 68(2):9:1--9:38, 2021.
\newblock \href {https://doi.org/10.1145/3433160} {\path{doi:10.1145/3433160}}.

\bibitem{BREU19983}
Heinz Breu and David Kirkpatrick.
\newblock Unit disk graph recognition is np-hard.
\newblock {\em Computational Geometry}, 9(1):3--24, 1998.
\newblock Special Issue on Geometric Representations of Graphs.

\bibitem{ChanH12}
Timothy~M. Chan and Sariel Har{-}Peled.
\newblock Approximation algorithms for maximum independent set of pseudo-disks.
\newblock {\em Discret. Comput. Geom.}, 48(2):373--392, 2012.
\newblock \href {https://doi.org/10.1007/s00454-012-9417-5}
  {\path{doi:10.1007/s00454-012-9417-5}}.

\bibitem{courcelle1990monadic}
Bruno Courcelle.
\newblock The monadic second-order logic of graphs. i. recognizable sets of
  finite graphs.
\newblock {\em Information and computation}, 85(1):12--75, 1990.

\bibitem{CyganFKLMPPS15}
Marek Cygan, Fedor~V. Fomin, Lukasz Kowalik, Daniel Lokshtanov, D{\'{a}}niel
  Marx, Marcin Pilipczuk, Michal Pilipczuk, and Saket Saurabh.
\newblock {\em Parameterized Algorithms}.
\newblock Springer, 2015.
\newblock \href {https://doi.org/10.1007/978-3-319-21275-3}
  {\path{doi:10.1007/978-3-319-21275-3}}.

\bibitem{danzer1986losung}
Ludwig Danzer.
\newblock Zur l{\"o}sung des gallaischen problems {\"u}ber kreisscheiben in der
  euklidischen ebene.
\newblock {\em Studia Sci. Math. Hungar}, 21(1-2):111--134, 1986.

\bibitem{DawarGKS06}
Anuj Dawar, Martin Grohe, Stephan Kreutzer, and Nicole Schweikardt.
\newblock Approximation schemes for first-order definable optimisation
  problems.
\newblock In {\em 21th {IEEE} Symposium on Logic in Computer Science {(LICS}
  2006), 12-15 August 2006, Seattle, WA, USA, Proceedings}, pages 411--420.
  {IEEE} Computer Society, 2006.
\newblock \href {https://doi.org/10.1109/LICS.2006.13}
  {\path{doi:10.1109/LICS.2006.13}}.

\bibitem{DBLP:journals/siamcomp/BergBKMZ20}
Mark de~Berg, Hans~L. Bodlaender, S{\'{a}}ndor Kisfaludi{-}Bak, D{\'{a}}niel
  Marx, and Tom~C. van~der Zanden.
\newblock A framework for exponential-time-hypothesis-tight algorithms and
  lower bounds in geometric intersection graphs.
\newblock {\em {SIAM} J. Comput.}, 49(6):1291--1331, 2020.
\newblock \href {https://doi.org/10.1137/20M1320870}
  {\path{doi:10.1137/20M1320870}}.

\bibitem{demaine2005subexponential}
Erik~D. Demaine, Fedor~V. Fomin, Mohammad~Taghi Hajiaghayi, and Dimitrios~M.
  Thilikos.
\newblock Subexponential parameterized algorithms on bounded-genus graphs and
  \emph{H}-minor-free graphs.
\newblock {\em J. {ACM}}, 52(6):866--893, 2005.
\newblock \href {https://doi.org/10.1145/1101821.1101823}
  {\path{doi:10.1145/1101821.1101823}}.

\bibitem{DemaineH05}
Erik~D. Demaine and Mohammad~Taghi Hajiaghayi.
\newblock Bidimensionality: new connections between {FPT} algorithms and ptass.
\newblock In {\em Proceedings of the Sixteenth Annual {ACM-SIAM} Symposium on
  Discrete Algorithms, {SODA} 2005, Vancouver, British Columbia, Canada,
  January 23-25, 2005}, pages 590--601. {SIAM}, 2005.
\newblock URL: \url{http://dl.acm.org/citation.cfm?id=1070432.1070514}.

\bibitem{DBLP:series/txcs/DowneyF13}
Rodney~G. Downey and Michael~R. Fellows.
\newblock {\em Fundamentals of Parameterized Complexity}.
\newblock Texts in Computer Science. Springer, 2013.

\bibitem{EisenstatKM12}
David Eisenstat, Philip~N. Klein, and Claire Mathieu.
\newblock An efficient polynomial-time approximation scheme for steiner forest
  in planar graphs.
\newblock In Yuval Rabani, editor, {\em Proceedings of the Twenty-Third Annual
  {ACM-SIAM} Symposium on Discrete Algorithms, {SODA} 2012, Kyoto, Japan,
  January 17-19, 2012}, pages 626--638. {SIAM}, 2012.
\newblock \href {https://doi.org/10.1137/1.9781611973099.53}
  {\path{doi:10.1137/1.9781611973099.53}}.

\bibitem{Eppstein00}
David Eppstein.
\newblock Diameter and treewidth in minor-closed graph families.
\newblock {\em Algorithmica}, 27(3):275--291, 2000.
\newblock \href {https://doi.org/10.1007/s004530010020}
  {\path{doi:10.1007/s004530010020}}.

\bibitem{ErlebachJS05}
Thomas Erlebach, Klaus Jansen, and Eike Seidel.
\newblock Polynomial-time approximation schemes for geometric intersection
  graphs.
\newblock {\em {SIAM} J. Comput.}, 34(6):1302--1323, 2005.
\newblock \href {https://doi.org/10.1137/S0097539702402676}
  {\path{doi:10.1137/S0097539702402676}}.

\bibitem{FominK10}
Fedor~V. Fomin and Dieter Kratsch.
\newblock {\em Exact Exponential Algorithms}.
\newblock Texts in Theoretical Computer Science. An {EATCS} Series. Springer,
  2010.
\newblock \href {https://doi.org/10.1007/978-3-642-16533-7}
  {\path{doi:10.1007/978-3-642-16533-7}}.

\bibitem{DBLP:journals/jacm/FominLS18}
Fedor~V. Fomin, Daniel Lokshtanov, and Saket Saurabh.
\newblock Excluded grid minors and efficient polynomial-time approximation
  schemes.
\newblock {\em J. {ACM}}, 65(2):10:1--10:44, 2018.

\bibitem{GibsonP10}
Matt Gibson and Imran~A. Pirwani.
\newblock Algorithms for dominating set in disk graphs: Breaking the
  log\emph{n} barrier - (extended abstract).
\newblock In Mark de~Berg and Ulrich Meyer, editors, {\em Algorithms - {ESA}
  2010, 18th Annual European Symposium, Liverpool, UK, September 6-8, 2010.
  Proceedings, Part {I}}, volume 6346 of {\em Lecture Notes in Computer
  Science}, pages 243--254. Springer, 2010.
\newblock \href {https://doi.org/10.1007/978-3-642-15775-2\_21}
  {\path{doi:10.1007/978-3-642-15775-2\_21}}.

\bibitem{Grohe03}
Martin Grohe.
\newblock Local tree-width, excluded minors, and approximation algorithms.
\newblock {\em Comb.}, 23(4):613--632, 2003.
\newblock \href {https://doi.org/10.1007/s00493-003-0037-9}
  {\path{doi:10.1007/s00493-003-0037-9}}.

\bibitem{gu2012improved}
Qian-Ping Gu and Hisao Tamaki.
\newblock Improved bounds on the planar branchwidth with respect to the largest
  grid minor size.
\newblock {\em Algorithmica}, 64(3):416--453, 2012.

\bibitem{HarPeledQ17}
Sariel Har{-}Peled and Kent Quanrud.
\newblock Approximation algorithms for polynomial-expansion and low-density
  graphs.
\newblock {\em {SIAM} J. Comput.}, 46(6):1712--1744, 2017.
\newblock \href {https://doi.org/10.1137/16M1079336}
  {\path{doi:10.1137/16M1079336}}.

\bibitem{HochbaumM85}
Dorit~S. Hochbaum and Wolfgang Maass.
\newblock Approximation schemes for covering and packing problems in image
  processing and {VLSI}.
\newblock {\em J. {ACM}}, 32(1):130--136, 1985.
\newblock \href {https://doi.org/10.1145/2455.214106}
  {\path{doi:10.1145/2455.214106}}.

\bibitem{hochbaum1997approximation}
D.S. Hochbaum and E.B.D.S. Hochbaum.
\newblock {\em Approximation Algorithms for NP-hard Problems}.
\newblock Computer science. PWS Publishing Company, 1997.
\newblock URL: \url{https://books.google.co.in/books?id=E2VRAAAAMAAJ}.

\bibitem{kedem1986union}
Klara Kedem, Ron Livne, J{\'a}nos Pach, and Micha Sharir.
\newblock On the union of jordan regions and collision-free translational
  motion amidst polygonal obstacles.
\newblock {\em Discrete and Computational Geometry}, 1(1):59--71, 1986.

\bibitem{Klein08}
Philip~N. Klein.
\newblock A linear-time approximation scheme for {TSP} in undirected planar
  graphs with edge-weights.
\newblock {\em {SIAM} J. Comput.}, 37(6):1926--1952, 2008.
\newblock \href {https://doi.org/10.1137/060649562}
  {\path{doi:10.1137/060649562}}.

\bibitem{KuhnWZ08}
Fabian Kuhn, Roger Wattenhofer, and Aaron Zollinger.
\newblock Ad hoc networks beyond unit disk graphs.
\newblock {\em Wirel. Networks}, 14(5):715--729, 2008.
\newblock \href {https://doi.org/10.1007/s11276-007-0045-6}
  {\path{doi:10.1007/s11276-007-0045-6}}.

\bibitem{LokshtanovPSSZ20}
Daniel Lokshtanov, Fahad Panolan, Saket Saurabh, Roohani Sharma, and Meirav
  Zehavi.
\newblock Covering small independent sets and separators with applications to
  parameterized algorithms.
\newblock {\em {ACM} Trans. Algorithms}, 16(3):32:1--32:31, 2020.
\newblock \href {https://doi.org/10.1145/3379698} {\path{doi:10.1145/3379698}}.

\bibitem{lokshtanov2022subexponential}
Daniel Lokshtanov, Fahad Panolan, Saket Saurabh, Jie Xue, and Meirav Zehavi.
\newblock Subexponential parameterized algorithms on disk graphs (extended
  abstract).
\newblock In {\em Proceedings of the 2022 Annual ACM-SIAM Symposium on Discrete
  Algorithms (SODA)}, pages 2005--2031. SIAM, 2022.

\bibitem{MustafaR10}
Nabil~H. Mustafa and Saurabh Ray.
\newblock Improved results on geometric hitting set problems.
\newblock {\em Discret. Comput. Geom.}, 44(4):883--895, 2010.
\newblock \href {https://doi.org/10.1007/s00454-010-9285-9}
  {\path{doi:10.1007/s00454-010-9285-9}}.

\bibitem{robertson1984graph}
Neil Robertson and Paul~D Seymour.
\newblock Graph minors. iii. planar tree-width.
\newblock {\em Journal of Combinatorial Theory, Series B}, 36(1):49--64, 1984.

\bibitem{smith1998geometric}
Warren~D Smith and Nicholas~C Wormald.
\newblock Geometric separator theorems and applications.
\newblock In {\em Proceedings 39th Annual Symposium on Foundations of Computer
  Science (FOCS)}, pages 232--243. IEEE, 1998.

\bibitem{thurston1979geometry}
William~P Thurston and John~Willard Milnor.
\newblock The geometry and topology of three-manifolds, 1979.

\bibitem{Leeuwen06}
Erik~Jan van Leeuwen.
\newblock Better approximation schemes for disk graphs.
\newblock In Lars Arge and Rusins Freivalds, editors, {\em Algorithm Theory -
  {SWAT} 2006, 10th ScandinavianWorkshop on Algorithm Theory, Riga, Latvia,
  July 6-8, 2006, Proceedings}, volume 4059 of {\em Lecture Notes in Computer
  Science}, pages 316--327. Springer, 2006.
\newblock \href {https://doi.org/10.1007/11785293\_30}
  {\path{doi:10.1007/11785293\_30}}.

\bibitem{vazirani2001approximation}
Vijay~V Vazirani.
\newblock {\em Approximation algorithms}, volume~1.
\newblock Springer, 2001.

\bibitem{williamson2011design}
David~P Williamson and David~B Shmoys.
\newblock {\em The design of approximation algorithms}.
\newblock Cambridge university press, 2011.

\bibitem{XuB06}
Jinbo Xu and Bonnie Berger.
\newblock Fast and accurate algorithms for protein side-chain packing.
\newblock {\em J. {ACM}}, 53(4):533--557, 2006.
\newblock \href {https://doi.org/10.1145/1162349.1162350}
  {\path{doi:10.1145/1162349.1162350}}.

\end{thebibliography}
